\pdfoutput=1
\documentclass[11pt]{article}
\usepackage{jheppub}

\usepackage{amsmath}
\usepackage{amssymb}
\usepackage{graphicx,color}
\usepackage{amsthm}
\usepackage{subfigure}

\newcommand{\bbold}{\fontencoding{U}\fontfamily{bbold}\selectfont}
\newcommand{\id}{\text{\bbold 1}}

\usepackage{ifpdf}

\usepackage[boxsize=2em,aligntableaux=center]{ytableau}

\usepackage{multirow}

\newcommand{\bC}{\mathbb{C}}
\newcommand{\bF}{\mathbb{F}}
\newcommand{\bH}{\mathbb{H}}

\newcommand{\bP}{\mathbb{P}}
\newcommand{\bR}{\mathbb{R}}

\newcommand{\bZ}{\mathbb{Z}}
\newcommand{\cA}{\mathcal{A}}
\newcommand{\cB}{\mathcal{B}}

\newcommand{\cH}{\mathcal{H}}
\newcommand{\cI}{\mathcal{I}}

\newcommand{\cN}{\mathcal{N}}
\newcommand{\cO}{\mathcal{O}}

\newcommand{\cR}{\mathcal{R}}

\newcommand{\cT}{\mathcal{T}}

\newcommand{\Tr}{\mathrm{Tr\,}}
\newcommand{\ov}{\overline}

\newcommand{\tA}{\tilde{A}}
\newcommand{\tB}{\tilde{B}}
\newcommand{\tN}{\tilde{N}}
\newcommand{\tX}{\tilde{X}}
\newcommand{\tY}{\tilde{Y}}
\newcommand{\tZ}{\tilde{Z}}

\newcommand{\sA}{\mathsf{A}}

\newcommand{\rd}[1]{\mathbf{#1}}

\newcommand{\yss}{\ytableausetup{boxsize=0.5em}}
\newcommand{\ysn}{\ytableausetup{boxsize=2em}}
\newcommand{\symm}{\yss\ydiagram{2}\ysn}
\newcommand{\asymm}{\yss\ydiagram{1,1}\ysn}
\newcommand{\fund}{\yss\ydiagram{1}\ysn}
\newcommand{\adj}{\ensuremath{\text{Adj}}}
\newcommand{\sing}{\ensuremath{\mathbf{1}}}
\newcommand{\ssymm}{\yss\scalebox{0.7}{$\symm$}\ysn}
\newcommand{\sasymm}{\yss\scalebox{0.7}{$\asymm$}\ysn}
\newcommand{\sfund}{\yss\scalebox{0.7}{$\fund$}\ysn}

\DeclareMathOperator{\diag}{diag}

\DeclareMathOperator{\Pf}{Pf}
\DeclareMathOperator{\Sym}{Sym}
\DeclareMathOperator{\gcf}{gcf}
\newcommand{\Sp}{U\!Sp}
\newcommand{\SO}{SO}
\newcommand{\SU}{SU}
\newcommand{\SL}{SL}
\newcommand{\U}{U}
\newcommand{\subSp}{\ensuremath{_{\rm Sp}}}

\newcommand{\subSU}{\ensuremath{_{\rm SU}}}
\newcommand{\symp}{\Omega}
\newcommand{\axiodil}{\tau_{\rm 10d}}

\renewcommand{\arraystretch}{1.2}

\newcommand{\alt}[2]{\texorpdfstring{#1}{#2}}

\def\Label#1{\label{#1}%
  \smash{\hbox to0pt{\raise1ex\hbox{\tiny[#1]}\hss}}}
\def\noLabels{\let\Label=\label}
\def\nobbibitem{\let\bbibitem=\bibitem}
\def\noBibitem{\let\Bibitem=\bibitem}

\newcommand{\be}{\begin{equation}}
\newcommand{\ee}{\end{equation}}
\newcommand{\bea}{\begin{eqnarray}}
\newcommand{\eea}{\end{eqnarray}}
\newcommand{\nn}{\nonumber}
\newcommand{\lp}{\left(}
\newcommand{\rp}{\right)}

\newcommand\varpm{\mathbin{\vcenter{\hbox{%
  \oalign{\hfil$\scriptstyle+$\hfil\cr
          \noalign{\kern-.3ex}
          $\scriptscriptstyle({-})$\cr}%
}}}}
\newcommand\varmp{\mathbin{\vcenter{\hbox{%
  \oalign{$\scriptstyle({+})$\cr
          \noalign{\kern-.3ex}
          \hfil$\scriptscriptstyle-$\hfil\cr}%
}}}}

\title{\centering New $\cN=1$ dualities from orientifold transitions\\
\Large{--- Part I: Field Theory ---}}

\author[a]{I\~naki Garc\'ia-Etxebarria,}
\author[b]{Ben Heidenreich,}
\author[c]{and Timm Wrase}

\affiliation[a]{Theory Group, Physics
  Department, CERN, CH-1211, Geneva 23, Switzerland}
\affiliation[b]{Department of Physics, Cornell University, Ithaca, NY
  14853, USA}
\affiliation[c]{Stanford Institute for Theoretical Physics, Stanford University, Stanford, CA 94305, USA}

\emailAdd{inaki@cern.ch}
\emailAdd{bjh77@cornell.edu}
\emailAdd{timm.wrase@stanford.edu}

\abstract{We report on a broad new class of $\cN=1$ gauge theory dualities which relate the worldvolume gauge theories of D3 branes probing different orientifolds of the same Calabi-Yau singularity. In this paper, we focus on the simplest example of these new dualities, arising from the orbifold singularity $\bC^3/\bZ_3$. We present extensive checks of the duality, including anomaly matching, partial moduli space matching, matching of discrete symmetries, and matching of the superconformal indices between the proposed duals. We then present a related duality for the $dP_1$ singularity, as well as dualities for the $\bF_0$ and $Y^{4,0}$ singularities, illustrating the breadth of this new class of dualities. In a companion paper, we show that certain infinite classes of geometries which include $\bC^3/\bZ_3$ and $dP_1$ all exhibit such dualities, and argue that their ten-dimensional origin is the $\SL(2,\bZ)$ self-duality of type IIB string theory.
}

\begin{document}
\noLabels 
\nobbibitem 
\noBibitem 

\makeatletter
\let\old@fpheader\@fpheader
\renewcommand{\@fpheader}{\old@fpheader\hfill
CERN-PH-TH/2012-293,
SU/ITP-12/33}
\makeatother

\maketitle

\newpage

\section{Introduction}

One of the most remarkable achievements of the study of supersymmetric gauge theories has been the discovery of strong/weak gauge theory dualities, and the correspondent increase in our understanding of (supersymmetric) strongly coupled gauge theories. A prototypical example of such dualities --- and indeed the most important of the $\mathcal{N}=1$ dualities --- is the duality, due to Seiberg~\cite{Seiberg:1994pq, Seiberg:1995ac}, between supersymmetric QCD with $N_C$ colors and $N_F$ flavors and supersymmetric QCD with $N_F - N_C$ colors, $N_F$ flavors, and additional gauge singlets interacting with the dual quarks via the superpotential. The duality, an infrared correspondence between two gauge theories which differ in the ultraviolet, allows the infrared behavior of supersymmetric QCD to be understood for all values of $N_F$ and $N_C$.

The success of Seiberg duality has motivated a thorough study of further dualities of this type, ranging from natural generalizations to $\SO$ and $\Sp$ gauge groups~\cite{Seiberg:1994pq, Intriligator:1995id, Intriligator:1995ne}, generalizations with adjoint matter and a superpotential~\cite{Kutasov:1995ve, Kutasov:1995np, Leigh:1995qp}, models with antisymmetric tensor matter~\cite{Berkooz:1995km, Pouliot:1995me, Csaki:1996eu, Terning:1997jj}, ``self-dual'' theories~\cite{Csaki:1997cu,Karch:1997jp}, to yet more complicated examples (see e.g.~\cite{Intriligator:1995ax, Brodie:1996xm}), in addition to the classifications of various types of confining gauge theories~\cite{Csaki:1996sm, Csaki:1996zb, Grinstein:1997zv, Grinstein:1998bu} where the confined phase has a weakly coupled dual description without a dual gauge  group.

Seiberg duality often admits a very natural and enlightening embedding in string theory, where it appears in the context of brane systems~\cite{Bershadsky:1996gx, Vafa:1997nx, Elitzur:1997fh, Uranga:1998vf}, the duality cascade~\cite{Klebanov:2000nc, Klebanov:2000hb,Strassler:2005qs}, toric duality~\cite{Beasley:2001zp, Feng:2001bn}, and geometric transitions~\cite{Dasgupta:2001fg, Cachazo:2001sg}. (Many of these are related manifestations of the same phenomenon, where Seiberg duality is realized as the effect of passing NS5 branes through each other in a particular T-dual picture~\cite{Feng:2001bn}.) String theory also supplies some contexts where Seiberg duality can be exact~\cite{Strassler:2005qs}. As such, the two fields have enjoyed a largely symbiotic relationship.

Another gauge theory duality of a different nature also enjoys a close
relationship to string theory. Montonen-Olive duality~\cite{Goddard:1976qe,Montonen:1977sn,Osborn:1979tq}, which relates $\mathcal{N}=4$ super-Yang Mills to itself at different couplings, is directly related to the $\SL(2,\bZ)$ self-duality of type IIB
string theory.\footnote{The term ``S-duality'' is sometimes used to
  refer to the entire $\SL(2,\bZ)$ self-duality. In this paper we will
  use it to refer specifically to the $\tau \to - 1/\tau$ element of
  the $\SL(2,\bZ)$ duality of type IIB string theory.} In particular,
the appearance of an $\SL(2,\bZ)$ Montonen-Olive duality in the
worldvolume gauge theory of D3 branes in a flat background follows
from the invariance of the D3 under $\SL(2,\bZ)$, which nonetheless
acts nontrivially on the worldvolume gauge field (as an
electromagnetic duality) and gauge coupling (as a strong/weak
duality), reproducing the action of Montonen-Olive duality on the
gauge theory.

Montonen-Olive duality is different from Seiberg duality in some
important ways. Unlike Seiberg duality, Montonen-Olive duality is
an exact duality, in the sense that it gives various superficially
distinct but quantum equivalent formulations of a single physical
theory, with each of the formulations most suitable for certain values
of the Yang-Mills coupling constant. There is no flow wherein distinct
gauge theories converge on the same infrared fixed point. Indeed, due
to maximal supersymmetry, there is no flow whatsoever, and when one
description is weakly coupled S-dual descriptions are necessarily
strongly coupled (at all energy scales).

In this paper, we construct $\mathcal{N}=1$ analogs of Montonen-Olive
duality.\footnote{
$\mathcal{N}=1$ examples of
  Montonen-Olive duality known in the literature include mass
  deformations of $\mathcal{N}=4$ theories (see
  e.g.~\cite{Argyres:1999xu,Wyllard:2007qw}) and of certain
  $\mathcal{N}=2$ theories with a similar $\SL(2,\bZ)$
  duality~\cite{Leigh:1995ep}. By contrast, our examples are chiral and are not obvious deformations of $\mathcal{N}>1$ theories. Recently,
  there has been a lot of work on $\cN=1$ dualities coming from
  wrapped $M5$-branes
  \cite{Gaiotto:2009we,Bah:2011vv,Bah:2012dg,Beem:2012yn,Gadde:2013fma,Maruyoshi:2013hja,Xie:2013gma,Maruyoshi:2013ega}. 
  While this is not obviously related to our work, it would be very interesting to search for connections.
  }
$\mathcal{N}=1$ gauge theories are interesting for many reasons:
unlike $\mathcal{N}=4$ gauge theories, they can exhibit chirality,
confinement, and dynamical supersymmetry breaking, among other
things. Our new class of $\mathcal{N}=1$ variants of Montonen-Olive
duality provide an interesting counterpoint to known examples of
Seiberg duality, while illuminating the dynamics of interesting gauge
theories via the duality. Moreover, our examples also serve to
illustrate which of the aforementioned features of Montonen-Olive
duality are due to extended supersymmetry, and which persist with less
supersymmetry.

Since Montonen-Olive duality arises from $\SL(2,\bZ)$ acting on the
worldvolume gauge theory of D3 branes in a flat background (with the
possible addition of an $O3$), a natural place to look for analogous
dualities with less supersymmetry is in the worldvolume gauge theory
of D3 branes probing a Calabi-Yau singularity. Since the geometry is
$\SL(2,\bZ)$ invariant, these gauge theories are expected to exhibit
an $\SL(2,\bZ)$ self-duality as well.\footnote{
  Since they often decouple and/or acquire a St\"uckelberg mass, it is
  common to ignore the $\U(1)$ factors in D-brane gauge groups when
  discussing the low energy effective theory. This can be confusing in
  the context of Montonen-Olive duality, since the group $\SU(N)$
  differs from its $SL(2,\bZ)$ dual $SU(N)/\bZ_N$ by a $\bZ_N$ factor
  coming from its center \cite{Goddard:1976qe}, and thus is not
  self-dual. However, this is fully consistent with the $\SL(2,\bZ)$
  self-duality of D3 branes, as the gauge group on $N$ D3 branes in a
  smooth background is actually $\U(N)$, which is self-dual (see for
  example \cite{Kapustin:2006pk}). These global subtleties do not
  affect the class of checks we will perform, so we will freely remove
  the $\U(1)$ factors when convenient, while still talking about
  self-dual theories. Nevertheless, these factors can in principle be
  detected by a more detailed analysis, and in that case we expect
  that the proper inclusion of the $\U(1)$ factors, as dictated by the
  brane construction, will play an important role.}  Unfortunately,
there are virtually no available checks of this conjecture. The class
of checks we perform in this paper, such as anomaly matching and
moduli space matching, are trivial and hence meaningless in the case
of a self-duality.

Fortunately, other types of Montonen-Olive duality are possible. By
placing $k$ D3 branes atop an O3 plane in flat space, one obtains an $\mathcal{N}=4$
$\SO(2k)$, $\SO(2k+1)$, or $\Sp(2k)$ gauge theory (depending on the
type of O3 plane). Whereas $\SO(2k)$ is again self-dual under
Montonen-Olive duality, $\SO(2k+1)$ and $\Sp(2k)$ are exchanged under
the duality due to the S-duality transformation properties of the
respective O3 planes~\cite{Witten:1998xy}. Thus, in order to construct
an $\mathcal{N}=1$ analog, we will consider the worldvolume
gauge theory of D3 branes probing an orientifolded Calabi-Yau singularity, where
$\SL(2,\bZ)$ can act nontrivially on the orientifold plane, leading to dual
theories with distinct gauge groups.

While such a construction generally involves collapsed O7 planes, rather than O3 planes, and the appearance of fractional branes at small volume further complicates the situation, we argue in a companion paper~\cite{transitions2} that S-duality nonetheless acts simply on the entire system. Analogously to the $\cN = 4$ case discussed above, we argue that the collapsed O7 planes undergo an ``orientifold transition'' at strong coupling, exchanging O7$^-$ and O7$^+$ planes while emitting / absorbing a number of fractional branes during the process. Understanding such a transition is one important motivation for our work, but we defer further details to~\cite{transitions2}.

There are numerous additional motivations for studying duality in this context. While worldvolume gauge theories on D3 branes probing (toric) singularities have been exhaustively studied, orientifolded singularities have received comparatively little attention. Systematic tools for the construction of many examples are available~\cite{Franco:2007ii}, whereas very few examples have been studied in any detail (see for example \cite{Franco:2010jv}).
Furthermore, the gauge theories we study are highly nontrivial chiral gauge theories with tree-level superpotentials, tensor matter, and a nontrivial flow. Depending on the singularity and the number of D3 branes, a range of interesting IR behavior arises. In particular in the limited sample of models we analyze, we find
a runaway superpotential, confinement with chiral symmetry breaking, a free magnetic phase, or a nontrivial superconformal fixed point.

These gauge theories are also interesting from the point of view of moduli stabilization, as the nonperturbative dynamics of these gauge theories for sufficiently low $N$ can lift D3 brane moduli and potentially K\"{a}hler moduli as well. Indeed, a number of interesting Calabi-Yau singularities correspond to rigid shrinking divisors, whereas blown-up versions of these have played an important role in stabilizing K\"{a}hler moduli in geometric compactifications of type IIB string theory \cite{Kachru:2003aw,Balasubramanian:2005zx}. Some recent results hint that an AdS/CFT description of the dynamics may be possible~\cite{Heidenreich:2010ad}, though pitfalls abound due to the necessity of low $N$ in this context to obtain gauge theories which are not approximately superconformal.

The outline of our paper is as follows. In~\S\ref{sec:MOduality}, we
review some basic facts about Montonen-Olive duality which illustrate
how it is distinct from Seiberg duality. In~\S\ref{sec:C3Z3fieldtheory}, we present the simplest example of a new
duality, relating two possible gauge theories for D3 branes probing
the orientifolded $\bC^3/\bZ_3$ singularity. We present several
nontrivial consistency checks and discuss an example of the
duality. In~\S\ref{sec:SCI}, we compute the superconformal indices for
the proposed dual gauge theories and show that they match (up to the
limits of our computational resources), a very nontrivial check of the
proposed duality. In~\S\ref{sec:infrared}, we discuss the infrared
physics of these gauge theories using Seiberg duality. In~\S\ref{sec:moreEx}, we discuss further examples of the duality coming
from different ten-dimensional geometries, with particular attention
to the $dP_1$ singularity. The corresponding gauge theories can be
blown down to recover the $\bC^3/\bZ_3$ gauge theories, and exhibit
interesting behavior at low $N$. We also briefly discuss dualities
which arise in the $\bF_0$ and $Y^{4,0}$ geometries, some of which
appear to have a different origin in string theory, unrelated to
$\SL(2,\bZ)$. We leave a detailed treatment of these dualities to a
future work. We present our conclusions in~\S\ref{sec4:conclusions}.

We provide several appendices for the reader's benefit. In
appendix~\ref{app:quiverfolds} we review the language of quiverfolds, a generalization of quiver gauge theories which arise naturally in the presence of orientifold planes.
In appendix~\ref{app:negativerankdual} we review a useful mathematical tool for anomaly matching.
In appendix~\ref{app:exactlydimensionless}, we show that holomorphic combinations of couplings which are invariant under all possible spurious and/or anomalous global symmetries are RGE invariant.
In appendix~\ref{app:coulombbranch}, we show how the string coupling can be related to the gauge and superpotential couplings of a D-brane gauge theory by moving out along the Coulomb branch.
In appendix~\ref{app:SCIdetails} we discuss some technical details of the computation of the superconformal index.
In appendix~\ref{app:Specht}, we relate the matching of certain baryonic directions in the moduli space of the prospectively dual $\bC^3/\bZ_3$ theories to a group theoretic conjecture and provide evidence for this conjecture.
Finally, in appendix~\ref{sec:SCI-identity}, we relate the matching of
the superconformal indices to a conjectural identity for elliptic hypergeometric integrals.

In companion papers~\cite{transitions2, transitions3}, we discuss the construction of these orientifold gauge theories using exceptional collections as well as details of their gravity duals, focusing on string theoretic arguments that the dual gauge theories are connected by ten-dimensional S-duality. We also discuss the nature of the orientifold transition which seems to govern the duality, and construct infinite families of geometries which exhibit similar dualities.

\section{Review of Montonen-Olive duality}
\label{sec:MOduality}

In this section, we review certain aspects of Montonen-Olive duality which will be important for our paper.

Rigid $\mathcal{N}=4$ gauge theories are characterized by their gauge group and by their holomorphic gauge coupling, which takes the form
\begin{align}
\tau_{\rm YM} = \frac{\theta_{\rm YM}}{2 \pi}+\frac{4 \pi i}{g_{\rm YM}^2}\,,
\end{align}
for an $\SU(N)$ gauge theory. Such gauge theories are easily realized in string theory; for instance, the world-volume gauge theory on $N$ D3 branes probing a smooth background is an $\cN =4$ $\U(N)$ gauge theory with holomorphic gauge coupling equal to the type IIB axio-dilaton, $\axiodil = C_0 + i e^{-\phi}$, though the extended supersymmetry may be broken by irrelevant operators in a smoothly curved background or by relevant operators in the presence of flux (see for example \cite{Polchinski:2000uf,Pilch:2000ue}).

Montonen-Olive duality relates such a gauge theory to a dual theory at
a different coupling under the action of the modular group,
$\SL(2,\bZ)$: \begin{align}\tau' = \frac{a \tau+b}{c
    \tau+d}\,.  \end{align} In particular, unless the modular
transformation is of the form $\tau \to \tau+n$ (enforcing the
periodicity of the theta angle), it is straightforward to check that
the original and dual descriptions cannot both be weakly coupled. The
string realization of this duality, in the case where the gauge theory
arises as the world-volume gauge theory on a stack of D3 branes, is
the $\SL(2,\bZ)$ self-duality of type IIB string theory.

It is important to bear in mind that Montonen-Olive duality is not literally a ``duality'' (a word whose root is ``two''): a weakly coupled $\cN = 4$ theory has not just one but an \emph{infinite number} of strongly-coupled dual descriptions. Alternately phrased, by deforming a weakly-coupled $\cN = 4$ gauge theory to strong coupling, we encounter an infinite number of phases with a weakly-coupled dual description.

\begin{figure}
  \begin{center}
    \includegraphics[width=.5\textwidth]{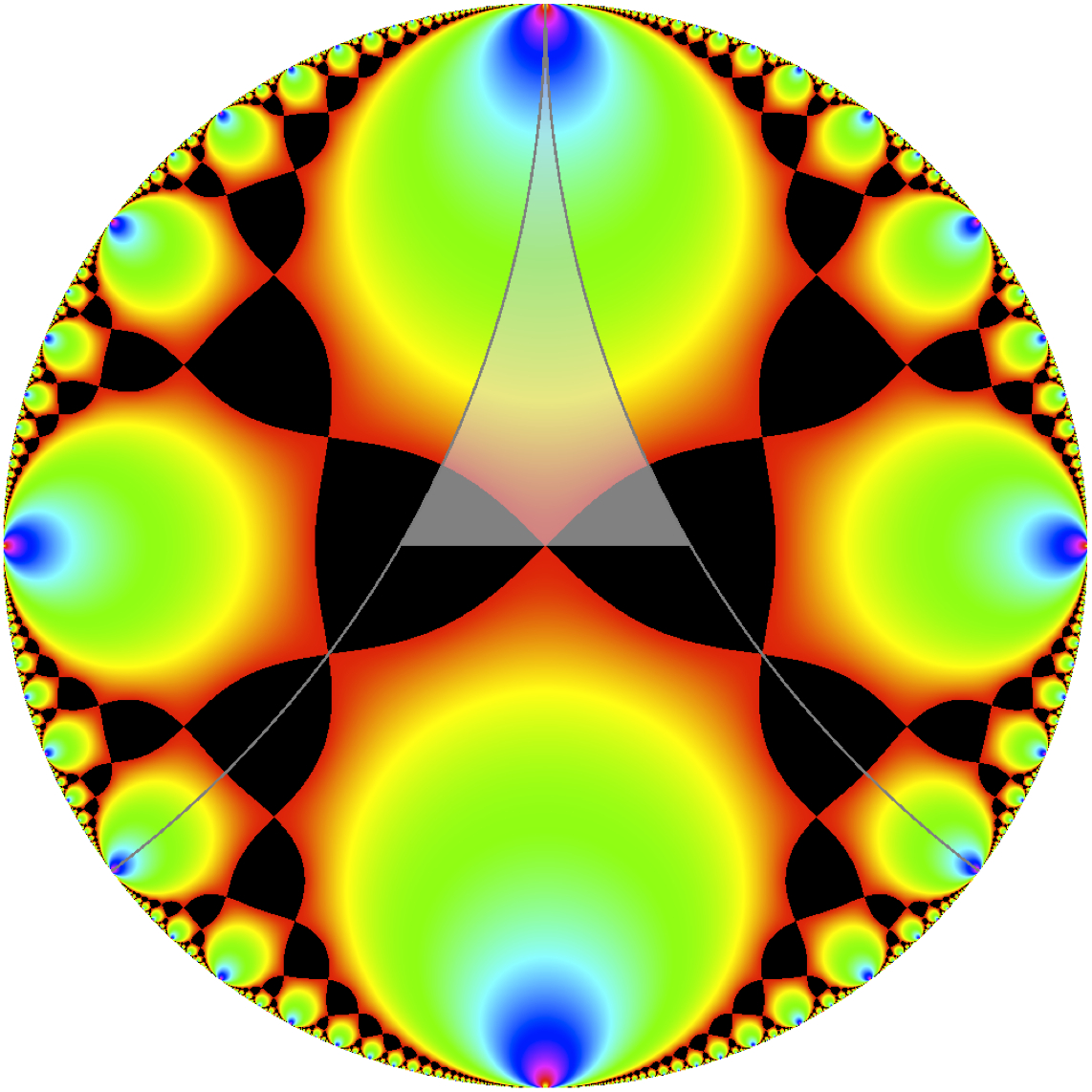}
  \end{center}
  \caption[The modular invariant $|j(\tau)|$ plotted across the upper half plane $\bH$]{The modular invariant $|j(\tau)|$ plotted across the upper half plane $\bH$, where $\bH$ is conformally mapped to the unit disk via $w = \frac{1+i \tau}{i+\tau}$, so that $\tau = +i \infty$ (weak coupling) lies at the top edge of the disk, $\tau = i$ (intermediate coupling) in the center and $\tau = 0$ (strong coupling) at the bottom, whereas the left and right edges correspond to $\tau = -1$ and $\tau = +1$ respectively, and the $\mathrm{Im\ } \tau = 0$ axis spans the perimeter. The black regions, corresponding to $|j(\tau)| < 12^3$, serve to divide the plane into an infinite number of disjoint colored regions, each containing a subregion with a weakly coupled dual description ($|j(\tau)| \to \infty$), colored blue/purple.  The superimposed curved grey lines illustrate the boundary of the region $|\mathrm{Re\ } \tau| < 1/2$, a fundamental domain under the identification $\tau \to \tau + 1$. The transformation $\tau \to -1/\tau$ is equivalent to inversion through the center of the disk, and the shaded triangular region is the canonical fundamental domain for $\SL(2,\bZ)$.}
\label{fig:SL2Zgraphic}
\end{figure}

To illustrate this intricate and fascinating behavior, we observe that
Klein's $j$-invariant $j(\tau)$ (see \cite{DLMF:KleinJ} for
definitions and basic properties) is approximately $e^{-2 \pi i \tau}$
at weak coupling, so that $|j(\tau)| \to \infty$ is a
modular-invariant definition of weak coupling. A plot of $|j(\tau)|$
on the upper half plane (conformally mapped to a disk) is shown in
figure~\ref{fig:SL2Zgraphic}. The infinite order of $\SL(2,\bZ)$ leads
to a fractal structure, as seen in the figure. Thus, the behavior of
these theories at strong coupling is very rich, with many dual weakly
coupled descriptions in the strong coupling limit, depending on the
exact value of the theta angle. The weakly coupled dual descriptions
become free as $\mathrm{Im\ }\tau \to 0$ for rational values of
$\theta/2 \pi$, and are therefore dense along the $\mathrm{Re\ }\tau$
axis.

For an $\SU(N)$ gauge group, all the dual descriptions have the same
perturbative gauge group. This is a consequence of the invariance
of the D3 brane under $\SL(2,\bZ)$. We now consider Montonen-Olive
duality for $\SO$ and $\Sp$ gauge groups. For an $\SO(2k)$ gauge
group, the duals likewise have the same gauge group; equivalently, the
O3$^-$ plane is $\SL(2,\bZ)$ invariant~\cite{Witten:1998xy}. For an
$\SO(2 k+1)$ gauge group, however, the dual descriptions have gauge
group $\Sp(2 k)$. In string theory, this corresponds to the fact that
the ($\tau \to -1/\tau$) S-dual of the $\widetilde{\rm O3}^-$ ---
another name for an O3$^-$ plus a single (pinned) D3
brane\footnote{The single additional D3 brane is ``pinned'' to the
  orientifold plane because it is its own orientifold image. A pinned
  brane arises whenever an odd number of (upstairs) branes is placed
  atop an orientifold plane, giving rise to an $\SO(2k+1)$ gauge
  group.} --- is an O3$^+$. Thus, S-duality maps
\begin{align}
\mathrm{O3}^- + (2 k+1) \mbox{\ D3's} \longrightarrow \mathrm{O3}^+ + 2 k \mbox{\ D3's}\,.
\end{align}
This is a well known example of what we will call an ``orientifold transition'',\footnote{The term ``orientifold transition'' was used in a different context in \cite{Hori:2005bk}. We do not mean to imply that our physical mechanism is the same, just that we also have a change in the orientifold type.} wherein strongly coupled orientifold planes recombine with branes to form a different, weakly coupled orientifold plane. Examples of this phenomenon with collapsed O7 planes and $\mathcal{N}=1$ supersymmetry are considered in~\cite{transitions2}, and play an important role in the new dualities discussed in this paper.

\begin{figure}
  \begin{center}
    \includegraphics[width=.5\textwidth]{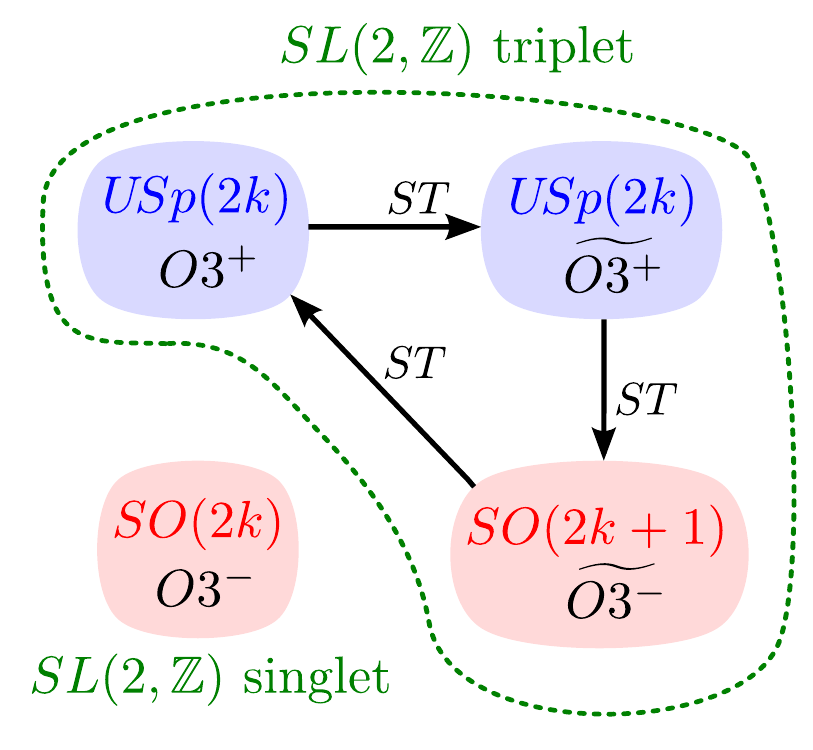}
  \end{center}
  \caption[The four gauge theories arising from D3 branes atop an O3 plane]{Summary of the four gauge theories that arise from placing D3 branes on top of the four different O3 planes.}
\label{fig:torsionclass31}
\end{figure}

The upshot of the previous paragraph is that Montonen-Olive duality
relates strongly coupled $\mathcal{N}=4$ gauge theories with
$\SO(2k+1)$ and $\Sp(2k)$ gauge groups to each other. This is not the
whole story, however. Because the O3$^+$ and $\widetilde{\rm O3}^-$
are related by S-duality, they must form some $\SL(2,\bZ)$
multiplet. However, the multiplet is as yet incomplete. To see this,
consider the $\SL(2,\bZ)$ generators $T: \tau \to \tau+1$ and $S: \tau
\to -1/\tau$. The $\widetilde{\rm O3}^-$ is $T$-invariant; therefore
$S T$ maps $\widetilde{\rm O3}^-$ to O3$^+$. However, since $(S T)^3 =
1$, it cannot be true that $S T$ maps O3$^+$ back to $\widetilde{\rm
  O3}^-$. We denote the $S T$ image of O3$^+$ as $\widetilde{\rm
  O3}^+$, where the three O3 planes form a triplet under
$\SL(2,\bZ)$~\cite{Witten:1998xy}. We summarize the resulting
structure in figure~\ref{fig:torsionclass31}.

The complete action of $\SL(2,\bZ)$ on the triplet is as follows: $S$
exchanges the O3$^+$ and $\widetilde{\rm O3}^-$, leaving the
$\widetilde{\rm O3}^+$ invariant, whereas $T$ exchanges the O3$^+$ and
$\widetilde{\rm O3}^+$, leaving the $\widetilde{\rm O3}^-$ invariant,
so that $S T$ cyclically permutes the three O3 planes. Since $T$ is a
perturbative duality, the $\widetilde{\rm O3}^+$ also gives rise to an
$\Sp(2 k)$ gauge group, and is perturbatively equivalent to the
O3$^+$, the two configurations being distinguished non-perturbatively
by their spectrum of BPS states \cite{Hanany:2000fq}. It is possible
to rephrase this by saying that the two different O3$^+$ planes give
rise to the same gauge theory at different theta angles. In
particular, $\tau \to \tau + 2$ leaves the O3 plane type invariant,
and defines the periodicity of the theta angle in the corresponding
gauge theory, whereas $\tau \to \tau + 1$ exchanges the two O3 plane
types. Thus, the gauge theories corresponding to $2 k$ D3 branes atop
an O3$^+$ and $\widetilde{\rm O3}^+$ can be identified with each other
upon shifting the theta angle by a half period.

\begin{figure}
  \begin{center}
    \includegraphics[width=.5\textwidth]{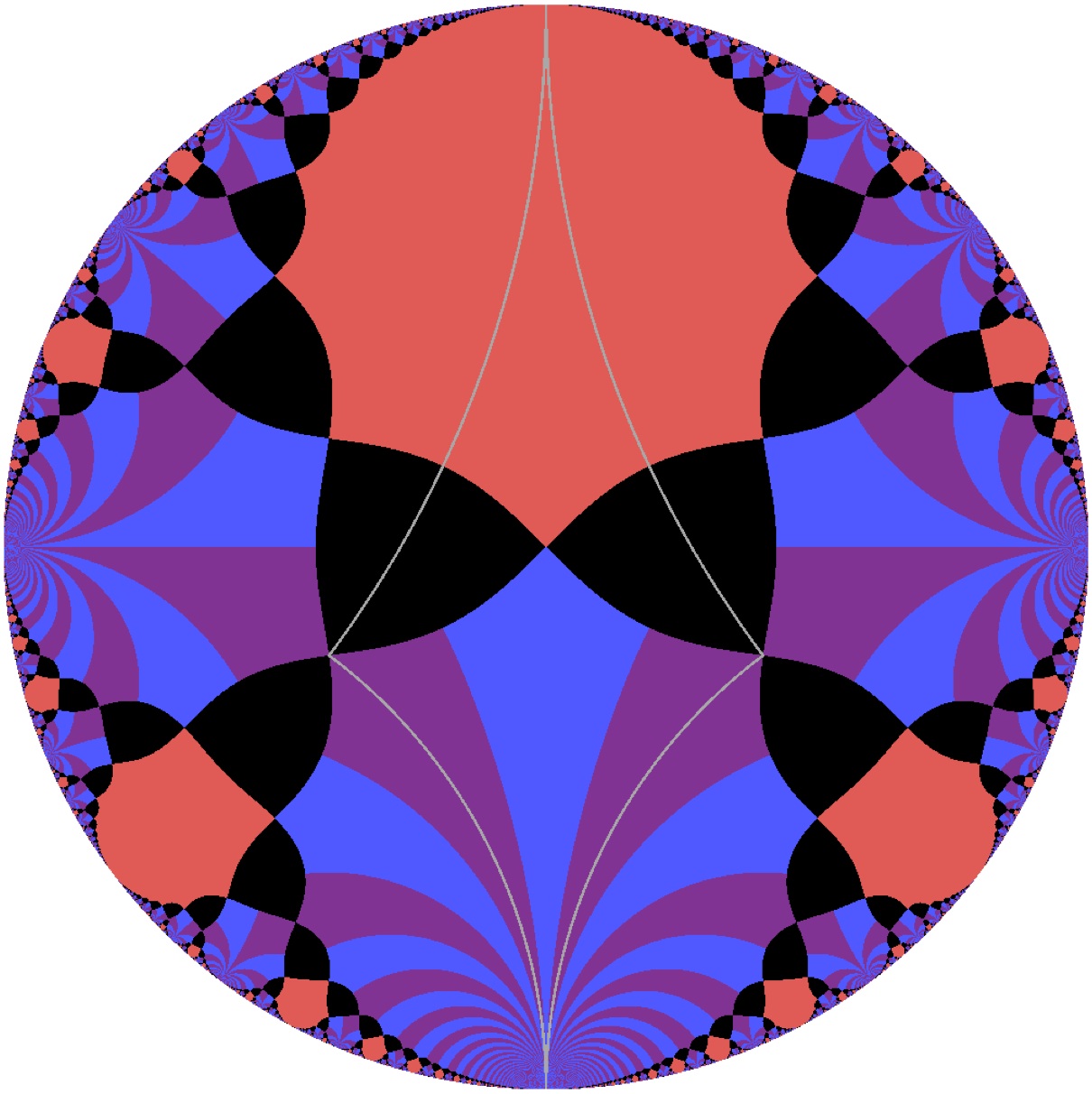}
  \end{center}
  \caption[The phase structure of $\mathcal{N}=4$ gauge theories as a function of $\tau$]{A schematic illustration of the phase structure of $\mathcal{N}=4$ $\SO(2k+1)$ and $\Sp(2k)$ gauge theories as a function of $\tau$, patterned after figure~\ref{fig:SL2Zgraphic}. The different colors indicate the type of the O3 plane (and hence the gauge group) in the dual weakly coupled phase for each value of $\tau$, where red, blue and purple correspond to an $\widetilde{\rm O3}^-$, O3$^+$ or $\widetilde{\rm O3}^+$, respectively, and the latter two possibilities are distinguished by requiring $-1/2 < \mathrm{Re\ }\tau \le 1/2$ in the dual theory. Thus, the red regions have a dual weakly coupled $\SO(2k+1)$ description, whereas the blue/purple striped regions have a dual weakly coupled $\Sp(2k)$ description. The thin grey lines outline a fundamental region for $\Gamma_0(2)$, the self-duality group for the $\SO(2k+1)$ theory. Note that the region where each dual theory is perturbative is most likely smaller than the colored region indicated here (see figure~\ref{fig:SL2Zgraphic}).}
\label{fig:phasediagram}
\end{figure}

To illustrate the nature of these dualities, we show how the weakly coupled description
changes as a function of the holomorphic gauge coupling in figure~\ref{fig:phasediagram}. As can be seen in the figure, each
gauge theory has additional self-dualities as well as the dualities
which relate the different theories. For example, the self-duality
group for $\SO(2 k+1)$ is the subgroup $\Gamma_0(2) \subset
\SL(2,\bZ)$ of elements $\bigl(\begin{smallmatrix} a&b\\
  c&d \end{smallmatrix} \bigr)$ for which $c$ is even (hence $a$ and
$d$ are odd, since $a d-b c = 1$), whereas for $\Sp(2k)$ it is the conjugate subgroup consisting of elements for which $b$ is even.

\section{Duality for \alt{$\bC^3/\bZ_3$}{C3/Z3}}
\label{sec:C3Z3fieldtheory}

In this section, we examine the simplest of our new $\cN=1$ dualities. In the $\cN=4$ examples discussed above, the six directions transverse to the D3 branes form a flat $\bR^6$ or equivalently $\bC^3$ transverse space, leading to gauge theories with maximal supersymmetry, where the $\SU(4) \cong \SO(6)$ R-symmetry is just the rotational isometry group of $\bR^6$. To obtain an $\cN=1$ theory at low energies, we must either switch on flux or introduce singularities. We choose to do the latter.

\begin{figure}
  \begin{center}
    \includegraphics[width=0.3\textwidth]{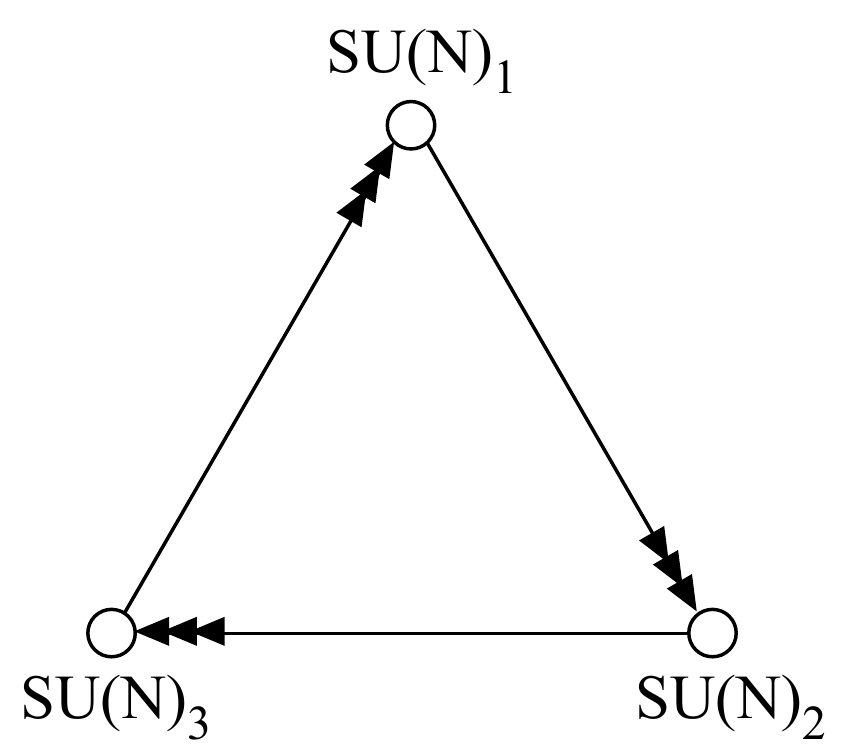}
  \end{center}
  \caption{The quiver gauge theory for $\bC^3/\bZ_3$.}
  \label{fig:dP0quiver}
\end{figure}

A simple and well-known example of such a transverse space is the
$\bC^3/\bZ_3$ orbifold, where the orbifold action on the transverse
complex coordinates is \begin{align}z^i \to e^{2 \pi i/3}
  z^i\,.  \end{align} The singularity can be resolved by blowing up a
$\bP^2$ exceptional divisor. Placing D3 branes at the singularity
leads to the $\cN=1$ $\SU(N)^3$ quiver gauge theory shown in
figure~\ref{fig:dP0quiver}. The orbifold reduces the isometry of the
transverse space to $\SU(3)\times\U(1)$, with the $\U(1)$ appearing as
an R-symmetry in the $\cN=1$ gauge theory.

We consider an orientifold of this configuration, since, as we argued
in the previous section, the $\SL(2,\bZ)$ dual descriptions of gauge
theories arising from D3 branes at singularities all have the same
gauge group and matter content. We choose the orientifold involution
$z^i \to -z^i$ which corresponds to an O7 plane wrapping the shrunken
$\bP^2$. As the resulting configuration is essentially a $\bZ_3$
orbifold of the $\cN = 4$ orientifolds considered in the previous
section, we will argue that the strong coupling behavior is closely
analogous. In this paper we focus on the characteristics of the
resulting gauge theories, deferring a detailed discussion of the
analogy between the gravity duals to~\cite{transitions2}.

In appendix~\ref{app:quiverfolds} we discuss in general how to
``orientifold'' a quiver gauge theory and apply this procedure to two
explicit examples. In particular in~\S\ref{subsec:C3Z3quiverfolds} we
study the orientifolds of the $\bC^3/\bZ_3$ orbifold theory for the
orientifold involution $z^i \to -z^i$. Counting $\SO(2N)$ and
$\SO(2N+1)$ as two separate cases, we find that there are three
possible gauge theories arising on D3 branes probing this orientifolded
singularity. They correspond to a shrunken $\bP^2$ that is wrapped by
an O7$^+$ plane or to an O7$^-$ plane with and without a pinned D3
brane respectively.

We will argue that two of these gauge theories are dual, whereas the third is
self-dual, analogous to the $\cN=4$ $\SO(2k+1)$, $\Sp(2k)$ and $\SO(2k)$ gauge theories discussed above.
The dualities studied here are merely the simplest
examples of a large class of $\cN=1$ dualities between orientifold
gauge theories, some of which we will study in detail in this paper as well as in~\cite{transitions2, transitions3}, where we will discuss many more examples.

The $\bC^3/\bZ_3$ orientifold we discuss here was, to the best of our
knowledge, first studied in
\cite{Angelantonj:1996uy,Lykken:1997gy,Lykken:1997ub,Kakushadze:1998tr}
and recently revisited from the dimer point of view in
\cite{Franco:2007ii,Franco:2010jv} and applied to the problem of moduli stabilization in~\cite{Heidenreich:2010ad, Cicoli:2012vw}.
Orientifolds of $\bC^3/\bZ_3$ and
related abelian orbifolds have also proven to be an interesting
testing ground for studying non-perturbative dynamics in string theory
\cite{Bianchi:2007fx,Bianchi:2007wy,Bianchi:2009bg,Bianchi:2009ij,Bianchi:2012ud}. For the involution discussed above, one finds $\SO(N-4)\times\SU(N)$ and
$\Sp(\tN+4)\times\SU(\tN)$ gauge theories for collapsed O7$^-$ and O7$^+$
planes, respectively. Both theories have a non-anomalous R-symmetry in
addition to a global $\SU(3)$ symmetry, corresponding to the
$\SU(3)\times\U(1)$ isometry of the transverse space. A careful
analysis reveals that both models also have a discrete ``baryonic''
$\bZ_3$ symmetry. The two models are\footnote{These two gauge theories are related by a negative rank duality as explained in appendix \ref{app:negativerankdual}.}
\begin{gather}
  \begin{tabular}{c|cc|ccc}
     & $\SO(N-4)$ & $\SU(N)$  & $\SU(3)$ & $\U(1)_R$ & $\bZ_{3}$\\ \hline
    $A^i$ & $\fund$ & $\ov{\fund}$ & $\fund$ & $\frac23+\frac2N$ & $\omega_{3 N}$ \\
    $B^i$ & \sing & $\asymm$ & \fund & $\frac23 - \frac4N$ & $\omega_{3 N}^{-2}$
  \end{tabular} \\[15pt]
  \begin{tabular}{c|cc|ccc}
     & $\Sp(\tilde{N}+4)$ & $\SU(\tilde{N})$ & $\SU(3)$ & $\U(1)_R$ & $\bZ_{3}$\\ \hline
    $\tilde{A}^i$ & $\fund$ & $\ov{\fund}$ & $\fund$ & $\frac23 - \frac{2}{\tilde{N}}$ & $\omega_{3 \tilde{N}}$ \\
    $\tilde{B}^i$ & \sing & $\symm$ & \fund & $\frac23 + \frac{4}{\tilde{N}}$ & $\omega_{3 \tilde{N}}^{-2}$
  \end{tabular}
\end{gather}
where $\omega_n \equiv e^{2\pi i/n}$, $\tN$ is even, and the
tree-level superpotentials are
\begin{align}
W = \frac{\lambda}{2} \epsilon_{ijk} \Tr A^i A^j B^k \,,\quad \tilde{W} = \frac{\tilde{\lambda}}{2} \epsilon_{ijk} \Tr \tilde{A}^i \tilde{A}^j \tilde{B}^k \,,
\end{align}
respectively, where $\lambda$ and $\tilde{\lambda}$ are superpotential couplings.

Note that we label the discrete symmetry as a $\bZ_3$ even though the cube of the generator is not the identity. This is because the cube of the generator lies within the $\bZ_N$ or $\bZ_{\tN}$ center of $\SU(N)$ or $\SU(\tN)$, so we obtain a $\bZ_3$ symmetry upon composing the generator with an element of $\SU(N)$ or $\SU(\tN)$ whose cube is the inverse of the cube of the generator. This latter $\bZ_3$ symmetry is equivalent to the discrete symmetry indicated in the charge table up to a constant gauge transformation.\footnote{In general, a discrete symmetry can be rewritten as a $\bZ_k$ discrete symmetry times a constant gauge transformation whenever the $k$th power of the generator lies within the gauge group.}

The $\SO(N-4)\times\SU(N)$ gauge theories have a classical moduli space which includes directions corresponding to moving D3 branes away from the singularity. The gauge group is then Higgsed down to $\SO(N-4-2k)\times\SU(N-2k)\times\U(k)$ where $k$ is the number of (downstairs) D3 branes removed from the O-plane, corresponding to the $\U(k)$ factor in the Higgsed gauge group. After integrating out massive matter, the $\U(k)$ decouples from the rest of the gauge group in the IR, giving a separate $\cN=4$ gauge theory corresponding to $k$ D3 branes probing a smooth region of the Calabi-Yau cone. Meanwhile, the remaining $\SO(N-4-2k)\times\SU(N-2k)$ reproduces the original gauge theory at a different rank $N' = N-2k$. Thus, we see that the moduli space of the $\SO(N-4)\times\SU(N)$ family of gauge theories falls into two disconnected components for even and odd $N$ respectively, where all even $N$ theories are connected by the above process, as are all odd $N$ theories. Similarly, all $\Sp(\tilde{N}+4)\times\SU(\tilde{N})$ theories are interconnected by an analogous motion in moduli space, where $\tilde{N}$ must be even for $\Sp(\tilde{N}+4)$ to exist.

In all, we have obtained three distinct families of gauge theories
corresponding to D3 branes probing the orientifolded $\bC^3/\bZ_3$
singularity, all corresponding to the same geometric orientifold
involution. Two of these theories, the $\SO(N-4)\times\SU(N)$ theories
for even and odd $N$, are distinguished from each other by the
presence or absence of a pinned D3 brane and its corresponding
half-integral D3 brane charge, while the
$\Sp(\tilde{N}+4)\times\SU(\tilde{N})$ theory corresponds to a compact
O7$^+$ plane rather than a compact O7$^-$ plane. Regardless of the
O-plane type, the seven-brane tadpole is cancelled locally by
(anti-)D7 branes, and the two configurations have the same
$\SL(2,\bZ)$ monodromy.\footnote{We argue in~\cite{transitions2} that the
  geometric duals are distinguished from each other by different
  (S-dual) choices of discrete torsion.}

The situation is closely analogous to the three gauge theories $\SO(2k)$, $\SO(2k+1)$, $\Sp(2k)$ appearing in the $\cN=4$ case, and we therefore hypothesize that one of the $\SO$ families enjoys an $\SL(2,\bZ)$ self-duality, whereas the other $\SO$ family and the $\Sp$ family are related by an $\SL(2,\bZ)$ duality. In the remainder of this section and in~\S\ref{sec:SCI}, we present strong field theoretic evidence for the latter duality, based on the matching of various computable infrared observables, and explore some of its properties.

We begin by discussing 't Hooft anomaly matching in~\S\ref{subsec:classicchecks}, leading to a more precise statement of the proposed duality. We then provide further evidence for the duality by a partial matching between the moduli spaces of the two theories. In~\S\ref{subsec:limitations}, we highlight an important limitation of our methods which is nonetheless linked to the nature of the duality, and in~\S\ref{subsec:N=5} we discuss a specific, finite $N$ example of the proposed duality.

We continue our discussion of these gauge theories in the following sections. In~\S\ref{sec:SCI}, we provide further strong evidence for the proposed duality by comparing the superconformal indices between the prospectively dual theories, and in~\S\ref{sec:infrared}, we discuss their infrared physics using Seiberg duality.

\subsection{Classic checks of the duality}\label{subsec:classicchecks}
As a precursor to anomaly matching, we note that the dual theories should have the same global symmetry groups. In particular, for $N$ or $\tilde{N}$ not divisible by three the baryonic $\bZ_3$ is equivalent to the $\bZ_3$ center of $\SU(3)$ composed with a constant gauge transformation, and therefore lies within the continuous symmetry group, whereas for $N=0\mod\ 3$ the $\bZ_3$ is distinct.\footnote{In this case the center of $\SU(3)$ lies within center of the $\SU(N)$ or $\SU(\tN)$ gauge group.} Moreover, there is an additional $\bZ_2$ ``color conjugation'' symmetry (see e.g.~\cite{Csaki:1997aw}) for the $\SO$ theory with even $N$, which comes from the outer automorphic group of $\SO(2n)$. In net, the global symmetry group for the $\SO$ theories is $\SU(3)\times\U(1)_R\times\bZ_{\gcf(6,N)}$, whereas for the $\Sp$ theories it is $\SU(3)\times\U(1)_R\times\bZ_{\gcf(3,\tilde{N})}$. Since the global symmetry groups must match between dual descriptions, this suggests that the ranks of the dual pair must be related as follows:
\begin{align}
N = \tilde{N} + 3 k \,,
\end{align}
for some odd integer $k$ to be determined.

The global anomalies for the two models are shown in table~\ref{tab:C3Z3anomalies}.\footnote{Here and in future we only write the $G^2 \bZ_k$ and $(\mathrm{grav})^2 \bZ_k$ discrete anomalies for $G$ nonabelian, as the remaining discrete ``anomalies'' need not match between two dual theories~\cite{Csaki:1997aw}, and do not appear as anomalies in the path integral measure~\cite{Csaki:1997aw, Araki:2008ek}.}
\begin{table}
\begin{center}
\begin{tabular}{ccc}
$\SO(N-4) \times \SU(N)$ theory: && $\Sp(\tilde{N}+4) \times \SU(\tilde{N})$ theory:\\
\begin{tabular}{|c|c|}
\hline
 $\SU(3)^3$ & $\frac{3}{2} (N-3) N$ \\
\hline
 $\SU(3)^2 \,\U(1)_R$ & $-\frac{1}{2} (N-3) N -6$ \\
\hline
 $\U(1)_R^3$ & $\frac{4}{3} (N-3) N-33$ \\
\hline
 $\U(1)_R$ & $-9$ \\
\hline
$\SU(3)^2 \, \bZ_3$ & 1\\
\hline
$\bZ_3$ & 1\\
\hline
\end{tabular} & &
\begin{tabular}{|c|c|}
\hline
 $\SU(3)^3$ & $\frac{3}{2} \tilde{N}(\tilde{N}+3) $ \\
\hline
 $\SU(3)^2 \, \U(1)_R$ & $-\frac{1}{2}  \tilde{N}(\tilde{N}+3)-6$ \\
\hline
 $\U(1)_R^3$ & $\frac{4}{3}\tilde{N}(\tilde{N}+3)-33$ \\
\hline
 $\U(1)_R$ & $-9$ \\
\hline
$\SU(3)^2 \, \bZ_3$ & 1\\
\hline
$\bZ_3$ & 1\\
\hline
\end{tabular}
\end{tabular}
\end{center}
\caption[The anomalies for the $\bC^3/\bZ_3$ orientifold gauge theories]{The anomalies for the $\bC^3/\bZ_3$ orientifold gauge theories. In our notation, $G^2 \bZ_k = \prod_i \eta_i^{2 T(r_i)}$ and $(\mathrm{grav})^2 \bZ_k = \prod_i \eta_i^2$, where $\eta_i$ is the multiplicative charge of $i$th Weyl fermion under the generator of the $\bZ_k$ discrete symmetry. For a discrete anomaly $\eta$, the Jacobian for the symmetry transformation in the path integral is $\eta^n$, where $n$ is the instanton number for the background in question; therefore, the anomaly vanishes iff $\eta = 1$.\label{tab:C3Z3anomalies}}
\end{table}
We see that the anomalies match between the two theories for $N=\tilde{N}+3$, in
agreement with the restriction from matching the global symmetry
groups discussed above. In \cite{transitions2}, it is shown that this
rank relation agrees with D3 charge conservation, as it must. Since $\tN$ is necessarily even, this is evidence
for a possible duality between the $\SO(N-4) \times \SU(N)$ theory for
odd $N$ and the $\Sp(N+1) \times \SU(N-3)$ theory. It will also follow
from the arguments of \cite{transitions2} that the $\SO(N-4) \times \SU(N)$
theory for even $N$ is self-dual.

We now consider the moduli spaces of the prospectively dual theories, which is classically equivalent to the affine variety parameterized by the holomorphic gauge invariant operators identified under the F-term conditions and classical constraints~\cite{Luty:1995sd}. In general, a holomorphic gauge invariant of the $\SO(N-4)\times\SU(N)$ theory takes the form
\begin{align}
\cO^{N_A, N_B} = A^{N_A} B^{N_B} \,,
\end{align}
for some particular choice of contraction of the gauge indices. Such operators may be classified as ``mesons'' or ``baryons'', depending on whether the $\SU(N)$ Levi-Civita symbol is irreducibly involved in the contraction of gauge indices or not, i.e. on whether the baryonic charge
\begin{align}
Q_A \equiv (N_A - 2 N_B)/N
\end{align}
is vanishing or not. The corresponding $\U(1)_A$ is anomalous, with an anomaly-free $\bZ_3$ subgroup that was identified above:
\begin{align}
Q_3 = \omega_3^{Q_A} \,,
\end{align}
where the $Q_A$ charge of a gauge invariant operator is necessary integral, since $\bZ_N \subset \SU(N)$ lies within the gauge group.

No $\SO(N-4)$ gauge invariants exist for the case $N_A = 1$ with $N>5$. Thus, baryonic operators can be further subdivided into those with $Q_A>0$, which can be ``factored''\footnote{We do not mean to imply that the gauge index contractions factorize in the indicated manner.} as
\begin{align}\label{eqn:ABfactor1}
\cO = (A^{N})^{n_1} (A A B)^{n_2} \,,
\end{align}
and those with $Q_A < 0$, which can be ``factored'' as
\begin{align}\label{eqn:ABfactor2}
\cO = (B^{N})^{n_1} (A A B)^{n_2} \,,
\end{align}
for integral powers $n_1$ and $n_2$.

We will focus on the ``irreducible'' baryons, of the form $\cO_k^{(A)} \equiv A^{k N}$ and $\cO_k^{(B)}  = B^{k N}$. These have $R$-charges
\begin{align}
Q_R (A^{k N}) = \frac{2 (N+3)}{3} k \, ,\quad Q_R (B^{k N}) = \frac{2 (N-6)}{3} k \,,
\end{align}
and in both cases the $\bZ_3$ charge $\omega_3^k$. ``Reducible'' baryons are similar, but with an additional $R$-charge of $+2$ for every factor of $(AAB)$ which appears.

The holomorphic gauge invariants of the $\Sp(\tilde{N}+4)\times\SU(\tilde{N})$ theory can be similarly classified, where now the irreducible baryons have $R$-charges
\begin{align}
Q_R (\tilde{A}^{k \tilde{N}}) = \frac{2 (\tilde{N}-3)}{3} k \,, \quad Q_R (\tilde{B}^{k \tilde{N}}) = \frac{2 (\tilde{N}+6)}{3} k \,,
\end{align}
with the $\bZ_3$ charge $\omega_3^k$, as before.

In general, mesons and reducible and irreducible baryons can all be intermixed in the duality relations between the two theories. However, in certain cases only one type of operator with the correct $R$-charge exists. In particular, this occurs in the following cases for the $\SO(N-4)\times\SU(N)$ theory:
\begin{enumerate}
\item
For $Q_R < 2 (N-6)$ and $Q_3 = \omega_3^0$, only mesonic operators are possible.
\item
For $Q_3 = \omega_3$ and $Q_3 = \omega_3^{-1}$, the minimum possible $R$-charges are $\frac{2 (N-6)}{3}$ and $\frac{4 (N-6)}{3}$, respectively, corresponding to the irreducible baryons $B^N$ and $B^{2 N}$.
\end{enumerate}
Similarly, for the $\Sp(\tilde{N}+4)\times\SU(\tilde{N})$ theory:
\begin{enumerate}
\item
For $Q_R < 2 (\tilde{N}-3)$ and $Q_3 = \omega_3^0$, only mesonic operators are possible.
\item
For $Q_3 = \omega_3$ and $Q_3 = \omega_3^{-1}$, the minimum possible $R$-charges are $\frac{2 (\tilde{N}-3)}{3}$ and $\frac{4 (\tilde{N}-3)}{3}$, respectively, corresponding to the irreducible baryons $\tilde{A}^{\tilde{N}}$ and $\tilde{A}^{2 \tilde{N}}$.
\end{enumerate}
This suggests the matching:
\begin{align}
B^N \longleftrightarrow \tilde{A}^{\tilde{N}} \,, \quad B^{2 N} \longleftrightarrow \tilde{A}^{2 \tilde{N}}\,,
\end{align}
between the $Q_3 = \omega_3$ and $Q_3 = \omega_3^{-1}$ operators of minimum possible $R$-charge in both theories. In particular, these operators must have the same $R$-charge, i.e.
\begin{align}
\frac{2 (N-6)}{3} = \frac{2 (\tilde{N}-3)}{3} \,,
\end{align}
which reproduces the rank relation $N = \tilde{N}+3$ that we saw from the anomaly matching conditions.

The $\SU(3)$ representations of these operators should also match. For
$\tA^{\tN}$, the $\SU(3)$ representation can be determined as follows:
the symplectic invariant contracts the $\tA$'s in pairs, and the
operator therefore factors as $(\tA^2)^{\tN/2}$. The $\tB$ F-term
condition implies that the nonvanishing component of the $\Sp(\tN+4)$
invariant $\tA^2$ transforms as
$(\ov{\asymm},\symm)_{4/3-4/\tilde{N}}$ under
$\SU(\tN)\times\SU(3)\times\U(1)_R$. Thus, $\tA^{\tN} =
(\tA^2)^{\tN/2}$ takes the form of a ``Pfaffian'' of $\tA^2$, which is
symmetric in its factors. The nonvanishing gauge-invariant component
of $\tA^{\tN}$ therefore transforms in the $\SU(3)$ representation
\be \label{eqn:ANtSU3rep}
\Sym^{\tN/2}(\symm) \equiv \underbrace{\symm \otimes_S \symm \otimes_S \ldots \otimes_S \symm}_{\tN/2} \,\,,
\ee
where $\otimes_S$ denotes the symmetric tensor product.

For $B^N$, the F-term conditions impose no additional
constraints. Using the computer algebra package \texttt{LiE}
\cite{LiE}, one can show that the gauge invariant component of $B^N$
also transforms in $\Sym^{\tN/2}(\symm)$ for $N=5,7,9,11$ and $13$ and $\tN =
N-3$, whereas checking that this holds for larger $N$ is too
computationally expensive using \texttt{LiE} directly. Using the more
efficient approach explained in appendix~\ref{sec:B21} we
have verified agreement up to $N=21$. It would be desirable to have an
argument for all $N$, and while we do not have a general proof, in
appendix~\ref{app:Specht} we show how agreement between the $SU(3)$
representations of $\tilde A^{\tilde N}$ and $B^N$ follows from a
certain conjectural mathematical identity involving representations of
the symmetric group, and we provide additional evidence for this
identity.

As a further check, we should be able to match the mesonic operators with $Q_R < 2 (N-6) = 2 (\tilde{N}-3)$ between the two theories. Such operators can be factored into products of single-trace operators of the form:
\begin{align}
\cO_n^{i_1 j_1 k_1 \ldots i_n j_n k_n} \equiv \Tr (A^{i_1})^T B^{j_1} A^{k_1} \ldots (A^{i_n})^T B^{j_n} A^{k_n} \,,
\end{align}
where the F-term conditions imply that $\cO_n$, with $Q_R = 2n$, is totally symmetric in its $3n$ $\SU(3)$ indices. A similar argument goes through for the $\Sp(\tilde{N}+4)\times\SU(\tilde{N})$ theory, resulting in the same spectrum of single-trace operators.

\subsection{Limitations from the perturbativity of the string coupling}
\label{subsec:limitations}

Before turning to specific examples of the duality, we briefly review some general obstructions to having a perturbative description of the D-brane gauge theories obtained from quantization of open strings. For a ``perturbative description'', we require that there exists some energy scale at which the gauge theory is weakly coupled, rather than weak coupling in the infrared.\footnote{As we shall see, all of the D-brane gauge theories considered above are strongly coupled in the infrared, so the latter requirement is too strong.} The nature of these obstructions will also serve to illustrate how our proposed duality differs from Seiberg duality.

The one-loop beta function for a supersymmetric gauge theory is given by~\cite{Terning:2006bq}:
\begin{equation} \label{eqn:betafunction1loop}
\beta(g) \equiv \frac{d g}{d \ln \mu} = - \frac{g^3}{16 \pi^2} \left( 3\, T(\adj) - \sum_i T(r_i) \right) \,,
\end{equation}
where $T(r)$ denotes the Dynkin index for the representation
$r$,\footnote{We employ the conventions $T(\fund)=\frac{1}{2}$ for
  $\SU$ and $\Sp$ gauge groups, and $T(\fund) = 1$ for $\SO$ gauge
  groups.}  $\adj$ denotes the adjoint representation, and the sum is
taken over all chiral superfields. If $g$ is taken to be the
holomorphic gauge coupling, then this result is exact, whereas the
corresponding exact result for the physical gauge coupling depends on
the anomalous dimensions of the chiral superfields.

The one-loop beta function coefficients (the term within parentheses
in~(\ref{eqn:betafunction1loop})) for the gauge group factors of the
$\SO$ theory are:
\begin{align}
b_{\SO} = -18\,, \quad b_{\SU} = 9 \, ,
\end{align}
whereas for those of the $\Sp$ theory they are
\begin{align}
\tilde{b}_{\Sp} = 9 \,, \quad \tilde{b}_{\SU} = -9 \, .
\end{align}
Since the beta functions for the two gauge group factors have opposite
signs, neither gauge theory is either IR free or asymptotically free,
and the perturbative description will be valid at most in a finite
range of energy scales. More precisely, a perturbative description at
any scale is only possible if there is a separation between the
dynamical scales, $\Lambda_{\SO} \gg \Lambda_{\SU}$ or
$\tilde{\Lambda}_{\SU} \gg \tilde{\Lambda}_{\Sp}$, along with a small
superpotential coupling ($\lambda \ll 1$ or $\tilde{\lambda} \ll 1$)
somewhere between these two scales. We will work in this limit. While
it is possible in principle to incorporate corrections which are
subleading in an expansion in small $\Lambda_{\SU}/\Lambda_{\SO}$ or
$\tilde{\Lambda}_{\Sp}/\tilde{\Lambda}_{\SU}$, this can be very
difficult in practice, and we will not attempt to do so.

Conversely, the duality we propose partly addresses the question of what happens to the gauge theory in the limit where the dynamical scales have an inverted hierarchy. To see why, note that the string coupling is given by
\begin{equation} \label{eqn:dP0axiodil}
\axiodil = \frac{1}{2 \pi i} \ln \left[ \lambda^{6 (N-2)} \Lambda_{\SO(N-4)}^{-18} \Lambda_{\SU(N)}^{18} \right] \,, \quad \axiodil= \frac{1}{\pi i} \ln \left[ \tilde{\lambda}^{3 (\tN+2)} \Lambda_{\Sp(\tN+4)}^{9} \Lambda_{\SU(\tN)}^{-9} \right] ,
\end{equation}
for the prospective dual theories, up to a multiplicative numerical factor within the square brackets. This result can be established in a variety of ways; for completeness, we present a Coulomb branch computation of it in appendix~\ref{app:coulombbranch}. The result is also intuitive: a perturbative gauge theory necessarily corresponds to a weak string coupling.

The duality we propose acts as a modular transformation on $\axiodil$, mapping any perturbative string coupling to a nonperturbative one. Conversely, deforming to strong string coupling and applying the duality, we obtain a dual description with a weak string coupling. Thus, since the string coupling is linked to the hierarchy $\Lambda_{\SU}/\Lambda_{\SO}$ or $\tilde{\Lambda}_{\Sp}/\tilde{\Lambda}_{\SU}$, the duality provides at least partial information about the behavior of these gauge theories with an inverted hierarchy $\Lambda_{\SU} \gg \Lambda_{\SO}$ or $\tilde{\Lambda}_{\Sp} \gg \tilde{\Lambda}_{\SU}$.\footnote{While the Lagrangian definition of the gauge theory may be insufficient in this case, in principle string theory provides a complete definition for any $N$ via the AdS/CFT correspondence, although this definition is impractical for computations except in the large $N$ limit.}

By contrast, Seiberg duality is generally used to understand the infrared behavior of a gauge theory which is perturbative at some scale, an illustration of the different natures of these two types of duality. While we can repeatedly apply Seiberg duality (together with deconfinement) to the individual gauge group factors, in our experience such an exercise never reproduces the prospective dual gauge theory,\footnote{Repeated application of Seiberg duality to these gauge theories requires a seemingly never-ending chain of deconfinements, leading to more and more gauge group factors. While one can imagine some of these factors eventually reconfining after several steps, we have not found this to be the case in our limited explorations of the matter.} providing further circumstantial evidence that the duality is not a Seiberg duality in the usual sense. Indeed, if we were able to do so, we would have to somehow reconcile the complicated gauge coupling relations which result from applying modular transformations to~(\ref{eqn:dP0axiodil}) with the algebraic relationships between dynamical scales predicted by Seiberg duality.

With these considerations in mind, we turn to a specific example of the proposed duality.

\subsection{Case study: the \alt{$\SU(5) \longleftrightarrow \Sp(6)\times\SU(2)$}{SU(5)<->Sp(6)xSU(2)} duality}
\label{subsec:N=5}

Since we are constrained to $N \ge 4$ and $\tilde{N} \ge 0$ to have gauge groups of non-negative rank, the lowest rank duality we expect to find is between the $\SU(5)$ and $\Sp(6)\times\SU(2)$ gauge theories:
\begin{equation}
\mbox{
\begin{minipage}{0.36\linewidth}
  \begin{center}
  \begin{tabular}{c|c|cc}
     & $\!\SU(5)\!$  & $\!\SU(3)\!\!\!$ & $\!\U(1)_R\!$ \\ \hline
    $\!A^i\!$ & $\ov{\fund}$ & $\fund$ & $16/15$ \\
    $\!B^i\!$ & $\asymm$ & \fund & $-2/15$
  \end{tabular}\\[12pt]
  $W = \frac{1}{2} \lambda\, \epsilon_{i j k} A^i_m A^j_n B^{m n;\, k}$ \,,
  \end{center}
  \end{minipage}}
  \longleftrightarrow
  \mbox{\begin{minipage}{0.45\linewidth}
  \begin{center}
  \begin{tabular}{c|cc|cc}
     & $\!\Sp(6)\!\!$ & $\!\SU(2)\!$ & $\!\SU(3)\!\!\!$ & $\!\U(1)_R\!$ \\ \hline
    $\!\tilde{A}^i\!$ & $\fund$ & $\fund$ & $\fund$ & $-1/3$ \\
    $\!\tilde{B}^i\!$ & \sing & $\symm$ & \fund & $8/3$
  \end{tabular}\\[12pt]
  $W = \frac{1}{2} \tilde{\lambda}\, \symp^{a b} \epsilon_{i j k} \tilde{A}^i_{a;\, m} \tilde{A}^j_{b;\, n} \tilde{B}^{m n;\, k}$ \,,
  \end{center}
  \end{minipage}}
\end{equation}
where $\Omega$ denotes the symplectic invariant. We characterize the classical moduli space of both theories, and show that both generate a runaway superpotential.

We discuss higher rank examples in~\S\ref{sec:infrared}.

\subsubsection{The \alt{$\Sp(6)\times\SU(2)$}{Sp(6)xSU(2)} theory}

The F-term conditions are
\begin{align}
\symp^{a b} \tilde{A}^{[i}_{a;\, m} \tilde{A}^{j]}_{b;\, n} = 0 \,, \quad \tilde{A}^{[j}_{b;\, n} \tilde{B}^{k];\, m n} = 0 \, .
\end{align}
The first condition implies that all nonvanishing $\Sp(6)$ holomorphic gauge invariants are built from
\begin{align}
\cA^{i j} = \symp^{a b} \epsilon^{m n} \tilde{A}^{i}_{a;\, m} \tilde{A}^{j}_{b;\, n} \,,
\end{align}
which transforms as a $\symm_{\,-2/3}$ under $\SU(3)\times\U(1)_R$. The remaining holomorphic gauge invariants are easily cataloged:
\begin{align}
\cB^{i j} = \epsilon_{m p} \epsilon_{n q} \tB^{i;\, m n} \tB^{j;\, p q}\,, \quad \cB = \frac{1}{6} \epsilon_{i j k} \epsilon_{n p} \epsilon_{q r} \epsilon_{s m} \tB^{i;\, m n} \tB^{j;\, p q} \tB^{k;\, r s} \,,
\end{align}
which transform as $\symm_{16/3}$ and $\sing_8$, respectively, and obey the constraint $\cB^2 = \frac{1}{2} \det \cB^{i j}$.

The second F-term condition implies a constraint relating $\cA^{i j}$ and $\cB^{i j}$. In particular,
\begin{align}
\cA^{i j} \cB^{k l} = \symp^{a b} \epsilon^{m n} \epsilon_{p r} \epsilon_{q s} \tilde{A}^{i}_{a;\, m} \tilde{A}^{j}_{b;\, n} \tB^{k;\, p q} \tB^{l;\, r s} = 2 \symp^{a b} \epsilon_{q s} \tilde{A}^{i}_{a;\, m} \tilde{A}^{j}_{b;\, n} \tB^{k;\, m q} \tB^{l;\, n s} \,,
\end{align}
where we apply the first F-term condition to simplify the right-hand side. The second F-term condition then implies the constraint
\begin{align}
\cA^{i [j} \cB^{k] l} = 0 \,.
\end{align}
Thus, the classical moduli space has three distinct branches:
\begin{enumerate}
\item
A branch with $\cB^{i j} = 0$, parameterized by $\cA^{i j} \ne 0$. For generic (full-rank) $\cA^{ij}$, $\Sp(6)\times\SU(2)$ breaks to a diagonal $\SU(2)$, whereas for rank-deficient $\cA^{i j}$, a larger gauge symmetry remains: $\Sp(4)\times\SU(2)$ for $\cA^{i j}$ rank one and $\SU(2)\times\SU(2)$ for $\cA^{i j}$ rank two.
\item
A branch with $\cA^{i j}=0$, parameterized by $\cB^{i j} \ne 0$ (and $\cB$). For $\cB^{i j}$ rank two or three, the $\SU(2)$ gauge factor is completely broken, whereas $\SU(2) \to \U(1)$ for $\cB^{i j}$ rank one.
\item
A branch with $\cA^{i j} = e^{i \phi} \cos \theta\, v^i v^j$ and $\cB^{i j} = e^{-i \phi} \sin \theta\, v^i v^j$. $\Sp(6)\times\SU(2)$ breaks to $\Sp(4)\times\U(1)$, except for when $\theta = 0$ or $\theta = \pi/2$, where this branch intersects branches 1 and 2, respectively.
\end{enumerate}

We now discuss quantum corrections to this picture. The $\Sp(6)$ gauge factor is asymptotically free, whereas the $\SU(2)$ gauge factor is infrared free. Thus, the infrared dynamics are primary governed by $\Sp(6)$ (to leading order in $\Lambda_{\Sp(6)} \ll \Lambda_{\SU(2)}$), which generates an ADS superpotential:\footnote{The same superpotential was obtained via a direct string computation in \cite{Bianchi:2007wy}.}
\begin{align}
W_{\rm ADS} = \frac{\Lambda_{\rm Sp}^9}{\det \tilde A} \,,
\end{align}
where $\tilde{A}$ is viewed as a $6\times6$ matrix over $\Sp(6)\times(\SU(2)\times\SU(3))$.

We now consider the effect of the ADS superpotential on the moduli space. It is helpful to rewrite $\tilde{B}$ in the form:
\begin{align}
\tilde{B}_{m n}^i = \frac{1}{2} \epsilon^{i j k} \hat{B}_{j_m k_n} \,,
\end{align}
where $\hat{B}$ transforms as $\ov{\asymm}$ under a (fictitious) $\SU(6) \supset \SU(2)\times\SU(3)$ flavor symmetry, which is broken by the constraint:
\begin{align}\label{eqn:Bcons}
\epsilon^{m n} \hat{B}_{j_m k_n} = 0 \,,
\end{align}
as well as the (weak) gauging of $\SU(2)$. We impose the constraint via a Lagrange multiplier field $M^{i j}$:
\begin{align}
W = \frac{1}{2} \tilde{\lambda}\, \symp^{a b} \tA_a^M \tA_b^N \hat{B}_{M N} + \frac{1}{2} \epsilon^{m n} \hat{B}_{i_m j_n} M^{i j} + \frac{\Lambda_{\rm Sp}^9}{\det \tilde A} \,,
\end{align}
where $M,N$ index the (fictitious) $\SU(6)$. The $\hat{B}$ F-term condition is then
\begin{align}\label{eqn:hatBFterm}
\tilde{\lambda}\, \symp^{a b} \tA_a^{i_m} \tA_b^{j_n} = -\epsilon^{m n} M^{i j} \,,
\end{align}
so $M^{i j}$ is related to the holomorphic gauge invariant $\cA^{i j}$. Finally, the $\tA$ F-term condition reads:
\begin{align}
\tilde{\lambda}\, \symp^{a b} \tA_b^N \hat{B}_{M N} = \frac{\Lambda_{\rm Sp}^9}{\det \tilde A}\, \left(\tA^{-1}\right)^a_M \,,
\end{align}
or
\begin{align}
\hat{B}_{M N} = -\frac{\Lambda_{\rm Sp}^9}{\det \tilde A}\, \left[\left(\tilde{\lambda}\, \tA^T \Omega \tA \right)^{-1}\right]_{M N} \,.
\end{align}
Applying (\ref{eqn:hatBFterm}), we obtain
\begin{align}
\hat{B}_{i_m j_n} = -\frac{\Lambda_{\rm Sp}^9}{\det \tA}\, \epsilon_{m n}\, M^{-1}_{i j} \,.
\end{align}
However, this is incompatible with the constraint (\ref{eqn:Bcons}). Therefore, supersymmetry is broken.

In particular, for generic $|\tA| \gg |\tilde{\lambda}^{-1/9} \Lambda_{\rm Sp}|$, the classical superpotential dominates, and the classical F-terms set $\tilde{B}=0$. We obtain a semiclassical ``moduli-space'' parameterized by $\cA^{ij}$, subject to a runaway scalar potential generated by the ADS superpotential:\footnote{In particular, branch 1 of the classical moduli space is approximately flat for large $\det \cA^{ij}$. While other approximately flat regions corresponding to the other branches of moduli space may exist, they are not semiclassical, in that the classical superpotential must be made to cancel the large vacuum energy arising from the ADS superpotential.}
\begin{align}
W_{\rm eff} \sim \frac{\Lambda_{\rm Sp}^9}{\det \cA} \,.
\end{align}

\subsubsection{The \alt{$\SU(5)$}{SU(5)} theory} \label{subsec:SU5}
The F-term conditions are:
\begin{align}
A^{[i}_a A^{j]}_b = 0 \,, \quad A^{[j}_a B^{k];\,a b}=0 \,.
\end{align}
We now characterize the classical moduli space. The first F-term constraint implies that
\begin{align}
\langle A^i_a \rangle = v^i u_a \,,
\end{align}
where we may choose $u_a u^{* a} = 1$ without loss of generality. Suppose that $\langle A^i_a \rangle \ne 0$. We gauge fix such that $u_a = (0,0,0,0,1)$. Thus, the second F-term constraint implies
\begin{align}
\langle B^{i; \hat{a} 5} \rangle = b^{\hat{a}} v^i\,, \quad B^{i; \hat{a}\hat{b}} = b^{i; \hat{a} \hat{b}}\,,
\end{align}
where $\hat{a}, \hat{b} = 1 \ldots 4$.

Due to the first F-term constraint, the only possible nonvanishing holomorphic gauge invariant involving $A$ is the following:
\begin{align}
\mathcal{O}^{i j k l} = A^i_a B^{j;\,a b}B^{k;\,c d} B^{l;\, e f} \epsilon_{b c d e f} \, .
\end{align}
However, applying the above gauge-fixed forms for $\langle A \rangle$ and $\langle B \rangle$, we find that this also vanishes. This suggests that $\langle A \rangle = 0$ once the D-term conditions are imposed, which can be verified by an explicit computation.\footnote{In fact, to show that $\langle \Phi \rangle = 0$ in all supersymmetric vacua for some field $\Phi$, it is sufficient to show that for every solution to the F-term conditions with $\langle\Phi\rangle \ne 0$, another solution with $\langle\Phi\rangle = 0$ exists with all holomorphic gauge invariants taking the same vevs. This is because the latter solution must be equivalent to the unique D-flat solution with the same holomorphic-gauge-invariant vevs under an extended complexified gauge transformation~\cite{Luty:1995sd}, but such a gauge transformation will never regenerate a vev for $\Phi$.}

Since the F-term conditions are then identically satisfied, the classical moduli space is the subset of that of the $\lambda = 0$ theory (without a superpotential) where $\langle A \rangle = 0$. This theory is s-confining, with the confined description~\cite{Csaki:1996sm, Csaki:1996zb}:
\begin{equation}
\begin{array}{c|c|cccc}
     & \SU(5)  & \SU(3) & \SU(3) & \U(1) & \U(1)_R \\ \hline
    A^i & \ov{\fund} & \fund & \rd1 & -3 & 16/15 \\
    B^i & \asymm & \rd1 & \fund & 1 & -2/15 \\ \hline \hline
    T_i^m = A^2 B & & \ov{\fund} & \fund & -5 & 2\\
    U^{i; m}_{\;\;n} = A B^3 & & \fund & \adj
     & 0 & 2/3\\
    V^{m n} = B^5     & & \rd1 & \symm & 5 & -2/3
\end{array}
\end{equation}
with the dynamical superpotential:
\begin{align}
W = \frac{1}{\Lambda^9} \left(\epsilon_{m n p}\, T_i^m U^{i;n}_{\;\;q} V^{p q} - \frac{1}{3} \epsilon_{i j k} \,U^{i;m}_{\;\;p} U^{j;n}_{\;\;m} U^{k;p}_{\;\;n}\right) \,,
\end{align}
up to an overall multiplicative factor, where
\begin{align}
  T^m_i &\equiv \frac{1}{2} \epsilon_{i j k} A^j_a A^k_b B^{a b ; m}\,,\qquad
  U^{i ; m}_{\;  \;  \; n} \equiv \frac{1}{12} \epsilon_{n p q}
  \epsilon_{b c d e f} A^i_a B^{a b ; p} B^{c d ; q} B^{e f ; m}\,, \nonumber \\
  V^{m n} &\equiv \frac{1}{160} \epsilon_{p q r} \epsilon_{a_1 b_1
  c_1 d_1 e_1} \epsilon_{a_2 b_2 c_2 d_2 e_2} B^{a_1 a_2 ; p} B^{b_1 c_1 ;
  q} B^{b_2 c_2 ; r} B^{d_1 e_1 ; m} B^{d_2 e_2 ; n}\,.
\end{align}
One feature of s-confining theories is that their classical and quantum
moduli spaces match. Thus, the above spectrum of gauge invariants
describes the classical moduli space of the $\lambda = 0$ theory,
subject to the classical constraints, which are equivalent to the
F-term conditions arising from the dynamical superpotential. Setting
$\langle A \rangle = 0$, we find $\langle T \rangle = \langle U
\rangle = 0$, with no remaining F-term constraints on $V$. Thus, the
classical moduli space of the $\lambda \ne 0$ theory is parameterized
by $V^{i j}$, which transforms as a $\symm_{\;-2/3}$ under the
$\SU(3)\times\U(1)_R$ preserved by the superpotential. This matches
branch 1 of the classical moduli space of the $\Sp(6)\times\SU(2)$
theory described above, and both are parameterized by the baryon
discussed in~\S\ref{subsec:classicchecks}.

To obtain a quantum description of this theory, we perturb the s-confining theory (without superpotential) by the classical superpotential $A A B$,\footnote{There may also be instanton corrections to the superpotential, due to the completely broken $\SO(N-4)$ gauge group. These are subleading for $g_s \ll 1$ and vevs $\sim \Lambda_{\SU}$, but could play a role for very large vevs.} which can now be written in the form:
$$
W_{\rm class} = \lambda T^i_i \,,
$$
which breaks $\SU(3)\times\SU(3)\times\U(1) \to \SU(3)_{\rm diag}$, but preserves $\U(1)_R$. One can show that the resulting F-term conditions cannot be solved, and therefore supersymmetry is broken~\cite{Lykken:1998ec}.

For simplicity, we restrict our attention to the case where $V^{i j}$
is full rank. We are then entitled to make the field redefinitions:
\begin{align}
T^i_j = \hat{T}^i_j + \frac{\lambda^2 \Lambda^{18}}{4 \det V} \delta^i_j \,, \quad U^{i j}_l = \hat{U}^{i j}_k - \frac{1}{2} \lambda\, \Lambda^9\, \epsilon^{i j k} V^{-1}_{k l} \,.
\end{align}
The resulting superpotential is
\begin{align}
W =
\frac{1}{\Lambda^9} \,\epsilon_{i j k} \left(\hat{T}_l^i \hat{U}^{l j}_m V^{m k}
 +\frac{1}{3}\hat{U}^{i l}_{n} \hat{U}^{j m}_{l} \hat{U}^{k n}_{m}\right)
  + \frac{\lambda}{2} \left(\hat{U}^{i k}_{l} \hat{U}^{l j}_{k}-\hat{U}^{i j}_{l} \hat{U}^{k l}_{k}\right)V^{-1}_{i j}
    +\frac{\lambda^3 \Lambda^{18}}{4 \det V} \,.
\end{align}
To show that there are no F-flat solutions, we first show that $\hat{T}=\hat{U}=0$ is the only solution to $\hat{T}$ and $\hat{U}$ F-term conditions for full-rank $V$. Note that in this case, $V^{i j}$ can always be brought to the form $V^{i j} \propto \delta^{i j}$ after a complexified $\SU(3)$ transformation. As the F-term conditions are appropriately covariant under this complexified symmetry transformation, it is sufficient to show that $\hat{T}=\hat{U}=0$ is the only solution for $V^{i j} = z\, \delta^{i j}$.

In this case, the $\hat{T}$ F-term conditions reduce to
\begin{align}
\epsilon_{i j k} \hat{U}^{l j}_k = 0\,,
\end{align}
so that $\hat{U}^{i j}_k = \hat{U}^{i k}_j$. The $\hat{U}$ F-term conditions are:
\begin{align}
0 = \frac{1}{\Lambda^9} \, \left(\epsilon_{i n m} \hat{T}_k^i \,z
 +\epsilon_{i j k} \hat{U}^{i l}_{n} \hat{U}^{j m}_{l} \right)
+ \frac{\lambda}{2 z} \left(\hat{U}^{m k}_{n}+\hat{U}^{n m}_{k}-\delta_{k n} \hat{U}^{i m}_{i}-\delta_{k m} \hat{U}^{i i}_{n}\right) \,.
\end{align}
Extracting the component which is symmetric in $n\leftrightarrow m$, we obtain
\begin{align}
\hat{U}^{(n m)}_{k}-\delta_{k (n} \hat{U}^{i i}_{m)} = 0
\end{align}
after applying the $T$ F-term condition. Contracting with $\delta_{k m}$ we find $\hat{U}^{i i}_j = 0$, so the above condition reduces to
\begin{align}
\hat{U}^{m n}_{k}+\hat{U}^{n m}_{k} = 0\,.
\end{align}
Together with the $\hat{T}$ F-term condition, this is sufficient to show that $\hat{U}=0$. The remaining components of the $\hat{U}$ F-term condition then imply that $\hat{T}=0$. By the above argument, these results apply for arbitrary (full-rank) $V$.

Having solved the $\hat{T}$ and $\hat{U}$ F-term conditions for $\hat{T}$ and $\hat{U}$, we may ``integrate out'' these fields, leaving the effective superpotential:
\begin{align}
W_{\rm eff} = \frac{\lambda^3 \Lambda^{18}}{4 \det V} \,,
\end{align}
for $V^{i j}$, which has no F-flat solutions and generates a runaway scalar potential, much like the $\Sp(6)\times\SU(2)$ theory.

\section{Matching of superconformal indices}
\label{sec:SCI}

In this section we discuss another very nontrivial test of the
proposed duality: the matching between the superconformal indices of
the two gauge theories. The discussion is inherently somewhat
technical in nature, and readers primarily interested in the gauge
theoretic consequences of the proposed duality may wish to skip ahead
to~\S\ref{sec:infrared}.

Superconformal indices for $\cN=1$ theories compactified on $S^3\times
\bR$ \cite{Romelsberger:2005eg,Kinney:2005ej}, while being a
relatively recent development, have already provided important
insights into the topic of dualities. In particular, equality of the
superconformal index provides very strong support for a number of
known and conjectured Seiberg dualities between $\cN=1$ theories
\cite{Romelsberger:2007ec,Dolan:2008qi,Spiridonov:2008zr,Spiridonov:2009za,Spiridonov:2010hh,Spiridonov:2011hf,Sudano:2011aa,Spiridonov:2012ww,Yamazaki:2012cp,Terashima:2012cx}
and S-dualities in $\cN=2$ \cite{Gadde:2009kb,Gadde:2010te,Gadde:2011uv,Gaiotto:2012uq} and
$\cN=4$ theories \cite{Gadde:2009kb,Spiridonov:2010qv}. It also proves to be a very useful tool in the study
of holography \cite{Kinney:2005ej}. In this section we will present
evidence for the agreement of the superconformal indices of the dual
pair of theories presented in~\S\ref{sec:C3Z3fieldtheory}. As we
will see momentarily, the agreement relies on extremely non-trivial
group-theoretical identities, providing very strong support for our
conjectured duality.

It is not our intention to give a detailed discussion of the
superconformal index here (we refer the interested reader to
\cite{Romelsberger:2005eg,Kinney:2005ej,Dolan:2008qi,Sudano:2011aa} for
very readable expositions of the topic), but we will briefly review in
this section the basic elements that enter into its computation in
order to settle notation. Consider a four dimensional theory
compactified on $S^3\times\bR$. The superconformal algebra has
generators $J_{\pm},J_3,\ov J_\pm, \ov J_3$ (associated to rotations
on the $S^3$), supersymmetry generators $Q_\alpha,\ov Q_{\dot\alpha}$,
translations $P_\mu$, special superconformal generators
$K_\mu,S_\alpha,\ov S_{\dot\alpha}$, superconformal dilatations $H$
and the $U(1)$ $R$-symmetry generator $R$. Define $Q=Q_1$, which
implies \cite{Kinney:2005ej} $Q^\dagger=S_1$. The superconformal
algebra then gives:
\begin{align}
  2\{Q^\dagger,Q\} = H - 2 J_3 - \frac{3}{2}R \equiv \cH \, .
\end{align}
The superconformal index is then defined by:
\begin{align}
  \label{eq:SCI-index-general}
  \Tr (-1)^F e^{-\beta\cH} M\, ,
\end{align}
with $F$ the fermion number operator, and $M$ any symmetry commuting
with $Q$ and $Q^\dagger$. Let us choose $M$ to be generated by $\cR\equiv R + 2J_3$, $\ov
J_3$, and gauge and flavor group elements $g,f$. Introducing
appropriate chemical potentials, the refined index is thus given by:
\begin{align}
  \label{eq:SCI-definition}
  \cI(t,x,f) = \int\!dg\, \Tr (-1)^F e^{-\beta\cH} t^\cR x^{2\ov J_3}
  f g\, ,
\end{align}
where we have integrated over the gauge group in order to count
singlets only (we will have more to say about this integration
below). It was argued in \cite{Kinney:2005ej} that, exactly as in the
case of the Witten index \cite{Witten:1982df}, the
index~\eqref{eq:SCI-definition} receives contributions only from
states annihilated by $Q$ and $Q^\dagger$, and thus the index does not
actually depend on $\beta$, which plays the role of a regulator
only.

In order to actually compute~\eqref{eq:SCI-definition} we follow the
prescription in \cite{Romelsberger:2007ec}, which gives a systematic
way of computing the superconformal index in terms of the fields in
the weakly coupled Lagrangian description of the theory, when one is
available. For a general weakly coupled theory $\cT$ with gauge group
$G$ and flavor group $F$, neither necessarily simple, and matter
fields $X_i$ with superconformal $R$-charge $r_i$ in the
representation $R^{i}_G\otimes R^i_F$ of $G\times F$, not necessarily
irreducible, one constructs the letter
\begin{align}
  \label{eq:SCI-letter}
  \begin{split}
    i_\cT(t, x, g, f) & =
    \frac{(2t^2-t(x+x^{-1}))\chi_{\adj}(g)}{{(1-tx)(1-tx^{-1})}}\\
    & \phantom{=} + \frac{\sum_i
      \left(t^{r_i} \chi_{R^i_G}(g)\,\chi_{R^i_F}(f) - t^{2-r_i}\chi_{\ov{
            R^i_G}}(g)\, \chi_{\ov{R^i_F}}(f)\right)}{(1-tx)(1-tx^{-1})} \, .
  \end{split}
\end{align}
Here $t,x,g,f$ are the same as
in~\eqref{eq:SCI-definition}. $\chi_R(g)$ denotes the character of $g$
in the representation $R$, and we denote with bars the complex
conjugate representations.
Once we have the letter~\eqref{eq:SCI-letter} for $\cT$, the
superconformal index $\cI_\cT$ is obtained by taking the plethystic
exponential, and integrating over the gauge group:
\begin{align}
  \label{eq:SCI-index}
  \cI_\cT(t,x,f) = \int dg \, \exp\left[\, \sum_{k=1}^\infty
    \frac{1}{k}i_\cT(t^k, x^k, g^k, f^k)\right]\, .
\end{align}
Here $dg$ denotes the Haar measure on the group $G$.\footnote{We refer
the reader to \cite{Bump,FultonHarris} for nice references to the
Lie group representation theory that we will need. We will give
explicit expressions for the Haar measure of the groups of interest
to us in~\S\ref{sec:SCI-large-N}.}

We will thus compute the index in a weakly coupled, non-conformal
description of the theory, and will assume that this gives the right
index for the theory at its (presumed) superconformal fixed point in
the IR. In the case of the theory compactified on $S^3$, one can argue
\cite{Gadde:2010en} that since the index is independent of the radius
$r$ of the $S^3$ it is independent of the dimensionless $r\Lambda$
coupling, and thus it is invariant under the RG flow and changes in
the coupling constant. In order for this to be the case, we need to
choose $M$ in~\eqref{eq:SCI-index-general} constant along the flow,
and in particular it should agree with the value of $M$ at the IR
superconformal fixed point. In particular, we need to choose the right
value of the superconformal $R$-charge --- determined using
a-maximization \cite{Intriligator:2003jj}, for instance --- in
constructing $M$.

Ideally, one would compute a closed form expression
for~\eqref{eq:SCI-index} in the two dual phases, and then show that
the two expressions agree for all $N$. Unfortunately we have not been
able to prove the equality of the resulting indices, but in~\S\ref{sec:SCI-t-expansion} and~\S\ref{sec:SCI-large-N} we will
provide very non-trivial evidence for the matching of the functions in
two particularly tractable limits.  The exact matching will thus
remain a well motivated conjecture about elliptic hypergeometric
integrals, which we formulate precisely in
appendix~\ref{sec:SCI-identity}.

\subsection{Expansion in \alt{$t$}{t}}
\label{sec:SCI-t-expansion}

The first limit corresponds to an expansion around
$t=0$. Expanding~\eqref{eq:SCI-index} is elementary, but the
integration over the gauge group requires some more advanced
technology. In particular, one needs to use orthogonality of the
characters under integration:
\begin{align}
  \label{eq:Schur}
  \int dg \, \chi_{R_i}(g) \chi_{\ov{R_j}}(g) = \delta_{ij}\, .
\end{align}
where $R_i$ and $R_j$ are irreps of $G$. When expanding the plethystic
exponential~\eqref{eq:SCI-index} one encounters expressions of the
form (we will deal with higher powers of $g$ momentarily):
\begin{align}
  \label{eq:generic-gauge-integral}
  \int dg\, \chi_{R_1}(g) \cdots \chi_{R_n}(g)\, .
\end{align}
By using the well know property of the characters
$\chi_{R_1}(g)\chi_{R_2}(g)=\chi_{R_1\otimes R_2}(g)$, and then
plugging the resulting expression into ~\eqref{eq:Schur} with the
second term being the character of the trivial representation (i.e.
just 1), we obtain that~\eqref{eq:generic-gauge-integral} just counts
the number of singlets in $R_1\otimes \ldots \otimes R_n$.

When expanding~\eqref{eq:SCI-index} we will also encounter terms of
the form $\chi_R(g^n)$. The act of decomposing such terms into
characters of irreducible representations with group element $g$ is
known as applying the $n$-th Adams operator $\sA_n$ to $R$. As an
example, consider the fundamental representation $\fund$ for $\SU(N)$,
which has character:\footnote{Characters for representations of Lie
  groups can be worked out systematically using the Weyl character
  formula, see for example \cite{Bump}.}
\begin{align}
  \chi_{\sfund}(g) = \sum_{i=1}^N t_i \, ,
\end{align}
where $t_i$ are the elements of $g$ on the maximal torus of
$\SU(N)$. Similarly, for the symmetric $\symm$ and antisymmetric
$\asymm$ representations we have:
\begin{align}
  \chi_{\ssymm}(g) & = \sum_{1\leq i \leq j \leq N} t_i t_j +
  \sum_{i=1}^N t_i^2\,,\\
  \chi_{\sasymm}(g) & = \sum_{1\leq i \leq j \leq N} t_i t_j \, .
\end{align}
It is thus clear that $\sA_2(\fund) = \symm - \asymm$, or in terms of
characters:
\begin{align}
  \chi_{\sfund}(g^2) = \chi_{\ssymm}(g) - \chi_{\sasymm}(g)\, .
\end{align}
Proceeding systematically in this way, one can decompose the
plethystic exponential, up to any order, into sums of products of
characters of irreps, which can then be easily integrated over the
gauge group. The flavor characters can be dealt with similarly, and we
will give the final results in terms of irreps. In the flavor case it
is particularly important to do the decomposition into irreducible
representations since there are important cancellations between terms,
we will give an example below.

When the problem is formulated in this way the rest of the computation
is conceptually straightforward, but doing this by hand quickly turns
impossible, and the aid of computer systems is required for doing any
non-trivial computations. We took advantage of the computer algebra
package \texttt{LiE} \cite{LiE} for doing the relevant group
decompositions and Adams operations, and the mathematics software
system \texttt{Sage} \cite{sage} for the polynomial
manipulations.\footnote{The actual code we used for the calculation
  can be downloaded from \url{http://cern.ch/inaki/SCI.tar.gz}}

With this technology in place, the actual computation of the indices
is straightforward, if lengthy. We obtain perfect agreement of the
indices between the two dual theories in~\S\ref{sec:C3Z3fieldtheory} up to the degrees that we
checked. In particular, for $\SO(3)\times \SU(7)\leftrightarrow
\Sp(8)\times \SU(4)$, we obtain the index:
  \begin{align}\label{eq:SCIforN=7}
    \cI_{\SO/\Sp}(t,x,f) & = 1 + t^{\frac{2}{3}}\bigl[\chi_{0,2}(f) +
    \chi_{4,0}(f)\bigr] \nn \\
    & \phantom{=}\, + t^{\frac{4}{3}}\bigl[2\chi_{0,4}(f) +
    2\chi_{2,0}(f) + \chi_{3,1}(f) + 2\chi_{4,2}(f) + \chi_{8,0}(f)
    \bigr] \nn \\
    & \phantom{=}\, + t^{\frac{5}{3}}(x+x^{-1})\bigl[\chi_{0,2}(f) +
    \chi_{4,0}(f)\bigr] \nn \\
    & \phantom{=}\, + t^2\bigl[4 + 3\chi_{0,6}(f) +
    \chi_{1,4}(f) + 5 \chi_{2,2}(f) + 3\chi_{3,3}(f) \nn \\
    & \phantom{=\, + t^2\bigl[} +2\chi_{4,1}(f) +
    3\chi_{4,4}(f) + \chi_{5,2}(f) + 4\chi_{6,0}(f) + \chi_{6,3}(f) \nn \\
    & \phantom{=\, + t^2\bigl[} + \chi_{7,1}(f) + 2\chi_{8,2}(f) + \chi_{12,0}(f)
    \bigr]  + \ldots
  \end{align}
where we have omitted terms of higher order in $t$. We have denoted
the $\SU(3)$ representation by its Dynkin labels, so for example the
representation with $(2,2)$ Dynkin labels can be described as
$\yss\ydiagram{4,2}\ysn$ in terms of ordinary Young tableaux. Notice that, as
we were indicating before, even at this relatively low order the
matching of the indices is very non-trivial, with rather complicated
character polynomials appearing. Furthermore, the agreement is only
obtained after some rather involved group theory cancellations. As a
particularly simple example, consider the $t^{\frac{2}{3}}$ term. In
the $\Sp(8)\times\SU(4)$ theory the relevant contribution after doing
the gauge integration is of the form:
\begin{equation}
  \cI_{\Sp}(x,t,f) =
  t^{\frac{2}{3}}\biggl[\frac{1}{8}\chi_{\sfund}^4(f) +
  \frac{1}{4}\chi_{\sfund}^2(f)\chi_{\sfund}(f^2) +
  \frac{3}{8}\chi_{\sfund}^2(f^2) +
  \frac{1}{4}\chi_{\sfund}(f^4)\biggr] + \ldots
\end{equation}
where we have ignored terms of other orders in $t$. On the other hand,
the corresponding expression for the $\SO(3)\times\SU(7)$ theory is
given by:
\begin{align}
  \begin{split}
    \cI_{\SO}(x,t,f) & =
    t^{\frac{2}{3}}\biggl[\frac{1}{140}\chi_{\sfund}^7(f) +
    \frac{1}{40}\chi_{\sfund}^5(f)\chi_{\sfund}(f^2)
    -\frac{1}{8}\chi_{\sfund}(f)\chi_{\sfund}^3(f^2) \\
    & \phantom{= t^{\frac{2}{3}}\biggl[} - \frac{1}{4}
    \chi_{\sfund}(f)\chi_{\sfund}(f^2)\chi_{\sfund}(f^4) +
    \frac{1}{10}\chi_{\sfund}^2(f)\chi_{\sfund}(f^5) \\
    & \phantom{= t^{\frac{2}{3}}\biggl[}
    + \frac{1}{10}\chi_{\sfund}(f^2)\chi_{\sfund}(f^5)
    + \frac{1}{7}\chi_{\sfund}(f^7)\biggr] + \ldots
  \end{split}
\end{align}
again ignoring terms of different degree in $t$. We clearly see that
both expressions look rather different, and only agree after using
some non-trivial group theoretical identities involving the Adams
operator.

One can proceed similarly for other ranks. As we have seen
in~\S\ref{subsec:N=5} reducing the rank leads to a runaway theory, so
we will restrict ourselves to larger ranks.\footnote{In terms of the
  index itself, setting $N\leq 5$ leads to negative
  R-charges for the chiral multiplets $B$ and $\tilde A$, and thus
  to negative powers of $t$ in the letter~\eqref{eq:SCI-letter}, which
  makes our method of computation ill-defined. The physical origin of
  this divergence is the same as in simpler examples with runaway
  superpotentials, such as SQCD with $0<N_f<N_c$, in which
  $a$-maximization gives rise to charge assignments allowing for gauge
  invariant operators (in our case $B^{N}\sim \tilde A^{N-3}$) with negative
  $R$-charge, which would violate unitarity if there were a
  superconformal vacuum without chiral symmetry breaking (assuming the absence of accidental $U(1)$
  symmetries in the IR).} In particular we have calculated the
superconformal index for $\SO(5)\times\SU(9)\leftrightarrow
\Sp(10)\times\SU(6)$ and $\SO(7)\times\SU(11)\leftrightarrow
\Sp(12)\times\SU(8)$ up to order $t^4$ and found in both cases perfect
agreement.

It is interesting to construct explicitly the states that are
annihilated by $\cH$ and therefore contribute to the superconformal
index \eqref{eq:SCI-definition}. They are the scalar components of the
chiral multiplets, the right-handed chiral fermions in the complex
conjugate representation as well as the gauginos and field strengths
of the gauge groups. The fields that contribute for $\SO(N-4)\times \SU(N)$ are shown in table~\ref{tab:SONindexfields}.
\begin{table}
\begin{center}
\begin{tabular}{|c|c|c|c|c|}
  \hline
  Field & $\SO(N-4)\times \SU(N)$ & $\SU(3)$ & $t$ exponent & $\SU(2)_{r}$\\\hline
  $A^i_{(l)}$ & $(\fund,\ov{\fund})$ & $\fund$ & $\frac{2}{3}+\frac{2}{N}+l$ & $l+1$ \\[1pt]\hline
  $B^i_{(l)}$ & $\lp\sing,\asymm\rp$ & $\fund$ &$\frac{2}{3}-\frac{4}{N}+l$ & $l+1$ \\[1pt]\hline
  $\bar{\psi}^A_{(l)}$ & $(\fund,\fund)$ & $\ov{\fund}$ &$\frac{4}{3}-\frac{2}{N}+l$ & $l+1$ \\[1pt]\hline 
  \rule{0pt}{17pt} $\bar{\psi}^B_{(l)}$ & $\lp\sing,\ov{\asymm}\rp$ & $\ov{\fund}$ &$\frac{4}{3}+\frac{4}{N}+l$ & $l+1$ \\[5pt]\hline
  $\lambda_{(l)}^{\SO}$ & $(\asymm,\sing)$ & \sing &$1+l$ & $l \oplus (l+2)$\\[1pt]\hline
  $F_{(l)}^{\SO}$ & $(\asymm,\sing)$ & \sing &$2+ l$ & $(l+1) \oplus (l+1)$\\[1pt]\hline
  $\lambda_{(l)}^{\SU}$ & $(\sing,\adj)$ & \sing &$1+l$ & $l \oplus (l+2)$\\[1pt]\hline
  $F_{(l)}^{\SU}$ & $(\sing,\adj)$ & \sing &$2+ l$ & $(l+1) \oplus (l+1)$\\[1pt]\hline
\end{tabular}
\end{center}
\caption[The fields which contribute to the superconformal index for $\SO(N-4)\times \SU(N)$]{The fields which contribute to the superconformal index for $\SO(N-4)\times \SU(N)$, where the $\SU(2)_r$ column denotes the representation under the $\SU(2)$ group generated by $\ov J_\pm, \ov J_3$.\label{tab:SONindexfields}}
\end{table}

The superconformal index counts gauge invariant
combinations of these fields and we can explicitly construct these
combinations to check our result \eqref{eq:SCIforN=7}. For the $\SO(3)\times \SU(7)$ theory, the gauge invariants which contribute at the lowest orders in the Taylor expansion about $t=0$ are shown in table~\ref{tab:SO3indexinvariants}.\footnote{For the
  operator $\lp B^i_{(0)} \rp^{21}$ a direct computation of the
  representation under the flavor group takes very long so that we
  devised a refined method that is explained in appendix
  \ref{sec:B21}.}
\begin{table}
\begin{center}
\begin{tabular}{|c|c|c|c|}
  \hline
  operator & $t$ exp. & $2\ov J_3$ & $\SU(3)$ character\\\hline
  \rule{0pt}{17pt}$\lp B^i_{(0)} \rp^7$ & $\frac{2}{3}$ & 0 & $\chi_{0,2}(f)+\chi_{4,0}(f)$ \\[4pt]\hline
  \rule{0pt}{17pt}$\lp B^i_{(0)} \rp^{14}$ & $\frac{4}{3}$ & 0 & $2\chi_{0,4}(f)+2\chi_{2,0}(f)+\chi_{3,1}(f)+2\chi_{4,2}(f)+\chi_{8,0}(f)$ \\[4pt]\hline
  \rule{0pt}{17pt}$\lp B^i_{(0)} \rp^6 B^i_{(1)}$ & $\frac{5}{3}$ & $\pm 1$ & $\chi_{0,2}(f)+\chi_{4,0}(f)$ \\[4pt]\hline
  $[\lambda^{\SO}_{(0)}]^2$ & $2$ & 0 & 1 \\[1pt]\hline
  $[\lambda^{\SU}_{(0)}]^2$ & $2$ & 0 & 1 \\[1pt]\hline
  $A_{(0)} \psi^A_{(0)}$ & $2$ & 0 & $1+\chi_{1,1}(f)$ \\[1pt]\hline
  $B_{(0)} \psi^B_{(0)}$ & $2$ & 0 & $1+\chi_{1,1}(f)$  \\[1pt]\hline
  $(A_{(0)})^2 B_{(0)}$ & $2$ & 0 & $1+\chi_{1,1}(f)$  \\[1pt]\hline
  \multirow{3}{*}{$\lp B^i_{(0)} \rp^{21}$} & \multirow{3}{*}{$2$} & \multirow{3}{*}{0} & $3 + 3\chi_{0,6}(f) + \chi_{1,1}(f) + \chi_{1,4}(f) + 5 \chi_{2,2}(f) $ \\
  & & & $+ 3\chi_{3,3}(f)+ 2 \chi_{4,1}(f) + 3\chi_{4,4}(f) + \chi_{5,2}(f)$ \\
  & & & $+4\chi_{6,0}(f)+ \chi_{6,3}(f)+ \chi_{7,1}(f) + 2\chi_{8,2}(f) + \chi_{12,0}(f)$ \\\hline
 \end{tabular}
\end{center}
\caption[The gauge-invariants contributing to the superconformal index for $\SO(3)\times \SU(7)$]{The gauge-invariants contributing to the superconformal index of the $\SO(3)\times \SU(7)$ theory at the lowest orders in the Taylor expansion about $t=0$, where $()^m$ denotes taking the $m$-th symmetrized tensor product and
$[]^{m}$ taking the $m$-th antisymmetrized tensor product.\label{tab:SO3indexinvariants}}
\end{table}
Taking into account the factor $(-1)^F$ we find perfect agreement with \eqref{eq:SCIforN=7}.

Likewise we can check the gauge invariant contributions for the
$\Sp(8)\times \SU(4)$ theory. We again find perfect agreement with
\eqref{eq:SCIforN=7} as is shown in detail in appendix
\ref{app:SCIdetails}.

\subsection{Large \alt{$N$}{N}}
\label{sec:SCI-large-N}

Using the tools given above, one can go as high in $N$ and $t$ as
desired, limited only by computing resources and patience. In this
section we will approach the computation of the index from a
complementary perspective, namely we will compute the index in the
dual pair of $\SO(N-4)\times\SU(N)$ and $\Sp(N+1)\times\SU(N-3)$
theories when $N\to\infty$, following the discussion in
\cite{Kinney:2005ej,Dolan:2008qi,Gadde:2010en}.

Let us start by taking the large $N$ limit of the Haar measures for
group integration over the $ABC$ Lie groups appearing in our
construction. Starting with $\SU(N)$, the explicit form of the
integral of a gauge invariant function $f(g)$ (such as a function of
group characters) over the group is given by \cite{Dolan:2008qi}:
\begin{align}
  \label{eq:SU(N)-Haar}
  \int dg f(g) = \frac{1}{N!}\oint \prod_{j=1}^{N-1} \frac{dx_i}{2\pi i x_i}
  \Delta(x) \Delta(x^{-1}) f(x)\, ,
\end{align}
with $\Delta(t) = \prod_{i < j}(t_i - t_j)$, and the integration can
be taken to be on the unit circle around $x_i=0$. Parameterizing
$x=e^{i\theta}$, the integral~\eqref{eq:SU(N)-Haar} can be
equivalently rewritten as:
\begin{align}
  \int dg f(g) = \frac{1}{N!} \frac{1}{(2\pi)^{N-1}}\int
  \prod_{j=1}^{N-1} d\theta_j \,\, \Delta(e^{i\theta})
  \Delta(e^{-i\theta}) f(\theta)\, .
\end{align}
Using now that $\sum_{n=1}^\infty x^n/n = -\log(1-x)$, this can be
conveniently rewritten as:
\begin{align}
  \label{eq:SU(N)-Haar-exp}
  \int dg f(g) = \frac{1}{N!} \frac{1}{(2\pi)^{N-1}} \int
  \prod_{k=1}^{N-1} d\theta_k \exp\left(-\sum_{n=1}^{\infty}
    \frac{1}{n} \sum_{i\neq j} e^{in(\theta_i - \theta_j)}\right)
  f(\theta)\, .
\end{align}
Let us point out in passing that this expression, modulo some constant
factors, can be also rewritten as
\begin{align}
  \int dg f(g) \propto \int \prod_{k=1}^{N-1} d\theta_k
  \exp\left(-\sum_{n=1}^{\infty} \frac{1}{n}
    \chi_{\adj}(e^{in\theta})\right)\, f(\theta) \, ,
\end{align}
where $\chi_{\adj}(e^{in\theta})$ denotes the character of $x^n$ in the
adjoint. This structure also applies to the $\Sp$ and $\SO$ cases we
analyze below.

In the large $N$ limit, we replace the sum over eigenvalues $\sum_i$
with a continuous integral $N \int d\alpha$. We also have that
$\theta$ becomes a continuous function $\theta(\alpha)$. It is
convenient to change the variable of integration to $\theta$ itself:
$\int d\alpha \to \int d\theta \rho(\theta)$, where we have denoted
the Jacobian $\rho(\theta)=d\alpha/d\theta$. Doing these changes, we
have that at large $N$:
\begin{align}
  \begin{split}
    \sum_{i\neq j} e^{in(\theta_i - \theta_j)} & = \biggl(\sum_i
      e^{in\theta_i}\biggr)\biggl(\sum_j e^{-in\theta_j}\biggr) - N \\
    & \to N^2\left(\int\!d\theta\rho\,e^{in\theta}\right)\left(\int
      d\theta \!\rho\,e^{-in\theta}\right) - N\, .
  \end{split}
\end{align}
In what follows we will drop constant terms (those independent of
$\rho$) for simplicity, we will account for their effect by fixing the
normalization of the final result. It is also convenient to introduce,
as in \cite{Dolan:2008qi}, $\rho_n = N\int\! d\theta\, \rho\,
e^{in\theta}$. With these changes, we have that:
\begin{align}
  \sum_{i\neq j} e^{in(\theta_i - \theta_j)} \to |\rho_n|^2\, ,
\end{align}
and the integration becomes simply a product of complex gaussian
integrals:
\begin{align}
  \begin{split}
    \int dg f(g) \to & \int \prod_{n=1}^{\infty} \left( \frac{i\,
        d^2\rho_n}{2\pi n} e^{-\frac{1}{n}|\rho_n|^2}\right) f(\rho) \\
    & \equiv \int \prod_{n=1}^{\infty} [d^2\rho_n] f(\rho) \equiv \int
    [d^2\rho] f(\rho) \, ,
  \end{split}
\end{align}
where we have introduce some convenient notation, and imposed unit
normalization for the large $N$ measure: $\int [d^2\rho_n]\, 1 = 1$.

We can proceed similarly for the other cases of interest to us. For
the $\Sp(2N)$ group, the Haar measure is given by:
\begin{align}
  \int dg f(g) = \frac{(-1)^N}{2^N\, N!} \oint \left( \prod_{j=1}^N
    \frac{dx_j}{2\pi i x_j} \bigl(x_j - x_j^{-1}\bigr)^2 \right)
  \Delta(x+x^{-1})^2 f(x) \, .
\end{align}
By an argument very similar to the above, we can rewrite this as
(ignoring constant prefactors):
\begin{align}
  \begin{split}
    \int dg f(g) \propto \int \prod_{j=1}^N d\theta_j
    \exp\biggl[-\sum_{n=1}^\infty \frac{1}{2n} \biggl\{& \biggl(\sum_k
      \bigl(e^{in\theta_k} + e^{-in\theta_k}\bigr)\biggr)^2 \\
      & + \sum_k
      \bigl(e^{2in\theta_k} + e^{-2in\theta_k}\bigr) \biggr\}\biggr] f(\theta)\, .
  \end{split}
\end{align}
It is thus natural to introduce $\gamma=d\alpha/d\theta$ as before,
and to define the real variable $\gamma_n\equiv N\int d\theta\,
\gamma(\theta)\, (e^{in\theta} + e^{-in\theta})$. The resulting
measure is again an infinite product of (real, in this case) Gaussian
integrals, which when properly normalized can be written as:
\begin{align}
  \begin{split}
    \int dg f(g) \to & \int \left[\prod_{n=1}^\infty \frac{d\gamma_n}{\sqrt{2\pi
          n}} \exp\biggl(-\frac{1}{2n}(\gamma_n + 1)^2\biggr)\right]f(\gamma)\\
    & \equiv \int \prod [d\gamma_n] f(\gamma) \equiv \int [d\gamma]
    f(\gamma)\, .
  \end{split}
\end{align}

Finally, for $\SO(2N+1)$, the process works very similarly to
$\Sp(2N)$. The integration over the gauge group is given by
\begin{align}
  \int dg f(g) = \frac{(-1)^N}{2^N\, N!} \oint \left( \prod_{j=1}^N
    \frac{dx_j}{2\pi i x_j} \bigl(\sqrt{x_j} - \sqrt{x_j}^{-1}\bigr)^2
  \right) \Delta(x+x^{-1})^2 f(x) \, ,
\end{align}
which at large $N$ becomes
\begin{align}
  \begin{split}
    \int dg f(g) \to & \int \left[\prod_{n=1}^\infty \frac{d\gamma_n}{\sqrt{2\pi
          n}} \exp\biggl(-\frac{1}{2n}(\gamma_n - 1)^2\biggr)\right]f(\gamma)\\
    & \equiv \int \prod [d\gamma_n] f(\gamma) \equiv \int [d\gamma]
    f(\gamma)\, ,
  \end{split}
\end{align}
where we have introduced $\gamma_n\equiv 1 + N\int d\theta\,
\gamma(\theta)\, (e^{in\theta} + e^{-in\theta})$. We have chosen to
shift the definition of $\gamma_n$ by 1 in order to make the argument
for the equality of the indices below more straightforward.

In order to rewrite the superconformal index
\eqref{eq:SCI-letter}, \eqref{eq:SCI-index} at large $N$, we need to find
out the large $N$ limit of the characters of the representations
appearing in our theory. Consider for instance the symmetric
representation of $\SU(N)$. Its character is given by:
\begin{align}
  \begin{split}
    \chi_{\ssymm}(x) & = \sum_{i<j} x_i x_j + \sum_{i=1}^N x_i^2 =
    \frac{1}{2}\left(\sum_{i\neq j} x_i x_j\right) + \sum_{i=1}^N
    x_i^2\\
    & = \frac{1}{2}\left(\biggl(\sum_{i} x_i\biggr)^2 + \sum_{i=1}^N
      x_i^2\right)\, .
  \end{split}
\end{align}
Introducing $\rho_n$ as before, this can be rewritten as:
\begin{align}
  \chi_{\ssymm}(x) \to \frac{1}{2}(\rho_1^2 + \rho_2)\, .
\end{align}
Other representations can be treated similarly, let us just quote the
results that we will need. For $\SU(N)$ we have:
\begin{align}
  \chi_{\SU,\,\sfund}(x^n) & = \sum_{i} x^n_i \;\;\;\;\to\; \rho_n \,,\\
  \chi_{\SU,\,\sasymm}(x^n) & = \sum_{i<j} x^n_i x^n_j - \sum_{i=1}^N
  x_i^{2n} \;\;\;\;\to\; \frac{1}{2}(\rho_n^2 - \rho_{2n}) \,,\\
  \chi_{\SU, \adj}(x^n) & = \sum_{i,j} x_i^n x_j^{-n} - 1 \;\;\;\; \to \;
  |\rho_n|^2 -1 \, .
\end{align}
For $\Sp(2N)$ we have:
\begin{align}
  \chi_{\Sp,\,\sfund}(x^n) & = \sum_{i}(x_i^n+x_i^{-n}) \;\;\;\; \to\; \gamma_n\,,\\
  \begin{split}
    \chi_{\Sp, \adj}(x^n) & = \sum_{i<j}(x_i^n x_j^n + x_i^n x_j^{-n}
    +
    x_i^{-n} x_j^n + x_i^{-n} x_j^{-n}) \\ &\phantom{=}+ \sum_i(x_i^{2n} + x_i^{-2n}) + N 
     \;\;\;\; \to \; \frac{1}{2}(\gamma_n^2 + \gamma_{2n})\,,
  \end{split}
\end{align}
and similarly for $\SO(2N+1)$:
\begin{align}
  \chi_{\SO,\,\sfund}(x^n) & = \sum_{i}(x_i^n+x_i^{-n}) + 1 \;\;\;\; \to\; \gamma_n\,,\\
  \begin{split}
    \chi_{\SO, \adj}(x^n) & = \sum_{i<j}(x_i^n x_j^n + x_i^n x_j^{-n}
    +
    x_i^{-n} x_j^n + x_i^{-n} x_j^{-n}) \\ &\phantom{=}+ \sum_i(x_i^{n} + x_i^{-n}) + N 
    \;\;\;\; \to \; \frac{1}{2}(\gamma_n^2 - \gamma_{2n})\, .
  \end{split}
\end{align}

The equality of the indices at large $N$ between the $\SO\times\SU$
and $\Sp\times\SU$ cases now follows from a simple redefinition of the
integration variables: $\rho_n\leftrightarrow -\rho_n$,
$\gamma_n\leftrightarrow -\gamma_n$. Indeed, under this change of
variables, for the measures of integration we have that $[d^2\rho]$
stays invariant, while $[d\gamma]_{\SO}$ gets exchanged with
$[d\gamma]_{\Sp}$. Similarly, we have that
$\chi_{\SO,\adj}\leftrightarrow \chi_{\Sp,\adj}$, the symmetric and
antisymmetric characters of $\SU$ get exchanged, and the character of
the bifundamental, given by $\rho_n\gamma_n$, stays invariant. This is
precisely the map between the two dual theories.

\medskip

It is also instructive to compare the result of the large $N$
computation in this section with the low $N$ computations in the
previous section. From the discussion in subsection
\ref{subsec:classicchecks}, baryons start contributing at order
$t^{\frac{2N}{3}-4}$, and thus disappear in the large $N$ limit of the
expressions above. The mesonic contributions have $N$ independent $t$
exponent, and survive the limit. This means in particular that the $t$
expansion becomes $N$ independent for large $N$. As an illustration,
for $N = 15$ we find that the direct low $N$ computation and the large
$N$ computation agree up to order 5 in the $t$ expansion, with the
result:
\begin{align}
  \begin{split}
    \cI_{\SO/\Sp}(t,x,f) =& 1 + t^2\bigl[-\chi_{1,1}(f) + 1 \bigr]\\
    & - t^3(x+x^{-1})\bigl[\chi_{1,1}(f)
    +\chi_{3,0}(f)\bigr]\\
    & - t^4(x^2 + x^{-2})\bigl[\chi_{1,1}(f)+\chi_{3,0}(f)\bigr]\\
    & + t^4\bigl[\chi_{0,3}(f) - 2\chi_{1,1}(f) +
    \chi_{6,0}(f) \bigr] \\
    & - t^5(x^3 + x^{-3})\bigl[\chi_{1,1}(f)+\chi_{3,0}(f)\bigr]\\
    & + t^5(x^1 + x^{-1})\bigl[\chi_{0,3}(f) + 2\chi_{2,2}(f) + 2\chi_{4,1}(f) +
    \chi_{6,0}(f) \bigr] + \ldots
    \end{split}
\end{align}

In addition to the physical arguments for the duality presented in the rest of this paper, we find the ``experimental'' evidence for the agreement of the indices presented in this and the previous subsection compelling enough to conjecture the equality of the indices for all values of $N$:
\begin{align}
  \label{eq:SCI-equality}
  \cI_{\Sp} = \cI_{\SO}\, .
\end{align}
In appendix \ref{sec:SCI-identity} we reformulate this equality in terms of elliptic hypergeometric integrals. This leads us to a conjecture about elliptic hypergeometric functions that could potentially be proven along the lines of \cite{Rains}.

\section{Infrared behavior} \label{sec:infrared}

We now discuss the infrared behavior of these gauge theories, and what
it implies about our proposed duality.

Before turning to specific examples where the infrared behavior can be
determined using Seiberg duality, we first note that the string
coupling~(\ref{eqn:dP0axiodil}) is constant along the RG flow, i.e.\
it is ``exactly dimensionless'' (its exact quantum-corrected
scaling-dimension vanishes). In the large $N$ limit, this result
follows from the no-scale structure of the supergravity dual. In appendix~\ref{app:exactlydimensionless}, we argue that this persists at finite $N$, and that the string coupling is neither perturbatively nor nonperturbatively renormalized (at the origin of moduli space).

The fact that $\axiodil$ is exactly dimensionless can have important consequences for the infrared behavior. Generically, this implies that the infrared fixed point is actually a fixed line parameterized by $\axiodil$. The string coupling therefore maps to an exactly marginal operator at the superconformal fixed point. This is to be expected: as we saw in~\S\ref{sec:MOduality}, an $\SL(2,\bZ)$ duality generally incorporates self-dualities relating each gauge theory to itself at different values of the couplings, whereas it has been suggested that the occurrence of self-dualities is closely tied to that of exactly marginal operators~\cite{Leigh:1995ep,Karch:1997jp}, with the corresponding deformation interpolating between the dual descriptions in the infrared.

Thus, in general the two fixed points reached by the dual theories in
their respective perturbative regimes will occur at different
locations along a line of fixed points parameterized by the string
coupling. Since the theories are connected by a continuous
deformation, the global anomalies, the superconformal index, and the
topology of the moduli space should match between the two fixed
points, provided that a discontinuous ``phase transition'' does not
occur in between; we have argued that these data do indeed match
in~\S\ref{sec:C3Z3fieldtheory} and~\S\ref{sec:SCI}.

In some cases, the infrared behavior may be different. In particular,
the string coupling, despite being exactly dimensionless along the
flow, does not always correspond to a deformation of the fixed
point. Instead, the flows may converge to a single fixed point; this
can happen when the string coupling becomes ill-defined at that point,
for instance when its constituent couplings approach some limit. As a
toy example, consider an $\SU(N)^2$ gauge group with $N_F$
$(\fund,\ov\fund)\oplus (\ov\fund, \fund)$ bifundamental ``flavors''
and no superpotential. If $N_F \ge 3$, then the two gauge theories are
infrared free, whereas
\begin{equation} \label{eqn:toymarginal}
\frac{1}{g_1^2} - \frac{1}{g_2^2} \,,
\end{equation}
is an exactly dimensionless coupling. However, while~(\ref{eqn:toymarginal}) is constant along the flow, as $g_1 \to 0$ the difference between the gauge couplings $g_1$ and $g_2$ also flows to zero, and in the deep infrared the theory is free, independent of the initial values of the couplings. In these cases, since the string coupling is irrelevant at the fixed point, the infrared physics should not depend on $\axiodil$, and the two fixed points should be the same, as in Seiberg duality.

We now consider specific examples. In~\S\ref{subsec:N=5}, we saw the both the $\SO$ and $\Sp$ theories have a dynamically generated runaway superpotential for $N=5$ ($\tN=2$). We now attempt to determine the infrared behavior of these gauge theories for larger values of $N$.

It turns out that the $\Sp$ theories are in general somewhat more
tractable than the $\SO$ theories, so we focus on the former,
extracting predictions for the IR behavior of the latter. We begin by
discussing the cases $N=7$ and $N=9$ in~\S\ref{subsec:N=7}
and~\S\ref{subsec:N=9}, respectively, where the infrared behavior can
be determined using known dualities. In~\S\ref{subsec:N>9}, we
speculate about the infrared behavior for $N > 9$.

\subsection{The \alt{$\Sp(8)\times\SU(4)$}{USp(8)xSU(4)} theory}
\label{subsec:N=7}

The prospective dual theories for $N=7$ are:
\begin{equation}
\mbox{
\begin{minipage}{0.43\linewidth}
  \begin{center}
  \begin{tabular}{c|cc|ccc}
     & \!\!$\SO(3)$\!\! & \!\!$\SU(7)$\!\!  & \!\!$\SU(3)$\!\! & \!\!$\U(1)_R$\!\! \\ \hline
    $A^i$ & $\fund$ & $\ov{\fund}$ & $\fund$ & $\frac{20}{21}$  \\
    $B^i$ & 1 & $\asymm$ & \fund & $\frac{2}{21}$
  \end{tabular}\\[12pt]
  $W = \frac{1}{2} \lambda\, \delta^{a b} \epsilon_{i j k} A^i_{a; m} A^j_{b; n} B^{m n;\, k}$\,,
  \end{center}
  \end{minipage}}
  \longleftrightarrow
  \mbox{
  \begin{minipage}{0.43\linewidth}
  \begin{center}
  \begin{tabular}{c|cc|ccc}
     & \!\!$\Sp(8)$\!\! & \!\!$\SU(4)$\!\! & \!\!$\SU(3)$\!\! & \!\!$\U(1)_R$\!\!\\ \hline
    $\tilde{A}^i$ & $\fund$ & $\ov{\fund}$ & $\fund$ & $\frac{1}{6}$ \\
    $\tilde{B}^i$ & 1 & $\symm$ & \fund & $\frac{5}{3}$
  \end{tabular}\\[12pt]
  $W = \frac{1}{2} \tilde{\lambda}\, \symp^{a b} \epsilon_{i j k} \tilde{A}^i_{a;\, m} \tilde{A}^j_{b;\, n} \tilde{B}^{m n;\, k}$\,.
  \end{center}
  \end{minipage}}
\end{equation}
We focus on the $\Sp(8)\times\SU(4)$ theory, showing that it has an infrared-free dual description with a quantum moduli space.

The IR dynamics of this theory are particularly easy to describe, as the $\Sp(8)$ factor is s-confining, leaving an $\SU(4) \cong \SO(6)$ gauge theory in the confined description which can be Seiberg dualized to obtain an IR free description.

The dynamics of the s-confined $\Sp(8)$ can be described in terms of the meson
\begin{align}
M^{I J} = \symp^{a b} \tilde{A}^I_a \tilde{A}^J_b \,,
\end{align}
with the superpotential
\begin{align}
W = \frac{1}{\Lambda_{\rm Sp}^9} \Pf M \,,
\end{align}
where the indices $I, J$ parameterize a fictitious $\SU(12) \subset \SU(4)\times\SU(3)$. $M$ decomposes into irreps $\Psi$ and $\Phi$ transforming as $(\ov\symm,\ov\fund)$ and $\left(\ov\asymm,\symm\right)$ under $\SU(4)\times\SU(3)$, respectively, where the superpotential now takes the form
\begin{align}
W \sim \frac{1}{\Lambda_{\rm Sp}^3} \left(\Phi^6 + \Phi^5 \Psi + \ldots \right)+\tilde{\lambda} \Lambda_{\rm Sp} \Psi \tilde{B} \,,
\end{align}
where we suppress the index structure for simplicity, and we absorb a factor of $\Lambda_{\rm Sp}^{-1}$ into the definition of $\Phi$ and $\Psi$ to make them dimension-one fields. Thus, $\Psi$ and $\tilde{B}$ acquire a mass and can be integrated out, leaving the superpotential
\begin{align}
W \sim \frac{1}{\Lambda_{\rm Sp}^3} \Phi^6 \, ,
\end{align}
where the remaining terms are exactly those generated by the Pfaffian for $\Psi = 0$.

It is now instructive to rewrite the gauge group $\SU(4)$ as $\SO(6)$, under which the $\Phi$ transform as a vector. The gauge-invariant meson $\Phi^2$ transforms as $\yss\ydiagram{4}\ysn_{\,+2/3}\oplus\ov\symm_{\,+2/3}$ under $\SU(3)\times\U(1)_R$, corresponding to the baryon $\tA^4$ in the original theory. In terms of this meson, the superpotential takes the form:
\begin{align}\label{eqn:SU4sup}
W \sim \frac{1}{\Lambda_{\rm Sp}^3} \left((\Phi^2)^3 + \det \Phi \right)\,,
\end{align}
where we suppress the index structure and numerical prefactors for simplicity, and $\det \Phi$ denotes the lone $\SO(6)$ baryon, which is automatically $\SU(3)$ invariant.

Applying Seiberg duality, we obtain the $\SO(4)$ gauge theory:
\begin{equation}
  \begin{array}{c|c|cc}
                            & \SO(4) & \SU(3)                           & \U(1)_R\\ \hline
    Q & \fund   & \ov\symm                      & \frac{2}{3} \\
    \cA             & \sing               & \yss\ydiagram{4}\ysn\oplus\ov\symm & \frac{2}{3}
\end{array}
\end{equation}
with the superpotential
\begin{align}
W \sim \lambda_1 \cA^3 + \lambda_2 \cA\, Q^2\,,
\end{align}
where $\cA = \frac{1}{\lambda_1^{1/3} \Lambda_{\rm Sp}} \Phi^2$. The
baryon $\det \Phi$ in the superpotential~(\ref{eqn:SU4sup}) maps to a
glueball $\epsilon^{i j k l} W_{i j} W_{k l}$ in the dual theory~\cite{Seiberg:1994pq},
which causes a splitting between the two gauge couplings, $\tau_1$ and
$\tau_2$, of the $\SU(2)\times\SU(2) \cong \SO(4)$ gauge
group. Performing scale matching at each
step in this chain of dualities, we find that
\begin{align} \label{eqn:gaugesplit}
\tau_1 - \tau_2 \sim e^{\pi i \axiodil/2} \,,
\end{align}
where $\axiodil$ is the ten-dimensional axio-dilaton, which is related to the other couplings by
\begin{align}\label{eqn:SU7axodil}
e^{\pi i \axiodil} = \lambda_1^2 \lambda_2^{-6} e^{-\pi i (\tau_1+\tau_2)} \,.
\end{align}
Thus, the splitting between the gauge couplings, (\ref{eqn:gaugesplit}), is nonperturbatively suppressed at weak string coupling.

We now consider the infrared behavior of this theory. Since the beta function coefficient of $\SO(4)$ vanishes, the theory has a free fixed point. We argue that this fixed point is attractive. The exact beta functions are:\footnote{The argument given here is somewhat of an oversimplification since $\cA$ is not an irrep, and therefore $\lambda_1$ and $\lambda_2$ correspond to more than one physical coupling. However, it is straightforward to account for the additional complications which arise in a more careful treatment.}
\begin{equation} \label{eqn:SU7beta}
\beta(g_i) = -\frac{3 g_i^3}{8 \pi^2} \frac{\gamma_Q}{1-g_i^2/4 \pi^2}\,, \quad \beta(\lambda_1) = \frac{3}{2} \lambda_1 \gamma_{\cA} \,, \quad \beta(\lambda_2) = \frac{1}{2} \lambda_2 (\gamma_{\cA} + 2 \gamma_Q)\,,
\end{equation}
where $\gamma_Q$ and $\gamma_{\cA}$ are the anomalous dimensions of $Q$ and $\cA$, which take the form
\begin{equation}
\gamma_Q = \frac{k_2 |\lambda_2|^2}{192 \pi^2} - \frac{3}{16 \pi^2} (g_1^2 + g_2^2) \,, \quad \gamma_{\cA} = \frac{k_1 |\lambda_1|^2}{112 \pi^2}+ \frac{k_2 |\lambda_2|^2}{336 \pi^2}\,,
\end{equation}
at the one-loop level, where we use the one-loop result (see e.g.~\cite{Barnes:2004jj})
\begin{align}\label{eqn:oneloopgamma}
\gamma_i \simeq \frac{n_i k_\lambda |\lambda|^2}{16 \pi^2 |r_i|} - \frac{g^2}{4 \pi^2} C(r_i)\,,
\end{align}
for a chiral superfield $\phi_i$ in the representation $r_i$ of the gauge group $G$, where $C(r) = |G| T(r)/|r|$ is the quadratic Casimir operator, $W = \lambda \prod_i \phi_i^{n_i}$ with $\sum_i n_i = 3$, and $k_{\lambda}$ is a positive real constant which depends on the index structure and normalization of the superpotential, which we will not need to compute.

A weakly-coupled flow can be approximated as follows. The gauge
coupling does not run at one loop, so we initially
treat it as a constant, whereas the superpotential couplings run to
the ``fixed point'' $k_1 |\lambda_1|^2 \sim 0$ and $k_2 |\lambda_2|^2
\sim 28 (g_1^2 + g_2^2)$. Thus,
\begin{equation} \label{eqn:gammaQend}
\gamma_Q \sim  -\frac{1}{24 \pi^2} (g_1^2 + g_2^2)\,,
\end{equation}
at the end of the one-loop flow. The remainder of the flow occurs more slowly, at the two-loop level, and can be approximated by substituting~(\ref{eqn:gammaQend}) into the beta function~(\ref{eqn:SU7beta}), giving
\begin{equation}
\beta(g_i) \simeq \frac{1}{(8 \pi^2)^2}\, g_i^3 (g_1^2 + g_2^2) \,,
\end{equation}
in the weak-coupling limit, where two-loop running can be treated adiabatically with respect to one-loop running. Thus, the gauge couplings (and hence $\lambda_2$) run to zero in the infrared, and the theory becomes free.

This is one example where the string coupling~(\ref{eqn:gaugesplit},
\ref{eqn:SU7axodil}) is an irrelevant deformation at the infrared
fixed point, as discussed previously. This is consistent because the
string coupling corresponds to a ratio of couplings which remains
constant as the flow approaches the infrared fixed point, and
therefore the exactly dimensionless coupling parameterizes a family of
flows, all of which converge to the same free fixed point.

While the chain of dualities we have employed to arrive at this infrared-free description is valid at weak string coupling, the above discussion suggests that the infrared fixed point is perturbatively independent of the string coupling. If this persists nonperturbatively, then the same $\SO(4)$ gauge theory should also describe the infrared behavior of the $\SO(3)\times\SU(7)$ gauge theory. It would be interesting to pursue this point further.

We now consider the moduli space of this theory. The F-term conditions take the form:
\begin{equation} \label{eqn:SU7schemFterm}
\cI_{I J K L M N} \cA^{K L} \cA^{M N} + \delta_{a b} Q^a_I Q^b_J = 0 \,, \quad  \cA^{I J} Q^b_J = 0\,,
\end{equation}
where $I$ and $J$ index the six components of a $\symm$ of $\SU(3)$, so that $\cA^{I J} = \cA^{J I}$, and $\cI_{I J K L M N}$ is an appropriate $\SU(3)$ invariant. The first equation fixes the $\SO(4)$ meson $Q^2$ in terms of $\cA^2$. Since the $\SO(4)$ baryon $Q^4$ obeys a classical constraint of the schematic form $(Q^4)^2 = (Q^2)^4$, its vev is fixed in terms of that of the meson $Q^2$ up to a sign, and therefore the classical moduli space is locally parameterized by the gauge invariant $\cA$, corresponding to the baryon discussed in~\S\ref{subsec:classicchecks}.

However, not all $\cA$ vevs can be extended to solutions to~(\ref{eqn:SU7schemFterm}). In particular, the complete F-term conditions imply the following constraints on $\cA$:
\begin{equation} \label{eqn:SU7cons}
\cI_{I K L M N P} \cA^{J K} \cA^{L M} \cA^{N P} = 0 \,, \quad \mathrm{cof\ } (\cI_{I J K L M N} \cA^{K L} \cA^{M N}) = 0\,,
\end{equation}
where $\mathrm{cof}$ denotes the matrix of cofactors, the first constraint arises upon contracting the first condition from~(\ref{eqn:SU7schemFterm}) with $\cA^{J P}$ and applying the second condition, and the second constraint follows from the classical constraint that $(Q^2)_{I J}$ has rank at most four.

One can show that the constraints~(\ref{eqn:SU7cons}) are necessary
and sufficient for a choice of $Q_I^a$ to exist which
satisfies~(\ref{eqn:SU7schemFterm}), and therefore characterize the
classical moduli space of the theory.\footnote{If $\cI_{I J K L M N}
  \cA^{K L} \cA^{M N}$ has (maximal) rank four, then the $Q^4$ baryon
  is nonvanishing, and the moduli space has two branches corresponding
  to the sign of $Q^4$ which are related by the spontaneously broken
  $\bZ_2$ outer automorphism of $\SO(4)$.} However, we have not yet
demonstrated that any nontrivial solutions to these equations
exist. Moreover, the quantum moduli space may differ from the
classical moduli space if, for instance, an F-flat $\cA$ vev with
$\langle Q \rangle = 0$ gives a mass to too many flavors, generating a
dynamical superpotential.

To address these issues, it is more convenient to write the superpotential as
\begin{equation}
W = \det \cA^{i j k l}+ c_1 \cA_{i j} \cA_{k l} \cA^{i j k l} + c_2 \det \cA_{i j} + \left(\cA^{i j k l} + \epsilon^{i k m} \epsilon^{j l n} \cA_{m n}\right) \delta_{a b} Q^a_{i j} Q^b_{k l} \,,
\end{equation}
in a non-canonically-normalized basis, where $\cA^{i j k l}$ and $\cA_{i j}$ denote the irreducible $\yss\ydiagram{4}\ysn$ and $\ov\symm$ components of $\cA$, respectively, and
\begin{equation}
\det M^{i_1 \ldots i_{2 p}} \equiv \frac{1}{d!} \epsilon_{i_{1 1} \ldots i_{1 d}} \ldots \epsilon_{i_{(2p) 1} \ldots i_{(2 p) d}} M^{i_{1 1} \ldots i_{(2 p) 1}} \ldots M^{i_{1 d} \ldots i_{(2 p) d}}\,,
\end{equation}
denotes an $\SU(d)$ invariant formed from $d$ copies of a $2 p$-index tensor $M$ which generalizes the determinant of a matrix. $c_1$ and $c_2$ are numerical prefactors corresponding to exactly marginal couplings, whose explicit values can be determined by relating $\cA^3$ to the Pfaffian superpotential generated by the s-confinement of $\Sp(8)$. An explicit computation gives $c_1 = -\frac{3}{4}$ and $c_2 = \frac{3}{2}$.

It is now straightforward to find directions in the classical moduli space. For instance $\langle \mathcal{A}^{1111} \rangle \ne 0$ with all other vevs vanishing satisfies the F-term conditions. As this gives a mass to only one $\SO(4)$ flavor, this suggests that this direction is part of the quantum moduli space, which is therefore nonempty. It would be interesting to better understand which parts of the moduli space defined by~(\ref{eqn:SU7cons}), if any, are lifted by quantum effects.

In summary, we find that the $\Sp(8)\times\SU(4)$ theory has a dual description with a free infrared fixed point and a quantum moduli space. Our proposed duality would seem to imply that the $\SO(3)\times\SU(7)$ theory has these features as well, and it would be interesting to check this in more detail to gain a better understanding of the proposed duality.

\subsection{The \alt{$\Sp(10)\times\SU(6)$}{USp(10)xSU(6)} theory}
\label{subsec:N=9}

The prospective dual theories for $N=9$ are:
\begin{equation}
\mbox{
\begin{minipage}{0.43\linewidth}
  \begin{center}
  \begin{tabular}{c|cc|ccc}
     & \!\!$\SO(5)$\!\! & \!\!$\SU(9)$\!\! & \!\!$\SU(3)$\!\! & \!\!$\U(1)_R$\!\! \\ \hline
    $A^i$ & $\fund$ & $\ov{\fund}$ & $\fund$ & $\frac{8}{9}$  \\
    $B^i$ & \sing & $\asymm$ & \fund & $\frac{2}{9}$
  \end{tabular}\\[12pt]
  $W = \frac{1}{2} \lambda\, \delta^{a b} \epsilon_{i j k} A^i_{a; m} A^j_{b; n} B^{m n;\, k}$\,,
  \end{center}
  \end{minipage}}
  \longleftrightarrow
  \mbox{
  \begin{minipage}{0.43\linewidth}
  \begin{center}
  \begin{tabular}{c|cc|ccc}
     & \!\!$\Sp(10)$\!\! & \!\!$\SU(6)$\!\! & \!\!$\SU(3)$\!\! & \!\!$\U(1)_R$\!\!\\ \hline
    $\tilde{A}^i$ & $\fund$ & $\ov{\fund}$ & $\fund$ & $\frac{1}{3}$ \\
    $\tilde{B}^i$ & \sing & $\symm$ & \fund & $\frac{4}{3}$
  \end{tabular}\\[12pt]
  $W = \frac{1}{2} \tilde{\lambda}\, \symp^{a b} \epsilon_{i j k} \tilde{A}^i_{a;\, m} \tilde{A}^j_{b;\, n} \tilde{B}^{m n;\, k}$\,.
  \end{center}
  \end{minipage}}
\end{equation}
We focus on the $\Sp(10)\times\SU(6)$ theory, showing that it has a line of infrared fixed points including a free fixed point.

We Seiberg-dualize the $\Sp(10)$ gauge group to obtain the theory
\begin{equation}
  \begin{array}{c|cc|ccc}
     & \Sp(4) & \SU(6) & \SU(3) & \U(1)_R\\ \hline
    \phi_i & \fund & \fund & \ov\fund & \frac{2}{3} \\
    \psi^{i j} & \sing & \ov\asymm & \symm & \frac{2}{3}
  \end{array}
\end{equation}
with the superpotential
\begin{align}
W = \frac{1}{2} \hat{\lambda} \, \symp_{a b}\, \phi_i^{a;\, m} \phi_j^{b;\, n} \psi^{i j}_{m n}\,,
\end{align}
after integrating out massive matter. The beta function coefficients for both gauge groups vanish, and the (exactly marginal) string coupling takes the form
\begin{align}\label{eqn:SU9axodil}
\axiodil = \frac{1}{\pi i} \ln \left[\hat{\lambda}^{24} e^{4 \pi i \tau\subSp} e^{2 \pi i \tau\subSU} \right]\,.
\end{align}
We find the exact beta functions
\begin{equation} \label{eqn:SU9beta}
\begin{split}
\beta(g\subSp) = -\frac{9 g\subSp^3}{16 \pi^2} \frac{\gamma_{\phi}}{1-3 g\subSp^2/8 \pi^2}\,, \quad \beta(g\subSU) &= -\frac{3 g\subSU^3}{8 \pi^2} \frac{\gamma_{\phi}+2\gamma_{\psi}}{1-3 g\subSU^2/4 \pi^2} \,,\\  \beta(\hat{\lambda}) &= \frac{1}{2} \hat{\lambda} (2 \gamma_{\phi} + \gamma_{\psi}) \,,
\end{split}
\end{equation}
where the anomalous dimensions $\gamma_{\phi}$ and $\gamma_{\psi}$ take the form
\begin{align}
\gamma_{\phi} \simeq \frac{k |\hat{\lambda}|^2}{576 \pi^2} - \frac{35 g\subSU^2}{48 \pi^2} - \frac{5 g\subSp^2}{16 \pi^2} \,, \quad
\gamma_{\psi} \simeq \frac{k |\hat{\lambda}|^2}{1440 \pi^2} - \frac{7 g\subSU^2}{6 \pi^2}\,,
\end{align}
at one loop, applying~(\ref{eqn:oneloopgamma}). As in~\S\ref{subsec:N=7}, we separate the flow into one-loop and higher-loop portions. At one loop, the gauge couplings do not run, and the superpotential coupling runs to the ``fixed point''
\begin{align}\label{eqn:SU9oneloopFP}
k |\hat{\lambda}|^2 \sim 30\, (21 g\subSU^2 + 5 g\subSp^2)\,.
\end{align}
Thus, after the one-loop running, we have
\begin{align}
\gamma_{\phi} \simeq \frac{5}{96 \pi^2} \left(7 g\subSU^2- g\subSp^2 \right) \,, \quad \gamma_{\psi} \simeq \frac{5}{48 \pi^2} \left(g\subSp^2 - 7 g\subSU^2 \right) \,,
\end{align}
Putting these into the beta functions for the gauge couplings, we obtain
\begin{align}
\beta(g\subSp) \simeq \frac{15 g\subSp^3}{2 (16 \pi^2)^2} \left(g\subSp^2 - 7 g\subSU^2 \right) \,, \quad \beta(g\subSU) \simeq \frac{15 g\subSU^3}{(16 \pi^2)^2} \left(7 g\subSU^2 - g\subSp^2 \right)\,,
\end{align}
under the same assumption of adiabaticity as before.

By inspection, we see that
$\frac{2}{g\subSp^2} + \frac{1}{g\subSU^2}$
is constant
along the two-loop flow under the stated assumptions. Indeed, this combination corresponds approximately to the exactly marginal coupling~(\ref{eqn:SU9axodil}) along this flow,
\begin{align}\label{eqn:SU9oneloopaxodil}
\frac{1}{8 \pi g_s} \sim \frac{2}{g\subSp^2} + \frac{1}{g\subSU^2} \,,
\end{align}
since the logarithm of the superpotential coupling, fixed by~(\ref{eqn:SU9oneloopFP}) in the adiabatic approximation, is small compared to $1/g^2$. Thus, the two-loop flow lines lie along contours of constant $\frac{2}{g\subSp^2} + \frac{1}{g\subSU^2}$, and converge on the fixed line $g\subSp^2 \simeq 7 g\subSU^2$ and $k |\hat{\lambda}|^2 \simeq 240 g\subSU^2$, with the final position along the fixed line dictated by the string coupling, as in~(\ref{eqn:SU9oneloopaxodil}).

Since the superpotential coupling and theta angles define one physical phase among them, there is a complex line of infrared fixed points parameterized by $\axiodil$, where weak string coupling corresponds to a weakly coupled gauge theory and vice versa. Thus, unlike the previous example, the string coupling corresponds to a marginal deformation at the infrared fixed point, and affects the physics there. As such, we cannot readily infer the complete infrared behavior of the prospectively dual $\SO(5)\times\SU(9)$ theory from the above treatment, as this corresponds to a portion of the infrared fixed line which is strongly coupled in the $\Sp(4)\times\SU(6)$ description.

\subsection{The infrared behavior for \alt{$N > 9$}{N > 9}}
\label{subsec:N>9}

While the $N= 7$ and $N=9$ examples treated in~\S\ref{subsec:N=7} and~\S\ref{subsec:N=9} are distinct in a number of ways, they both share the feature that the infrared physics is perturbatively accessible in some dual description, i.e.\ that there is a weakly coupled dual description, at least for certain values of the string coupling. We now ask whether this can hold more generally, for $N>9$.

At any free fixed point, all the fundamental chiral superfields will have dimension one and the corresponding superconformal R-charge $+2/3$. If we assume that no accidental $\U(1)$ symmetries appear along the flow, then the superconformal R-charge of gauge invariant operators can be determined via a-maximization~\cite{Intriligator:2003jj}, whereas the assumption of a free fixed point requires that the R-charge of such an operator be an integer multiple of $2/3$.

Indeed, since an arbitrary gauge invariant of the $\SO$ theory takes
the form~(\ref{eqn:ABfactor1}) or~(\ref{eqn:ABfactor2}), it is easy to
check that all such operators have R-charge $Q_R = \frac{2}{3} n$ for
$n>0$ and $N \ge 7$, whereas a similar argument applies to the $\Sp$
theory for $\tN \ge 4$. This is suggestive and nontrivial evidence for
a free fixed point, which we have already shown to occur for the cases
$\tN = 4, 6$.

If such a fixed point exists, the $\U(1)_R^3$ and $\U(1)_R$ anomalies further constrain its form. In particular, a collection of $N_{\chi}$ chiral superfields with $Q_R = +2/3$ interacting via a gauge group $G$ have the following anomalies
\begin{align}
\U(1)_R^3 = |G| - \frac{1}{27} N_{\chi} \,, \quad \U(1)_R = |G| - \frac{1}{3} N_{\chi} \,.
\end{align}
Therefore,
\begin{align}
|G| = \frac{1}{8} \left(9 \U(1)_R^3 - \U(1)_R \right) \,, \quad N_{\chi} = \frac{27}{8} \left(\U(1)_R^3 - \U(1)_R \right)\,.
\end{align}
Thus,
\begin{align}\label{eqn:Gchicount}
|G| = \frac{3}{2} N (N-3) - 36 \,,\quad N_{\chi} = \frac{9}{2} N (N-3) - 81 \,,
\end{align}
for the case at hand. Conservation of the superconformal R-charge implies that the semi-simple component of $G$ must have vanishing beta function coefficient, whereas any $\U(1)$ factors must decouple.

Even if we assume that $G$ is semisimple, for large $N$ there are many possible product gauge groups which can reproduce the dimension formula~(\ref{eqn:Gchicount}). One possibility, which explains the pattern for all $N\ge 7$, is
\begin{align}
G = \left[ \Sp(4)\times\SU(6)\right]^{\frac{N-7}{2}} \times \SO(4)^{\left(\frac{N-9}{2}\right)^2}\,.
\end{align}
However, for any fixed $N$, there remain many possible spectra for this gauge group with vanishing beta function coefficients. While there are many further consistency checks one can apply to any specific candidate description, no obvious candidate presents itself. Moreover, the possibilities are yet broader if we allow for accidental $\U(1)$ symmetries.

Should such an infrared description be found, it would be interesting to understand if it has a direct string theory interpretation, e.g.\ in terms of branes.
We leave further study of the infrared behavior of these theories to a future work.

\section{Further examples}
\label{sec:moreEx}

So far we have focused on a single example of a new $\mathcal{N}=1$
$\SL(2,\bZ)$ duality which arises on the worldvolume of D3 branes
probing the orientifolded $\bC^3/\bZ_3$ singularity. While this
example is closely analogous to the known $\mathcal{N}=4$ examples,
making the parallels easier to grasp, it is but one example of a
previously unexplored class of dualities of this type. In this
section, we aim to briefly illustrate the breadth of this class, and
also to point out other new dualities which arise from D3 branes
probing orientifolded singularities but which appear to be of a
different origin. We focus on a few simple examples, and defer further
examples to~\cite{transitions2, transitions3}.

We begin by discussing the Calabi-Yau cone over $dP_1$ (a real cone over $Y^{2,1}$),\footnote{See~\cite{Gauntlett:2004yd,Martelli:2004wu} for more on the infinite class of Sasaki-Einstein manifolds known as the $Y^{p,q}$.} which provides a simple, non-orbifold example of the $\SL(2,\bZ)$ dualities we have focused on. The resulting gauge theories are related to the $\bC^3/\bZ_3$ theories by Higgsing, and exhibit interesting infrared physics. We discuss anomaly matching, moduli space matching, and Higgsing for all $N$, before treating a specific example where the quantum moduli spaces can be shown to match exactly.

We then briefly discuss two other non-orbifold examples given by the
Calabi-Yau cones over $Y^{2,0}$ and $Y^{4,0}$,\footnote{The real cone
  over $Y^{2,0}$ is the same as the Calabi-Yau cone over the zeroth
  Hirzebruch surface $\bF_0 \equiv \bP^1 \times \bP^1$.} both of which
exhibit different, more complicated patterns of dualities.

\subsection{Complex cone over \alt{$dP_1$}{dP1}}
\label{sec:dP1}

We begin by considering the complex cone over the first del Pezzo surface
$dP_1$, which can be obtained by blowing up $\bP^2$ at a point. We are
interested in orientifolds of this configuration corresponding to a
compact O7 plane wrapping the del Pezzo base. As shown in figure~\ref{fig:dP1quiverfold}, only one involution of the parent quiver exists which satisfies rule I of appendix~\ref{app:quiverfolds}, up to the choice of fixed element signs.
\begin{figure}
  \begin{center}
    \includegraphics[width=0.8\textwidth]{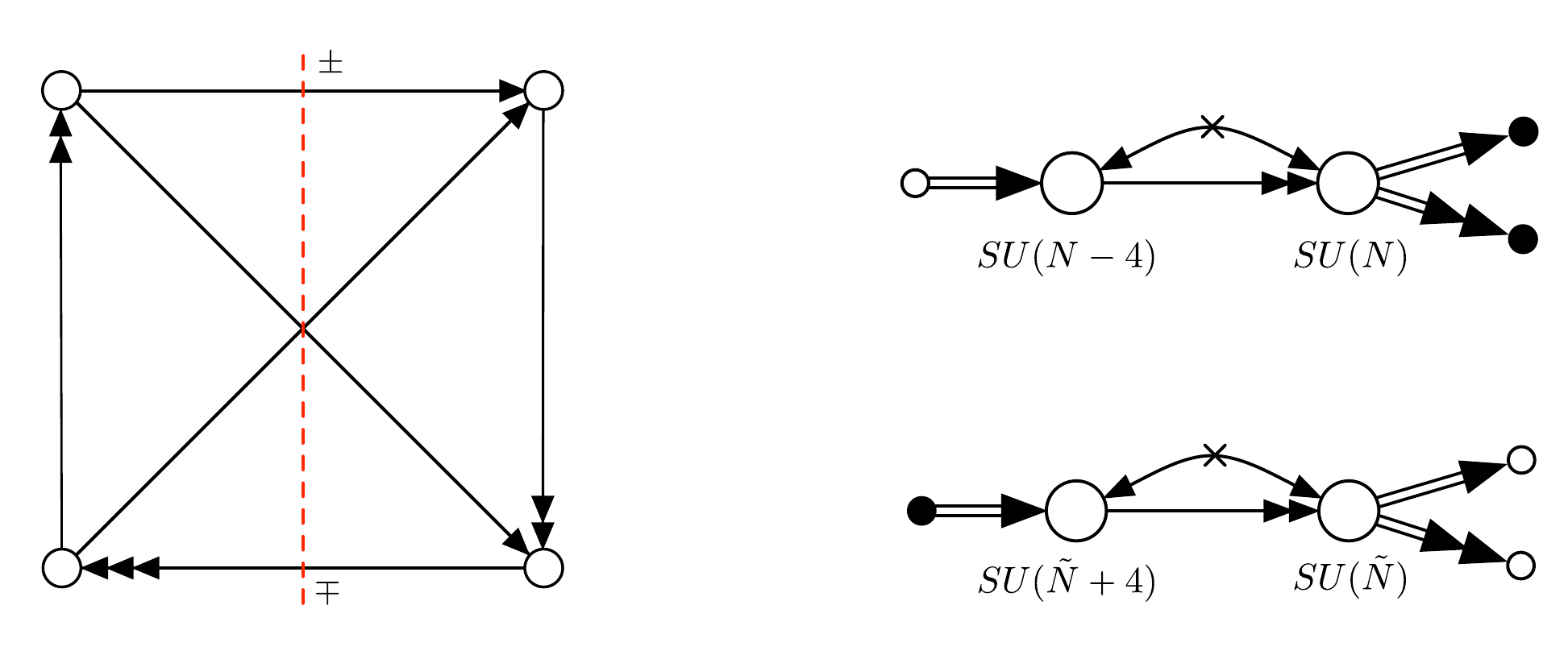}
  \end{center}
  \caption[The quiver gauge theory for $dP_1$ and its orientifold]{The left side shows the quiver gauge theory for $dP_1$,
    with the involution of interest indicated by the dashed line. The
    resulting quiverfold (see appendix~\ref{app:quiverfolds}) theories
    for the two sign choices are shown on the right.}
  \label{fig:dP1quiverfold}
\end{figure}
Moreover, only two choices for these signs lead to theories which can be anomaly free without the addition of noncompact ``flavor'' D7 branes. As we show in~\cite{transitions3} using brane tiling methods, these involutions also satisfy rule II and lead to superpotentials which inherit the $\SU(2)\times\U(1)_{X}\times\U(1)_{R}$ geometric flavor symmetries of the parent theory, as expected for a compact O7 plane.

The two possible sign choices lead to the orientifold gauge theory
\begin{align}
  \label{eq:dP1thyA}
  \begin{array}{c|cc|ccccc}
    &\SU(N-4) & \SU(N) & \SU(2) &\U(1)_X & \U(1)_B & \U(1)_R\\
    \hline
    A^i & \fund & \ov\fund & \fund & \frac{N-2}{N-4} & -\frac{2(N-1)}{N(N-4)} & -\frac{8}{N(N-4)}\\
    Y & \ov\fund & \ov\fund & {\bf 1} & -\frac{N-2}{N-4} & \frac{(N+2)}{N(N-4)} & \frac{N^{2}-8}{N(N-4)}\\
    Z & \ov\symm & {\bf 1} & {\bf 1} & -\frac{N}{N-4} & \frac{3}{N-4} & \frac{N}{N-4}\\
    B^i & {\bf 1} & \asymm & \fund & 0 & \frac{1}{N} & \frac{N-4}{N}\\
    X & {\bf 1} & \asymm & {\bf 1} & -1 &\frac{1}{N} & \frac{N-4}{N}
  \end{array}
\end{align}
with superpotential
\begin{align}
W=\epsilon_{ij} \Tr[B^i A^j Y + X A^i Z A^j]\,,
\end{align}
as well as the theory\footnote{Up to charge conjugation of the global
  $\U(1)_B$ this theory is the negative rank dual of the first theory
  (see appendix \ref{app:negativerankdual} for details on negative
  rank duality).}
\begin{align}
  \label{eq:dP1thyB}
  \begin{array}{c|cc|ccccc}
    &\SU(\tN+4) & \SU(\tN) & \SU(2) &\U(1)_X & \U(1)_B & \U(1)_R\\
    \hline
    \tA^i & \fund & \ov\fund & \fund & \frac{\tN+2}{\tN+4} & -\frac{2(\tN+1)}{\tN(\tN+4)} & -\frac{8}{\tN(\tN+4)}\\
    \tY & \ov\fund & \ov\fund & {\bf 1} & -\frac{\tN+2}{\tN+4} & \frac{(\tN-2)}{\tN(\tN+4)} & \frac{\tN^{2}-8}{\tN(\tN+4)}\\
    \tZ & \ov\asymm & {\bf 1} & {\bf 1} & -\frac{\tN}{\tN+4} & \frac{3}{\tN+4} & \frac{\tN}{\tN+4}\\
    \tB^i & {\bf 1} & \symm & \fund & 0 & \frac{1}{\tN} & \frac{\tN+4}{\tN}\\
    \tX & {\bf 1} & \symm & {\bf 1} & -1 &\frac{1}{\tN} & \frac{\tN+4}{\tN}
  \end{array}
\end{align}
with superpotential
\begin{align}
W=\epsilon_{ij} \Tr[\tB^i \tA^j \tY + \tX \tA^i \tZ \tA^j]\,,
\end{align}
where in either case the gauge indices are cyclically contracted.
Henceforward, for want of a better label we refer to these theories as ``Theory A'' and ``Theory B'', respectively. For ease of presentation we have chosen a basis for the R-symmetry which does not satisfy a-maximization, as the superconformal R-charges are irrational.

The global anomalies for both theories are shown in table~\ref{tab:dP1anomalies}.
\begin{table}
\begin{equation*}
  \label{eq:dP1-anomalies}
  \begin{array}{|c|c|c|}
    \multicolumn{3}{r}{\text{Theory A} \qquad \qquad \text{Theory B}\qquad} \\ \hline
    \SU(2)^3 & (-1)^{\frac32 N(N-3)} & (-1)^{\frac32 \tilde{N}(\tilde{N}+3)} \\ \hline
    \SU(2)^2 \, \U(1)_X & N(N-2)& \tN(\tN+2)\\ \hline
    \SU(2)^2 \, \U(1)_B & -\frac32(N-1) & -\frac32(\tN+1)\\ \hline
    \U(1)_X^2 \, \U(1)_B & -(N-1) & -(\tilde{N}+1)\\ \hline
    \U(1)_X \, \U(1)_B^2 & 2 & 2\\ \hline
    \SU(2)^2 \, \U(1)_R & -N(N-2)-6 & -\tilde{N}(\tilde{N}+2)-6\\\hline
    \U(1)_X^2 \, \U(1)_R & -2N(N-2)-4 & -2\tilde{N}(\tilde{N}+2)-4\\\hline
    \U(1)_B^2 \, \U(1)_R & -8 & -8\\\hline
    \U(1)_X \, \U(1)_B \, \U(1)_R & 4(N-1) & 4(\tilde{N}+1)\\\hline
    \U(1)_X \, \U(1)_R^2 & 2N(N-2) & 2\tilde{N}(\tilde{N}+2)\\\hline
    \U(1)_B \, \U(1)_R^2 & -4(N-1) & -4(\tilde{N}+1)\\\hline
    \U(1)_R^3 & -34 & -34\\\hline
    \U(1)_R & -10 & -10\\\hline
  \end{array}
\end{equation*}
\caption[The anomalies for the two $dP_1$ theories]{The nonvanishing anomalies for theory A and theory B, where we use a multiplicative notation for the $\SU(2)^3$ Witten anomaly analogous to that used for discrete symmetries in~\S\ref{subsec:classicchecks}. The $\U(1)_B^3$, $\U(1)_X^3$, $\U(1)_B$, and $\U(1)_X$ anomalies vanish in both theories.\label{tab:dP1anomalies}}
\end{table}
We see that the anomalies match between the two theories for $\tilde{N}=N-2$ provided that $N$ is odd. For even $N$ the $\SU(2)^3$ Witten anomalies do not match, and the theories are not dual.\footnote{One can also show that holomorphic gauge invariants do not match between the two theories. For instance, the B-theory operator $\tZ^{(\tN+4)/2}$, defined for even $\tN$, has no dual in the A-theory.}

For completeness, we also present the $a$-maximizing superconformal
R-charge, which is a linear combination:\footnote{As remarked in
  \cite{Bertolini:2004xf}, it is important to do $a$-maximization over
  the symmetries preserved by the superpotential, as we do here. It is
  easy to verify that in the large $N$ regime our results agree with
  those in \cite{Bertolini:2004xf}, as they should since the
  orientifold corrections are subleading in this limit
  \cite{Aharony:1999ti}.}
\begin{align}
\U(1)_R^{\rm (sc)} = \U(1)_R + a_X \U(1)_X + a_B \U(1)_B\,.
\end{align}
A-maximization in theory A gives
\begin{align}
a_B = \frac{a_X^2-8a_X+4}{4(a_X-4)}(N-1)\,,
\end{align}
where $a_X$ is a solution to the quartic equation
\begin{align}
0=(N-1)^2 (a_X^2-4)[3a_X^2-16a_X+4]+16(2a_X+1)(a_X-4)^2\,,
\end{align}
in the range $a_X \in \left(-\frac{1}{2},\frac{2}{3}(4-\sqrt{13})\right)$, whereas exactly one solution lies in this range for any $N>1$. For example, we obtain approximately
\begin{align}
\begin{array}{c|cccccccccccc}
\!N\! & 2 & 3 & 4 & 5 & 6 & 7 & 8 & 9 & 10 & 11 & 12 \\ \hline
\!a_X\! & \!-0.431\! & \!-0.270\! & \!-0.113\! & 0 & \!0.074\! & \!0.122\! & \!0.155\! & \!0.178\! & \!0.194\! & \!0.207\! & \!0.216\!
\end{array}
\end{align}
The result for $N=5$ is exact, giving $a_X=0$ and $a_B=-1$. For large $N$, $a_X$ asymptotically approaches $\frac{2}{3}(4-\sqrt{13})\approx 0.263$. The same considerations apply in theory B upon replacing $N-1\to\tN+1$.

In the following sections, we explore the prospective duality for odd $N$. We will also have more to say about the case of even $N$ in the next section.

\subsubsection{Moduli space and Higgsing}

We begin by considering the mapping between the moduli spaces of the two theories, which is equivalently expressed as a map between the holomorphic gauge invariants subject to the F-term conditions.

To obtain this map, we consider certain ``minimal'' operators, i.e.\ operators whose $\U(1)$ charges cannot be obtained as the sum of the $\U(1)$ charges of two or more nonvanishing operators. Operators of this type can only mix with other minimal operators under the duality, whereas generically no two minimal operators share the same $\U(1)$ charges, leading to a unique matching between the minimal operators of the dual theories.

To find minimal operators, we begin by classifying irreducible ``gauge-invariant'' monomials in the fundamental fields, i.e.\ formal products of the fields (disregarding gauge-indices) which are neutral under the $\bZ_{N-4}\times\bZ_N$ or $\bZ_{\tN+4}\times\bZ_{\tN}$ gauge-group center, and which cannot be factored into two or more gauge-invariant pieces. The resulting finite list generates all gauge-invariant monomials, a subset of which will correspond to actual gauge-invariant operators. Using this classification, it is possible to show that certain candidate operators are minimal.

Using these methods, we obtain the following minimal operators in theory A for odd $N$:
\begin{align}
  \label{eq:dP1spectrumA}
  \begin{array}{c|ccc}
  & \U(1)_B & \U(1)_X' & \U(1)_R' \\ \hline
  A^2 (Y|AZ)^2 (B|X)^2 & 0 & [-2,2] & 4 \\
  B^2 X (B|X)^{N-3} & 1 & [-\frac{N-3}{2}, \frac{N-3}{2}] & N-5 \\
  Z^{N-4-2k} (Y^2 X)^{2k} & 3 & \frac{N-3}{2}-4k & N-3+4k \\
  A^{N-4} B (B|X)^{N-3} & -1 & [-\frac{N-3}{2}, \frac{N-3}{2}] & N-5 \\
  A^{N} (Y|AZ)^4 (B|X)^2 & -2 & [-3,3] & 8 \\
  A^{p (N-4)} (B|X)^{N-2p} & 1-2p & [-\frac{N+1-2p}{2}, \frac{N-1-2p}{2}] & N-5
  \end{array}
\end{align}
where $2\le p \le \frac{N-1}{2}$, $(x|y)^n$ denotes a monomial of degree $n$ in $x$ and $y$, and we employ a slightly different basis for the $\U(1)$ charges:
\begin{align}
\U(1)_X' = \U(1)_X + \frac{N-1}{2} \U(1)_B \,, \quad \U(1)_R' = \U(1)_R - \U(1)_B \,.
\end{align}
A similar analysis in theory B for odd $\tN$ gives
\begin{align}
  \label{eq:dP1spectrumB}
  \begin{array}{c|ccc}
  & \U(1)_B & \U(1)_X' & \U(1)_R' \\ \hline
  \tA^2 (\tY|\tA\tZ)^2 (\tB|\tX)^2 & 0 & [-2,2] & 4 \\
  \tY \tZ^2 (\tY|\tA\tZ)^{\tN-1}& 1 & [-\frac{\tN-1}{2}, \frac{\tN-1}{2}] & \tN-3 \\
  \tZ^{\tN+3-2k} (\tY^2 \tX)^{2k+1} & 3 & \frac{\tN-1}{2}-4k & \tN-1+4k \\
  \tA^{\tN+1} \tZ (\tY|\tA\tZ)^{\tN-1} & -1 & [-\frac{\tN-1}{2}, \frac{\tN-1}{2}] & \tN-3 \\
  \tA^{\tN+6} (\tY|\tA\tZ)^2 (\tB|\tX)^4 & -2 & [-3,3] & 8 \\
  \tA^{p (\tN+2)-2} (\tY|\tA\tZ)^{\tN+2-2p} & 1-2p & [-\frac{\tN+3-2p}{2}, \frac{\tN+1-2p}{2}] & \tN-3
  \end{array}
\end{align}
Thus, the spectrum of minimal operators appears to match between the two theories for $\tN = N-2$, a highly nontrivial check of the proposed duality.\footnote{It would be instructive to also compute the $\SU(2)$ representations of these operators.}

Several comments are in order. Firstly, while this may not comprise a complete list of minimal operators, one can show that all of the listed operators are minimal, and that no other minimal operators share the same $\U(1)$ charges, so the matching is reliable. Secondly, to obtain this matching, it is necessary to carefully account for the structure of the gauge-index contraction as well as the F-term conditions. For example, consider the operator $Z^{N-4-p} (Y^2 X)^{p}$. Each $X$ factor must appear in the combination $X^{m n} Y^a_m Y^b_n$, which is therefore antisymmetric in the $\SU(N-4)$ indices. Since $Z^{ab}$ is symmetric, a gauge invariant index contraction exists if and only if an even number of $X Y^2$ factors appear, i.e. if and only if $p$ is even. By contrast, in the operator $\tZ^{\tN+4-p} (\tY^2 \tX)^{p}$ the symmetry properties are reversed, and an even number $\tZ$ factors must appear, i.e. $p$ must be \emph{odd} (since $\tN$ is odd).

These particular operators are also interesting in that they correspond to Higgsing to the $dP_0$ theories studied in~\S\ref{sec:C3Z3fieldtheory}. In particular, the operator $Z^{N-4-2k} (Y^2 X)^{2k}$ corresponds to Higgsing the A theory $\SU(N-4)\times\SU(N)$ to $\SO(N-4-2k)\times\SU(N-2k)$, whereas the operator $\tZ^{\tN+3-2k} (\tY^2 \tX)^{2k+1}$ corresponds to Higgsing the B theory $\SU(\tN+4)\times\SU(\tN)$ to $\Sp(\tN+3-2k)\times\SU(\tN-2k-1)$.\footnote{Note that from this viewpoint, the $dP_0$ theories enjoy an unbroken $\bZ_3$ baryonic symmetry precisely because they are obtained by turning on a vev for an operator with $Q_B =3$.} Consistent with the proposed operator mapping, we observe that the resulting theories are related by the duality proposed in~\S\ref{sec:C3Z3fieldtheory}. This is another nontrivial consistency check.

At this point, it is also instructive to consider the behavior of the
even-$N$ theories under Higgsing. Turning on a vev for $Z^{N-4-2k}
(Y^2 X)^{2k}$ once again corresponds to Higgsing theory A to
$\SO(N-4-2k)\times\SU(N-2k)$, where now the resulting theory is
conjectured to be self-dual under S-duality, suggesting that the A
theory for even $N$ is also self-dual. However, things are quite
different in the even-$\tN$ B theory. Here, the simplest Higgsing,
corresponding to the operator $\tZ^{(\tN+4)/2}$, breaks
$\SU(\tN+4)\times\SU(\tN)$ to $\Sp(\tN+4)\times\SU(\tN)$, where now
the resulting theory is \emph{not} a singlet under S-duality,
inconsistent with self-duality for the parent theory, while on the
other hand there is no candidate dual for the parent theory.

We hypothesize that the even-$\tN$ B theory is inconsistent in string
theory, potentially due to an uncanceled K-theory (discrete)
tadpole. We hope to verify this through explicit computation of the K-theory
tadpoles in future work.

Having discussed some generic features of the proposed odd-$N$ duality, we next discuss a particularly tractable example with a deformed quantum moduli space.

\subsubsection{Case study: the \alt{$\SU(5)\longleftrightarrow\SU(7)\times\SU(3)$}{SU(5) <-> SU(7)xSU(3)} duality}

The lowest rank example of the proposed duality between the A and B theories is for $N=5$. This example turns out to be particularly tractable, and we now show that the dual theories have biholomorphic quantum-deformed moduli spaces. The $\SU(7)\times\SU(3)$ theory turns out to be somewhat more intuitive, so we begin by discussing this theory, after which we briefly explain how to show that the $\SU(5)$ theory has the same moduli space.

\paragraph{Theory B}

We consider the B-theory $\SU(7) \times \SU(3)$:
\begin{equation}
  \begin{array}{c|cc|cccc}
    & \SU(7) & \SU(3) & \SU(2) & \U(1)_B &
    \U(1)_X' & \U(1)_R'\\
    \hline
    \tilde{A}^i & \ov\fund & \fund & \fund & - \frac{8}{21} & -
    \frac{1}{21} & 0\\
    \tilde{Y} & \fund & \fund & \rd1 & \frac{1}{21} & -
    \frac{13}{21} & 0\\
    \tilde{Z} & \asymm & \rd1 & \rd1 & \frac{3}{7} &
    \frac{3}{7} & 0\\
    \tilde{B}^i & \rd1 & \ov\symm &
    \fund & \frac{1}{3} & \frac{2}{3} & 2\\
    \tilde{X} & \rd1 & \ov\symm &
    \rd1 & \frac{1}{3} & - \frac{1}{3} & 2
  \end{array}
\end{equation}
with the superpotential:
\begin{equation}
  W = \tilde{\lambda}\, \epsilon_{i j} \Tr [ \tilde{B}^i  \tilde{A}^j
  \tilde{Y}] + \frac{1}{2 \tilde{\mu}} \epsilon_{i j} \Tr [ \tilde{X}
  \tilde{A}^i  \tilde{Z}  \tilde{A}^j]\,.
\end{equation}
All possible $\SU(7)$ gauge invariants are products of the following:
\begin{eqnarray}
  \mathcal{Y}^{I m} &\equiv& \tilde{A}^I_a  \tilde{Y}^{a ; m}\,,\quad
  \mathcal{Q}^m \equiv \frac{1}{48} \epsilon_{a b c d e f g} \tilde{Y}^{a
  ; m}  \tilde{Z}^{b c}  \tilde{Z}^{d e}  \tilde{Z}^{f g} \,, \\
  \mathcal{Z}^{I J} &\equiv& \tilde{A}^I_a  \tilde{A}^J_b  \tilde{Z}^{a b} \,,\quad \,
  \Phi \equiv \frac{1}{48} \epsilon_{m n p} \epsilon_{a b c d e f g}
  \tilde{Y}^{a ; m}  \tilde{Y}^{b ; n}  \tilde{Y}^{c ; p}  \tilde{Z}^{d e}
  \tilde{Z}^{f g}\,,\nn
\end{eqnarray}
where $a, b, \ldots$ index $\SU(7)$, $m, n, \ldots$ index $\SU(3)$, and $I, J, \ldots$ index a fictitious $\SU(6) \supset \SU(3) \times \SU(2)$. There is a classical constraint:
\begin{equation}
  \frac{1}{2} \epsilon_{m n p}  ( \mathrm{Pcf} \mathcal{Z})_{I J}
  \mathcal{Y}^{I m} \mathcal{Y}^{J n} \mathcal{Q}^p = ( \mathrm{Pf} \mathcal{Z})
  \Phi\,,
\end{equation}
where we define
\bea
  \mathrm{Pf} M &\equiv& \frac{1}{2^n n!} \epsilon_{i_1 j_1 \ldots i_n j_n}
  M^{i_1 j_1} \ldots M^{i_n j_n} \,, \\ ( \mathrm{Pcf} M)_{i j} &\equiv&
  \frac{1}{2^{n - 1}  ( n - 1) !} \epsilon_{i j i_2 j_2 \ldots i_n j_n}
  M^{i_2 j_2} \ldots M^{i_n j_n}\,,\nn
\eea
for a $2 n \times 2 n$ antisymmetric matrix $M^{i j}$ and ``Pcf'' stands for
``Pfaffian cofactor'', since for $M$ invertible it takes the form $\mathrm{Pcf}
M = ( \mathrm{Pf} M)  [ M^{- 1}]^T$, much like $\mathrm{cof} M = ( \det M)  [ M^{-
1}]^T$ for an arbitrary invertible matrix $M$.

The classical constraint is quantum modified to \cite{Poppitz:1995fh}
\begin{equation}
  \frac{1}{2} \epsilon_{m n p}  ( \mathrm{Pcf} \mathcal{Z})_{I J}
  \mathcal{Y}^{I m} \mathcal{Y}^{J n} \mathcal{Q}^p - ( \mathrm{Pf} \mathcal{Z})
  \Phi = \Lambda_{\SU(7)}^{14} \label{eqn:SU7modcons} \,.
\end{equation}
This equation describes the quantum moduli space of the $\SU(7)$ gauge
theory when we take the $\SU(3)$ gauge coupling and superpotential
couplings to zero.

We now account for the finiteness of these couplings. In particular, the
superpotential couplings give a mass to certain components of $\mathcal{Y}$
and $\mathcal{Z}$, so that on the moduli space we must have
\begin{equation}
  \mathcal{Y}^{m_i n} = \epsilon^{m n p} \mathcal{Y}^i_p \,, \quad
  \mathcal{Z}^{m_i n_j} = \epsilon^{m n p} \mathcal{Z}^{i j}_p\,,
\end{equation}
where $m_i = m + 3 ( i - 1)$ indexes the fictitious $\SU(6)$. Since
the LHS of (\ref{eqn:SU7modcons}) contains only $\SU(3)$ baryons built
from $\SU(3)$ fundamentals, $\SU(3)$ is completely broken
everywhere in the moduli space, leading to a confined description where the
effect of gauging $\SU(3)$ is to remove 8 Higgsed degrees of freedom
and their superpartners.

Thus, the moduli space is parameterized by the operators:
\begin{equation}
  \begin{array}{c|c|cccc}
    & \SU(3) & \SU(2) & \U(1)_B & \U(1)_X' &
    \U(1)_R'\\
    \hline
    \mathcal{Y}^{_i}_m=( \tilde{Y}  \tilde{A}^i) & \ov\fund &
    \rd2 & - \frac{1}{3} & - \frac{2}{3} & 0\\
    \mathcal{Z}^{i j}_m=( \tilde{Z}  \tilde{A}^i  \tilde{A}^j) &
    \ov\fund & \rd3 & - \frac{1}{3} & \frac{1}{3} & 0\\
    \mathcal{Q}^m=( \tilde{Z}^3  \tilde{Y}) & \fund &
    \rd1 & \frac{4}{3} & \frac{2}{3} & 0\\
    \Phi=( \tilde{Z}^2  \tilde{Y}^3) & \rd1 &
    \rd1 & 1 & - 1 & 0
  \end{array}
\end{equation}
subject to the gauging of $\SU(3)$ and the quantum-modified
constraint.{\footnote{Note that this spectrum has an $\SU(3)^3$ gauge
anomaly, but this is fine because $\SU(3)$ is completely broken on the
moduli space. Adding an $\SU(7)$ flavor to the original theory and
s-confining leads to an anomaly free spectrum for $\SU(3)$. Upon adding a
mass for the additional flavor, one obtains a tadpole in the s-confined
description, whereupon the additional fields are set to zero by the F-term
conditions, leaving the moduli given here. }} Therefore, the dimension of the
moduli space is:
\begin{equation}
  \dim \mathcal{M}= 19 - 8 - 1 = 10\,.
\end{equation}
Since all operators are neutral under $\U(1)_R'$, there is an unbroken $\U(
1)_R'$ everywhere in the moduli space.

A complete list of $\SU(3)$ gauge invariants formed from these four
fields is:
\begin{equation}
  \begin{array}{c|cccc}
    & \SU(2) & \U(1)_B & \U(1)_X' & \U(1)_R'\\
    \hline
    \mathcal{Y}^2 \mathcal{Z} & \rd3 & - 1 & - 1 & 0\\
    \mathcal{Y}\mathcal{Z}^2 & \rd4 \oplus \rd2 & - 1 &
    0 & 0\\
    \mathcal{Z}^3 & \rd1 & - 1 & 1 & 0\\
    \mathcal{Z}\mathcal{Q} & \rd3 & 1 & 1 & 0\\
    \mathcal{Y}\mathcal{Q} & \rd2 & 1 & 0 & 0\\
    \Phi & \rd1 & 1 & - 1 & 0
  \end{array}
\end{equation}
Since there are a total of 16 invariants, still subject to one modified
constraint, we conclude that there are five further ``classical'' constraints
relating these $\SU(3)$ composites. To make these constraints
explicit, we define:
\begin{equation}
  \bar{\mathcal{Q}}^A_m \equiv \{ \mathcal{Y}^i_m, \mathcal{Z}^{\alpha}_m \}\,,
\end{equation}
where $A$ indexes a fictitious $\SU(5) \supset \SU(2) \times
\SO(3)$, and $\mathcal{Z}^{\alpha}_m \equiv \frac{1}{2}
\sigma^{\alpha}_{i j} \mathcal{Z}^{i j}_m$ with the $\SU(2) \cong
\SO(3)$ conventions:
\begin{equation}
\renewcommand{\arraystretch}{0.7}
  \sigma^{\alpha \; j}_i = \left\{ \left(\begin{array}{cc}
    0 & 1\\
    1 & 0
  \end{array}\right), \left(\begin{array}{cc}
    0 & - i\\
    i & 0
  \end{array}\right), \left(\begin{array}{cc}
    1 & 0\\
    0 & - 1
  \end{array}\right) \right\} , \;\;
   \sigma^{i j}_{\alpha} = \epsilon^{i k}
  \sigma^{\alpha \; j}_k \,, \;\; \sigma^{\alpha}_{i j} = \sigma^{\alpha \; k}_i
  \epsilon_{k j} \,, \;\; \epsilon^{12} = \epsilon_{12} = +1 \,. \label{eqn:SU2conventions}
\end{equation}
The classical constraints then take the form:
\begin{equation}
  {}[ \bar{\mathcal{Q}}^3]_{A B}  [ \mathcal{Q} \bar{\mathcal{Q}}]^B = 0 \,, \quad \epsilon^{A B C D E}  [ \bar{\mathcal{Q}}^3]_{A B}  [
  \bar{\mathcal{Q}}^3]_{C D} = 0\,.
\end{equation}
Although both equations appear to have five components, examining small
fluctuations about a background with $\bar{\mathcal{Q}} \neq 0$ satisfying
these constraints gives only three independent constraints from the second
equation, and a further two from the first, for a total of five constraints,
as expected.

We define:
\bea
  \Psi &\equiv& \det \mathcal{Z}^{\alpha}_m \,, \quad \Psi^i_{\alpha} \equiv
  \mathcal{Y}^i_m  [ \mathrm{cof} \mathcal{Z}]^m_{\alpha} \,, \quad \Psi^{\alpha} \equiv \frac{1}{2} \epsilon_{i j} \epsilon^{m n p}
  \mathcal{Y}^i_m \mathcal{Y}^j_n \mathcal{Z}^{\alpha}_p \,, \\
  \Phi^i &\equiv& \mathcal{Q}^m \mathcal{Y}_m^i \,, \quad \,\Phi^{\alpha} \equiv \mathcal{Q}^m
  \mathcal{Z}_m^{\alpha}\,.
\eea
In terms of these gauge invariants, the classical constraints becomes:
\begin{equation}
  \Psi \Phi^j - \Psi^j_{\alpha} \Phi^{\alpha} = 0 \,, \;\;
  \epsilon_{i j} \Psi^i_{\alpha} \Phi^j + \epsilon_{\alpha \beta \gamma}
  \Psi^{\beta} \Phi^{\gamma} = 0 \,,\label{eqn:SU7cons1}
\end{equation}
and
\begin{equation}
  \Psi^i_{\alpha} \Psi^{\alpha} = 0 \,, \;\; \Psi \Psi^{\alpha} -
  \frac{1}{2} \epsilon^{\alpha \beta \gamma} \epsilon_{i j}
  \Psi^i_{\beta} \Psi^j_{\gamma} = 0 \,.\label{eqn:SU7cons2}
\end{equation}
After a somewhat longer computation, we find that the quantum modified
constraint takes the form:
\begin{equation}
  i \Phi^i \Psi^j_{\alpha} \sigma^{\alpha}_{i j} - \Phi_{\alpha} \Psi^{\alpha}
  - 2 i \Phi \Psi = \Lambda^{14}_{\SU(7)} \,.\label{eqn:SU7cons3}
\end{equation}
Together, (\ref{eqn:SU7cons1}, \ref{eqn:SU7cons2}, \ref{eqn:SU7cons3})
completely describe the deformed moduli space of the quantum theory.

The maximal unbroken flavor symmetry is $\SU(2) \times U (1)_{B + X}'
\times U (1)_R'$, which is attained when we take $\Phi$ and $\Psi$ to be
nonvanishing with all other fields vanishing. We can then solve the
constraints to eliminate $\Phi^i$, $\Psi^{\alpha}$, and $\Phi$:
\begin{equation}
  \Phi^i = \frac{1}{\Psi} \Psi^i_{\alpha} \Phi^{\alpha} \,, \quad
  \Psi^{\alpha} = \frac{1}{2 \Psi} \epsilon^{\alpha \beta \gamma}
  \epsilon_{i j} \Psi^i_{\beta} \Psi^j_{\gamma} \,, \quad \Phi =
  \frac{i}{2 \Psi} \Lambda^{14}_{\SU(7)}\,,
\end{equation}
whereupon the remaining constraints are trivially satisfied. The light modes
along this line are therefore:
\begin{equation}
  \begin{array}{c|ccc}
    & \SU(2) & \U(1)_{B + X}' & \U(1)_R'\\
    \hline
    \Psi^i_{\alpha} & \rd4 \oplus \rd2 & - 1 & 0\\
    \Psi & \rd1 & 0 & 0\\
    \Phi^{\alpha} & \rd3 & 2 & 0
  \end{array}
\end{equation}
One can check that the global $\SU(2)\times\U(1)_{B+X}'\times\U(1)_R'$ anomalies match those of the original description, as expected.

\paragraph{Theory A}

We now consider the dual theory:\footnote{Reference~\cite{Lin:2011vd} discusses a similar theory in the context of dynamical supersymmetry breaking.}
\begin{equation}
  \begin{array}{c|c|cccc}
    & \SU(5) & \SU(2) & \U(1)_B & \U(1)_X' & \U(1)_R'\\
    \hline
    A^i & \ov\fund & \fund & - \frac{8}{5} & - \frac{1}{5} & 0\\
    Y & \ov\fund & \rd1 & \frac{7}{5} & - \frac{1}{5} &
    2\\
    Z & \rd1 & \rd1 & 3 & 1 & 2\\
    B^i & \asymm & \fund & \frac{1}{5} & \frac{2}{5} & 0\\
    X & \asymm & \rd1 & \frac{1}{5} & - \frac{3}{5} & 0
  \end{array}
\end{equation}
with the superpotential:
\begin{eqnarray}
  W & = & \lambda \; \epsilon_{i j} Y_m A^i_n B^{j m n} + \frac{1}{2 \mu}
  \epsilon_{i j} Z A^i_m A^j_n X^{m n}\,.
\end{eqnarray}
As reviewed in~\S\ref{subsec:SU5}, taking $W \rightarrow 0$, the
$\SU$(5) gauge theory has an s-confined description:
\begin{equation}
  \begin{array}{c|c|cccc}
    & \SU(5) & \SU(3)_a & \SU(3)_b & \U(1)^{(s)}_B & \U(1)_R^{( s)}\\
    \hline
    \mathcal{A}^I & \ov\fund & \fund & \rd1 & - \frac{3}{5}
    & 2 / 3\\
    \mathcal{B}^I & \asymm & \rd1 & \fund & \frac{1}{5} & 0\\
    Z & \rd1 & \rd1 & \rd1 & 3 & 2\\
    \hline \hline
    T^I_J =\mathcal{A}^2 \mathcal{B} &  & \ov\fund & \fund & - 1 & 4
    / 3\\
    U^{I ; J}_{\;  \; K} =\mathcal{A}\mathcal{B}^3 &  & \fund &
    \adj & 0 & 2 / 3\\
    V^{I J} =\mathcal{B}^5 &  & \rd1 & \symm & 1 &
    0
  \end{array}
\end{equation}
with the dynamical superpotential:
\begin{equation}
  W = \frac{1}{\Lambda^9}  \left( \epsilon_{I J K} T^I_L U^{L ; J}_{\;  \;
  \; M} V^{M K} - \frac{1}{3} \epsilon_{I J K} U^{I ; L}_{\;  \;  \; N}
  U^{J ; M}_{\;  \;  \; L} U^{K ; N}_{\;  \;  \; M} \right)\,,
\end{equation}
where $I, J, \ldots = 1, 2, 3$, and we omit an unimportant $\U(1)$ global
symmetry under which only the additional singlet $Z$ is charged.

Thus, deforming the resulting theory by the tree-level superpotential, we
obtain:
\begin{equation}
  W = \frac{1}{\Lambda^9}  \left( \epsilon_{I J K} T^I_L U^{L ; J}_{\;  \;
  \; M} V^{M K} - \frac{1}{3} \epsilon_{I J K} U^{I ; L}_{\;  \;  \; N}
  U^{J ; M}_{\;  \;  \; L} U^{K ; N}_{\;  \;  \; M} \right) + \lambda T^i_i +
  \frac{1}{\mu} Z \; T^3_3 \,,
\end{equation}
where $i, j, \ldots = 1, 2$. The superpotential partially breaks the flavor
symmetries of the pure s-confining theory. In particular, $\SU(3)_a
\times \SU(3)_b \rightarrow \SU(2) \times U (1)_a \times U (
1)_b$ where $\U(1)_a$ and $\U(1)_b$ denote the $\mathrm{diag} (1 / 3, 1 / 3,
- 2 / 3)$ elements of each $\SU(3)$, and the unbroken $\U(1)$ linear
combinations are:
\begin{equation}
  U (1)_R' = U (1)_R^{( s)} - 2 U (1)_a \,, \quad U (1)_B = U (
  1)_B^{( s)} - 3 U (1)_a \,, \quad U (1)_X' = U (1)_a + U (1)_b \,.
\end{equation}
The F-term conditions now read:
\begin{align}
  \frac{1}{\Lambda^9} \epsilon_{I J K} U^{L ; J}_{\;  \;  \; M} V^{M K} +
  \lambda \delta_I^i \delta_i^L + \frac{1}{\mu} Z \delta^3_I \delta^L_3 &=
  0 \,, \qquad T^3_3 = 0 \,, \nonumber\\
  \epsilon_{I J K} T^I_L V^{M K} - \epsilon_{I L K} U^{I ; M}_{\;  \;
  \; N} U^{K ; N}_{\;  \;  \; J} &= 0 \,, \qquad \epsilon_{I J K}
  T^I_L U^{L ; J}_{\;  \;  \; M} = 0 \, .
\end{align}
It is straightforward to show, using the Gr\"{o}bner basis
algorithm,{\footnote{We use the \texttt{Elimination[]} function of the
\texttt{Stringvacua} package~\cite{Gray:2008zs}, which uses \texttt{SINGULAR} \cite{DGPS} for
computations.}} that solutions to these equations must satisfy:
\begin{equation}
  T^I_J = 0 \,, \quad Z = 0 \,, \quad U^{3 ; I}_{\;  \; J} = 0 \,, \quad U^{( i ;
  j)}_{\;\;\;\; 3} = 0 \,, \quad U^{( i ; j)}_{\;\;\;\; i} = 0 \,, \quad U^{i ; 3}_{\;\; i} = 0 \,.
\end{equation}
We decompose the non-vanishing fields as follows:
\bea
  U^{i ; j}_{\;\; 3} &=& \psi \epsilon^{i j} \,, \quad U^{i ; 3}_{\;  \; j}
  = \sigma^{\alpha i}_j \psi^{\alpha} \,, \quad V^{i j} = \sigma^{i j}_{\alpha}
  \phi^{\alpha} \,, \quad V^{i 3} = \phi^i \,, \nn \\
  V^{3 3} &=& \phi \,, \quad \quad \, U^{i ; j}_{\;  \; k} = \frac{1}{3} \psi_{\alpha}^j \sigma^{\alpha i}_k -
  \frac{1}{3} \sigma^{i j}_{\alpha} \psi^{\alpha}_k - \psi_{\alpha}^i
  \sigma^{\alpha j}_k \,,
\eea
where $\psi_k^{\alpha} \equiv \psi^j_{\alpha} \epsilon_{j k}$, our
remaining conventions are given in (\ref{eqn:SU2conventions}), and the fields
transform as
\begin{equation}
  \begin{array}{c|cccc}
    & \SU(2) & \U(1)_B & \U(1)_X' & \U(1)_R'\\
    \hline
    \psi=A B^3 & \rd1 & - 1 & 1 & 0\\
    \psi^i_{\alpha}=A B^2 X & \rd4 \oplus \rd2 &
    - 1 & 0 & 0\\
    \psi^{\alpha}=A B X^2 & \rd3 & - 1 & - 1 & 0\\
    \phi^{\alpha}=B^4 X & \rd3 & 1 & 1 & 0\\
    \phi^i=B^3 X^2 & \rd2 & 1 & 0 & 0\\
    \phi=B^2 X^3 & \rd1 & 1 & - 1 & 0
  \end{array}
\end{equation}
under the global symmetries.

The F-term conditions involving nonvanishing fields are:
\begin{equation}
  \epsilon_{i J K} U^{l ; J}_{\;  \;  \; M} V^{M K} + \lambda \Lambda^9
  \delta^l_i = 0 \,, \quad \epsilon_{j k} U^{i ; j}_{\;  \;  \; M} V^{M k} = 0
  \,, \quad \epsilon_{i k} U^{i ; M}_{\;  \;  \; N} U^{k ; N}_{\;  \;  \; J} =
  0\,.
\end{equation}
Applying the above decomposition and simplifying, we eventually obtain:
\begin{eqnarray}
  \psi \phi^i - 2 \psi_{\alpha}^i \phi^{\alpha} & = & 0 \,, \quad
  \epsilon_{i j} \psi^i_{\alpha} \phi^j + i \epsilon_{\alpha \beta
  \gamma} \phi^{\beta} \psi^{\gamma} = 0 \,, \nonumber\\
  \psi_{\alpha}^i \psi^{\alpha} & = & 0 \,, \quad\; \psi \psi^{\alpha} -
  i \epsilon^{\alpha \beta \gamma} \epsilon_{i j} \psi_{\beta}^i
  \psi_{\gamma}^j = 0 \,, \nonumber\\
  \sigma^{\alpha}_{i j} \psi^i_{\alpha} \phi^j + \psi \phi + \psi_{\alpha}
  \phi^{\alpha} & = & - \lambda \Lambda^9 \, .
\end{eqnarray}
Upon replacing:
\begin{equation}
\psi \rightarrow \frac{2}{m^5} \Psi \,, \;\;\; \psi^{\alpha} \rightarrow
   \frac{i}{m^3} \Psi^{\alpha} \,, \;\;\; \psi^i_{\alpha} \rightarrow
   \frac{1}{m^4} \Psi^i_{\alpha} \,, \;\;\;  \phi \rightarrow - \Phi \,, \; \phi^i
   \rightarrow \frac{1}{m} \Phi^i \,, \;\;\; \phi^{\alpha} \rightarrow
   \frac{1}{m^2} \Phi^{\alpha} \,,
\end{equation}
for some mass scale $m$, we recover the exact constraint equations for the
moduli space of theory B for $\Lambda^{14}_{\SU(7)} = - i \lambda m^5
\Lambda_{\SU(5)}^9$. Thus, the moduli spaces are biholomorphic.

\subsection{Complex cone over \alt{$\bF^0$}{F0}}
\label{sec:F0}

\begin{figure}
  \begin{center}
   \subfigure[Phase I\label{fig:F0phaseI}]{\includegraphics[width=0.25\textwidth]{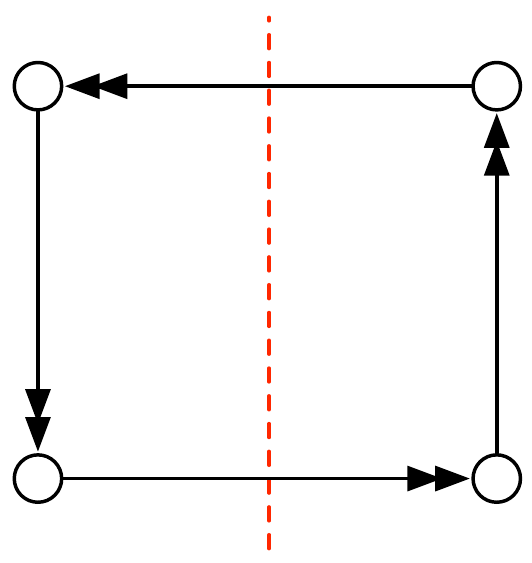}}\hspace{5mm}
   \subfigure[Phase II\label{fig:F0phaseII}]{\includegraphics[width=0.25\textwidth]{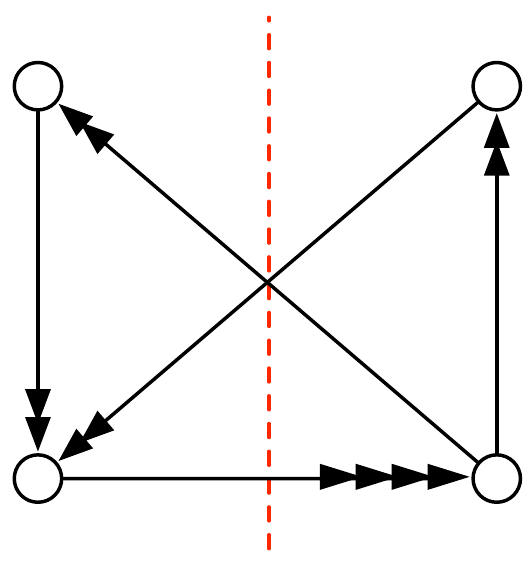}}\hspace{5mm}
   \subfigure[The resulting quiverfolds\label{fig:F0quiverfolds}]{\includegraphics[width=0.35\textwidth]{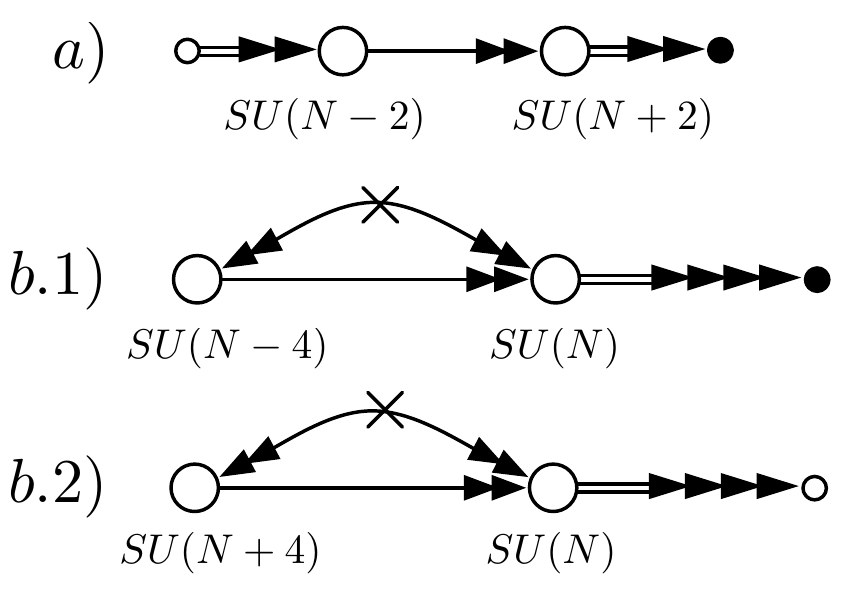}}
  \end{center}
  \caption[The two Seiberg-dual phases of $\bF_0$ and their orientifolds]{\subref{fig:F0phaseI}--\subref{fig:F0phaseII} The two Seiberg-dual quiver gauge theories for $\bF_0$. The red dashed lines indicate orientifold involutions compatible with the $\SU(2)\times\SU(2)$ isometry of the base. \subref{fig:F0quiverfolds} The $\SU(2)\times\SU(2)$-preserving anomaly-free quiverfolds that result from orientifolding these theories.}
  \label{fig:F0quivers}
\end{figure}

We now consider the Calabi-Yau cone over $\bF_0=\bP^1\times\bP^1$, a $\bZ_2$ orbifold of the conifold which is the same as the real cone over $Y^{2,0}$. As shown in figure~\ref{fig:F0phaseI}--\subref{fig:F0phaseII}, there are two different toric\footnote{In this context, a toric quiver gauge theory is one for which the number of arrows entering and exiting each node is the same.} quiver gauge theories (``phases'') which describe D3 branes probing this singularity. These theories, which we denote by phase I and phase II, are related by Seiberg duality on one of the nodes.

In the $\bC^3/\bZ_3$ and $dP_1$ examples studied previously, there was
only one toric phase, and we found a duality relating two different
orientifolds of that phase which differed by exchanging $\SO$ and
$\Sp$ groups and symmetric and antisymmetric tensor matter, a
``negative rank duality'' as explained
in appendix~\ref{app:negativerankdual}. We argue in~\cite{transitions2}
that these dualities relate to the $\SL(2,\bZ)$ self-duality of type
IIB string theory. Interestingly, negative rank duality also partly
``explains'' the pattern of $\mathcal{N}=4$ dualities between $\SO$
and $\Sp$ theories, suggesting that it may have some physical
interpretation relating to Montonen-Olive duality and its
$\mathcal{N}=1$ analogues.

By contrast, for the $\bF_0$ orientifolds we now study, the negative
rank duals are either trivially equivalent or related by Seiberg
duality. Instead, we find a nontrivial duality between orientifolds of
the two different phases. Although the two phases are related by
Seiberg duality in the parent theory, the resulting orientifolds are
not obviously related in this way, giving yet another new field theory
duality.\footnote{We cannot eliminate the possibility that a chain of
  deconfinements, dualizations, and reconfinements might relate the
  two theories via known dualities, but we have not been able to find such a
  chain.}

As in our previous examples, we wish to consider orientifolds corresponding to compact O7 planes wrapping the base $\bF_0$. This is equivalent to the requirement that the orientifold preserves the $\SU(2)\times\SU(2)$ isometry of the base. Only the involutions pictured in figure~\ref{fig:F0phaseI}--\subref{fig:F0phaseII} do so, and of the fixed-element sign choices compatible with $\SU(2)\times\SU(2)$ invariance, only one choice for phase I and two for phase II lead to anomaly-free theories, giving the theories pictured in figure~\ref{fig:F0quiverfolds}.\footnote{See~\cite{transitions3} for a more detailed, brane tiling-based derivation of these orientifolds.}

Notice that the sole orientifold of phase I is its own negative rank dual, whereas the two orientifolds of phase II are related by negative rank duality. As we shall see, the latter two theories are Seiberg dual upon dualizing the left-node. We now discuss the orientifolds of each phase in turn, providing evidence for a duality between the orientifolds of the different phases.

\subsubsection*{Phase I}

We obtain the orientifold theory:
\begin{align}
  \label{eq:F0-phaseI+-quiverFT}
  \begin{array}{c|cc|ccccc}
    &\SU(N-2) & \SU(N+2) & \SU(2)_1 & \SU(2)_2 & \U(1)_B & \U(1)_R & \bZ_2\\
    \hline
    A^i & \fund & \ov\fund & \fund & {\bf 1} & \frac{N}{N^2-4} & \frac12 -\frac{6}{N^2-4} & \omega_{2(N-2)}\\
    B^i & \ov\symm & {\bf 1} & {\bf 1} & \fund & -\frac{1}{N-2} & \frac12 +\frac{3}{N-2} & \omega_{2(N-2)}^{-2}\\
    C^i & {\bf 1} & \asymm & {\bf 1} & \fund & -\frac{1}{N+2} & \frac12 -\frac{3}{N+2} & 1
  \end{array}
\end{align}
with superpotential given by \be W = \epsilon_{ij} \epsilon_{kl} \Tr
\lp A^i B^k A^j C^l \rp \,.  \ee For odd $N$, the $\bZ_2$ discrete
symmetry is gauge equivalent\footnote{Here and in future by ``gauge
  equivalent'' we mean that the two generators are related by
  composition with a (constant) gauge transformation, see the
  discussion at the beginning of~\S\ref{sec:C3Z3fieldtheory}.} to the
$\bZ_2$ center of $\SU(2)_1$, whereas for even $N$ it is a distinct
global symmetry. Thus, the global symmetry group is actually $\SU(2)_1
\times \SU(2)_2 \times \U(1)_B \times \U(1)_R \times \bZ_{\gcf(2,N)}$.

The global anomalies for this theory are shown in table~\ref{tab:F0-anomaliesI}.
\begin{table}
\begin{center}
\subtable[Phase I\label{tab:F0-anomaliesI}]{
$\begin{array}{|c|c|} \hline
    \SU(2)_{1/2}^3 & (-1)^N \\ \hline
    \SU(2)_{1/2}^2 \, \U(1)_B & \pm N \\ \hline
    \SU(2)_{1/2}^2 \, \U(1)_R & -\tfrac12 (N^2+8) \\\hline
     \U(1)^2_B \, \U(1)_R & -2 \\\hline
    \U(1)_R^3 & \tfrac32 N^2 - 34 \\\hline
    \U(1)_R & -10 \\\hline
    \SU(2)_{1}^2 \, \bZ_2 & (-1)^N \\ \hline
    \SU(2)_{2}^2 \, \bZ_2 & -(-1)^N \\ \hline
    \bZ_2 & 1 \\ \hline
  \end{array}$
  }\hspace{5mm}
\subtable[Phase II\label{tab:F0-anomaliesII}]{
$\begin{array}{|c|c|}
    \hline
    \SU(2)_{1/2}^3 & (-1)^N \\ \hline
    \SU(2)_{1/2}^2 \, \U(1)_B & \pm N \\ \hline
    \SU(2)_{1/2}^2 \, \U(1)_R & -\tfrac12 (N^2+8) \\\hline
    \U(1)^2_B \, \U(1)_R & -2 \\\hline
    \U(1)_R^3 & \tfrac32 N^2 -34 \\\hline
    \U(1)_R & -10 \\\hline
    \SU(2)_{1}^2 \, \bZ_{4} & -1 \\ \hline
    \SU(2)_{2}^2 \, \bZ_{4} &  -(-1)^N \\ \hline
    \bZ_{4} & 1 \\ \hline
  \end{array}$
}
\end{center}
\caption[The anomalies for the orientifolds of the different phases of $\bF_0$]{The nonvanishing anomalies for the orientifolds of the different phases of $\bF_0$, where we use a multiplicative notation for discrete and Witten anomalies as before (see~\S\ref{subsec:classicchecks}). The $\U(1)_B$, $\U(1)_B^3$, and $\U(1)_B \, \U(1)_R^2$ anomalies all vanish.}
\end{table}
Note that the anomalies are invariant under $N \rightarrow -N$ combined with a charge conjugation of $\U(1)_B$. This invariance corresponds to taking the negative rank dual as explained in appendix \ref{app:negativerankdual}. While it led to two different gauge theories in the previous examples, one can check that in this case it maps the above theory to itself.

\subsubsection*{Phase II}
We obtain the orientifold theory:
\begin{align}
  \label{eq:F0-phaseII+-quiverFT}
  \begin{array}{c|cc|ccccc}
    & \SU(N-4) & \SU(N) & \SU(2)_1 & \SU(2)_2 & \U(1)_B & \U(1)_R & \bZ_4\\
    \hline
    A^i & \fund & \ov\fund & \fund & {\bf 1} & \frac{1}{N-4} & \frac12+\frac{2}{N} & \omega_{4 N}^{-1}\, \omega_{4 (N-4)}^{-1}\\
    B^i & \ov\fund & \ov\fund & {\bf 1} & \fund & -\frac{1}{N-4} & \frac12+\frac{2}{N} & \omega_{4 N}^{-1}\, \omega_{4 (N-4)}\\
    C^{i;j} & {\bf 1} & \asymm & \fund & \fund & 0 & 1-\frac{4}{N} & \omega_{4 N}^{2}
  \end{array}
\end{align}
with superpotential
\be
W = \epsilon_{ij} \epsilon_{kj} \Tr \lp A^i C^{j;k} B^l \rp,
\ee
as well as the theory
\begin{align}
  \label{eq:F0-phaseII--quiverFT}
  \begin{array}{c|cc|ccccc}
    & \SU(N+4) & \SU(N) & \SU(2)_1 & \SU(2)_2 & \U(1)_B & \U(1)_R & \bZ_4\\\hline
    \tilde{A}^i & \fund & \ov\fund & \fund & {\bf 1} & \frac{1}{N+4} & \frac12-\frac{2}{N} & \omega_{4 N}\, \omega_{4 (N+4)}\\
    \tilde{B}^i & \ov\fund & \ov\fund & {\bf 1} & \fund & -\frac{1}{N+4} & \frac12-\frac{2}{N} & \omega_{4 N}\, \omega_{4 (N+4)}^{-1}\\
    \tilde{C}^{i;j} & {\bf 1} & \symm & \fund & \fund & 0 & 1+\frac{4}{N} & \omega_{4 N}^{-2}
  \end{array}
\end{align}
with superpotential
\be
\tilde{W} = \epsilon_{ij} \epsilon_{kj} \Tr \lp \tilde{A}^i \tilde{C}^{j;k} \tilde{B}^l \rp.
\ee
For odd $N$, the $\bZ_4$ discrete symmetry is gauge-equivalent to the $\bZ_2$ center of $\SU(2)_1$, whereas for $N=4k+2$ the $\bZ_2 \subset \bZ_4$ subgroup is gauge-equivalent to the $\bZ_2$ center of $\SU(2)_1$, and for $N=4k$ the $\bZ_4$ is a distinct global symmetry. Thus, the global symmetry group is $\SU(2)_1 \times \SU(2)_2 \times \U(1)_B \times \U(1)_R \times \bZ_{\gcf(4,N)}$.

It is straightforward to check that these two theories, which are
related by negative rank duality, are also related by Seiberg
dualizing the $\SU(N\pm4)$ gauge group factor and integrating out
massive matter.

\subsubsection*{Relationship between the two phases}

The global anomalies for these theories are shown in table~\ref{tab:F0-anomaliesII}, where for simplicity we do not display the anomalies of the Seiberg dual theories separately; one can verify that they match as expected. More importantly, we see that the phase I and phase II orientifolds have matching anomalies for odd $N_{\text{phase I}} = N_{\text{phase II}}$. For even $N$ the global symmetry groups do not match, and the theories are not dual.\footnote{Although the global symmetry groups match for $N=2k+2$, by removing a single D3 brane we reduce $N \to N-2$, after which the global symmetry groups no longer match, so there is no duality for even $N$. One can also show this by constructing holomorphic gauge invariants in one theory with no dual in the other theory, for example the phase I operator $C^{\frac{N+2}{2}}$.}

\medskip

It is interesting to understand better the nature of this prospective
duality between orientifolds of the two phases. We will present
evidence in~\cite{transitions3} that the embeddings in string theory for
the two phases are related as in realizations of ordinary Seiberg
duality, so we can expect the nature of the duality relating the two phases
phases to be an infrared duality closely analogous to it.
However, we emphasize that this duality is not obviously derivable from known examples of Seiberg duality.
In~\cite{transitions3}, we also argue that the
action of IIB S-duality on the D-brane configuration describing each
phase reproduces the field theory dualities inside each phase that we
just studied: it is a self-duality in phase I and it exchanges the two
theories in phase II.

\subsection{The real cone over $Y^{4,0}$}

Before concluding, we present one final example of new dualities relating the world-volume gauge theories of D3 branes probing a Calabi-Yau singularity. Much like the $\bF_0$ example studied above, this example exhibits interesting new patterns of dualities which appear distinct from the $\bC_3/\bZ_3$ and $dP_1$ examples discussed previously.

\begin{figure}
  \begin{center}
   \subfigure[Phase I\label{fig:Y40phaseI}]{\includegraphics[height=0.25\textwidth]{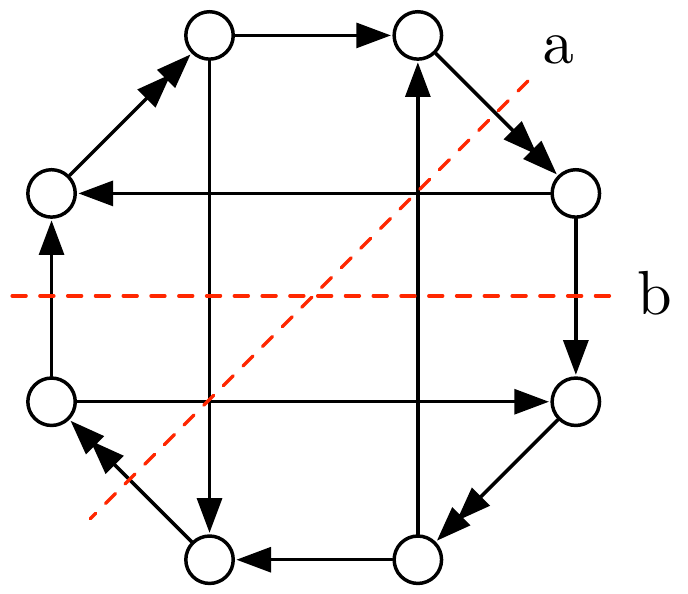}}\hspace{5mm}
   \subfigure[Phase II\label{fig:Y40phaseII}]{\includegraphics[height=0.25\textwidth]{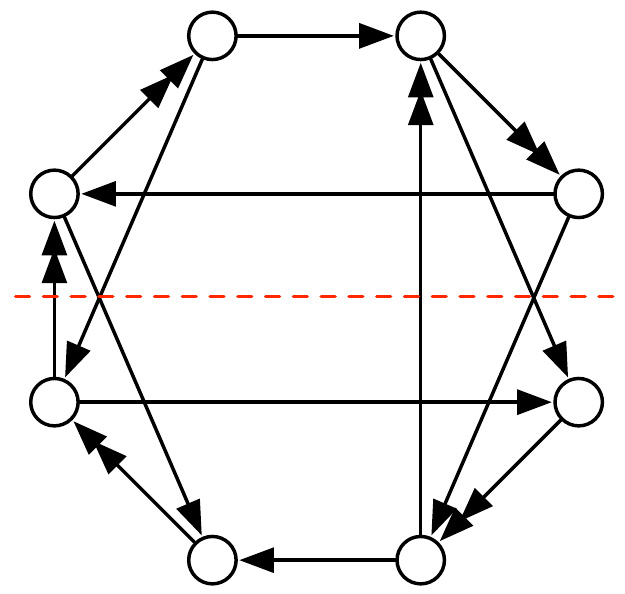}}\hspace{5mm}
  \end{center}
  \caption[Two of the five Seiberg-dual toric quiver gauge theories for $Y^{4,0}$]{Two of the five Seiberg-dual toric quiver gauge theories for $Y^{4,0}$. The red dashed lines indicate the orientifold involutions we will consider.}
  \label{fig:Y40quivers}
\end{figure}

We consider the real cone over $Y^{4,0}$ which, like the cone over $Y^{2,0}$ considered above, is an orbifold of the conifold. There are five toric quiver gauge theories which describe D3 branes probing this singularity,\footnote{See~\cite{Benvenuti:2004dy,Benvenuti:2004wx} for a classification of the toric phases of D3 branes probing a $Y^{p,q}$ singularity.} all of which are Seiberg dual. We focus on the two phases pictured in figure~\ref{fig:Y40quivers} and on the involutions also pictured there.\footnote{Two of the remaining three phases also admit involutions, and several of the resulting orientifold theories are manifestly Seiberg dual to those considered here.}

\begin{figure}
  \begin{center}
   \includegraphics[width=\textwidth]{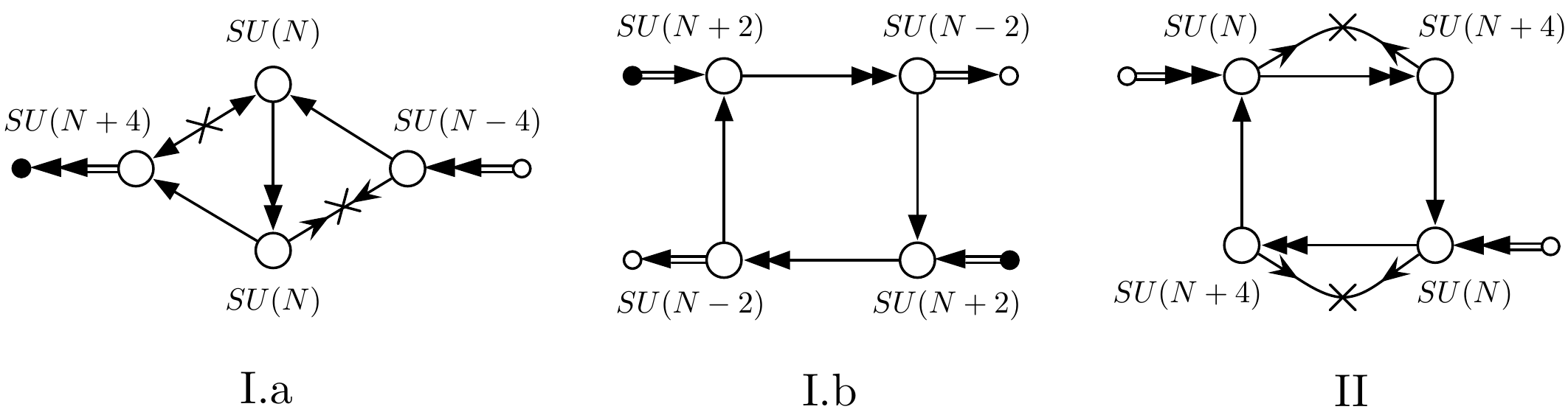}
  \end{center}
  \caption[Quiverfolds for three anomaly-free orientifolds of $Y^{4,0}$]{Quiverfolds for the anomaly-free orientifold gauge theories we will consider, arranged by the parent quiver and involution used to generate them (see figure~\ref{fig:Y40quivers}). The phase II quiverfold has a negative rank dual which is not pictured, as it is manifestly Seiberg dual to the quiverfold which is shown. The phase I quiverfolds are ``self-dual'' under negative rank duality.}
  \label{fig:Y40quiverfolds}
\end{figure}

For simplicity, we restrict our attention to the anomaly-free orientifolds pictured in figure~\ref{fig:Y40quiverfolds}. One can show that these orientifolds correspond to compact O7 planes, and preserve the full $\SU(2)\times\U(1)_X\times\U(1)_R$ isometry group of $Y^{4,0}$. We now briefly discuss each of the three theories in turn, after which we illustrate a potential duality between them using anomaly matching.

\subsubsection*{Phase I, Involution a} We obtain the orientifold theory:
\begin{align}
  \label{eq:Y40-phaseIa-quiverFT}
  	\begin{array}{c|cccc|cccc}
		 & \SU(N+4) & \SU(N) & \SU(N) & \SU(N-4) & \SU(2) & \U(1)_B & \U(1)_X & \U(1)_R \\ \hline
		A^i & \asymm & \sing & \sing & \sing & \fund & \frac{1}{N+4} & 0 & \frac{1}{2}-\frac{6}{N+4}  \\
		S^i & \sing & \sing & \sing & \ov\symm & \fund & \frac{1}{N-4} & 0 & \frac{1}{2}+\frac{6}{N-4}  \\
		P_{12} & \ov\fund & \ov\fund & \sing & \sing & \sing & -\frac{N+2}{N(N+4)} & -\frac{N-4}{N} & \frac{1}{2}+\frac{3}{N+4}  \\
		P_{13} & \ov\fund & \sing & \fund & \sing & \sing & -\frac{N+2}{N(N+4)} & \frac{N-4}{N} & \frac{1}{2}+\frac{3}{N+4} \\
		P_{23}^i & \sing & \fund & \ov\fund & \sing & \fund & \frac{1}{N} & 0 & \frac{1}{2} \\
		P_{24} & \sing & \ov\fund & \sing & \fund & \sing & -\frac{N-2}{N(N-4)} & \frac{N+4}{N} & \frac{1}{2}-\frac{3}{N-4}  \\
		P_{34} & \sing & \sing & \fund & \fund & \sing & -\frac{N-2}{N(N-4)} & -\frac{N+4}{N} & \frac{1}{2}-\frac{3}{N-4}
	\end{array}
\end{align}
with superpotential
\begin{equation}
W = \epsilon_{i j} A^i P_{12}^{} P_{23}^j P_{13}^{} + \epsilon_{i j} S^i P_{24}^{} P_{23}^j P_{34}^{} \,,
\end{equation}
where there is an additional discrete $\bZ_{\gcf(4,N)}$ symmetry for even $N$, which we omit from the charge table for simplicity, as it will not play a large role in our analysis.

\subsubsection*{Phase I, Involution b} We obtain the orientifold theory:
\begin{align}
  \label{eq:Y40-phaseIb-quiverFT}
	\begin{array}{c|cccc|cccc}
		 & \SU(N+2) & \SU(N-2) & \SU(N+2) & \SU(N-2) & \SU(2) & \U(1)_B & \U(1)_X & \U(1)_R \\ \hline
		T_1 & \ov\asymm & \sing & \sing & \sing & \sing & -\frac{1}{N+2} & \frac{N+4}{N+2} & \frac{1}{2}-\frac{7}{N+2} \\
		T_2 & \sing & \symm & \sing & \sing & \sing & -\frac{1}{N-2} & -\frac{N-4}{N-2} & \frac{1}{2}+\frac{7}{N-2} \\
		T_3 & \sing & \sing & \ov\asymm & \sing & \sing & -\frac{1}{N+2} & -\frac{N+4}{N+2} & \frac{1}{2}-\frac{7}{N+2} \\
		T_4 & \sing & \sing & \sing & \symm & \sing & -\frac{1}{N-2} & \frac{N-4}{N-2} & \frac{1}{2}+\frac{7}{N-2} \\
		P_{12}^i & \fund & \ov\fund & \sing & \sing & \fund & \frac{N}{N^2-4} & -\frac{2N}{N^2-4} & \frac{1}{2}-\frac{14}{N^2-4} \\
		P_{23} & \sing & \fund & \ov\fund & \sing & \sing & -\frac{N}{N^2-4} & \frac{N^2}{N^2-4} & \frac{1}{2}+\frac{14}{N^2-4} \\
		P_{34}^i & \sing & \sing & \fund & \ov\fund & \fund & \frac{N}{N^2-4} & \frac{2N}{N^2-4} & \frac{1}{2}-\frac{14}{N^2-4} \\
		P_{41} & \ov\fund & \sing & \sing & \fund & \sing & -\frac{N}{N^2-4} & -\frac{N^2}{N^2-4} & \frac{1}{2}+\frac{14}{N^2-4}
	\end{array}
\end{align}
with superpotential
\begin{equation}
W = \frac{1}{2} \epsilon_{i j} T_1^{} P_{12}^i P_{12}^j T_2^{} +\frac{1}{2} \epsilon_{i j} T_3^{} P_{34}^i P_{34}^j T_4^{} + \epsilon_{i j} P_{12}^i P_{23}^{} P_{34}^j P_{41}^{} \,.
\end{equation}
As before, there is an additional discrete $\bZ_{\gcf(4,N)}$ symmetry for even $N$.

\subsubsection*{Phase II} We obtain the orientifold theory:
\begin{align}
  \label{eq:Y40-phaseII-quiverFT}
	\begin{array}{c|cccc|cccc}
		 & \SU(N+4) & \SU(N) & \SU(N+4) & \SU(N) & \SU(2) & \U(1)_B & \U(1)_X & \U(1)_R \\ \hline
		P_{12} & \fund & \ov\fund & \sing & \sing & \sing & -\frac{1}{N+4} & -\frac{N+2}{N+4} & \frac{1}{2}+\frac{2(N+8)}{N(N+4)} \\
		P_{23} & \sing & \fund & \fund & \sing & \sing & -\frac{1}{N+4} & -\frac{N+6}{N+4} & \frac{1}{2}-\frac{2(3N+8)}{N(N+4)} \\
		P_{34} & \sing & \sing & \fund & \ov\fund & \sing & -\frac{1}{N+4} & \frac{N+2}{N+4} & \frac{1}{2}+\frac{2(N+8)}{N(N+4)} \\
		P_{41} & \fund & \sing & \sing & \fund & \sing & -\frac{1}{N+4} & \frac{N+6}{N+4} & \frac{1}{2}-\frac{2(3N+8)}{N(N+4)} \\
		X_2^i & \sing & \ov\symm & \sing & \sing & \fund & 0 & 1 & 1+\frac{8}{N} \\
		X_4^i & \sing & \sing & \sing & \ov\symm & \fund & 0 & -1 & 1+\frac{8}{N} \\
		T_{41}^i & \ov\fund & \sing & \sing & \fund & \fund & \frac{1}{N+4} & -\frac{2}{N+4} & \frac{1}{2}-\frac{2(N+8)}{N(N+4)} \\
		T_{23}^i & \sing & \fund & \ov\fund & \sing & \fund & \frac{1}{N+4} & \frac{2}{N+4} & \frac{1}{2}-\frac{2(N+8)}{N(N+4)}
	\end{array}
\end{align}
with superpotential
\begin{equation}
W = \epsilon_{i j} X_2^i P_{23}^{} T_{23}^j+\epsilon_{ij} X_4^i P_{41}^{} T_{41}^j + \epsilon_{i j} P_{12}^{} T_{23}^i P_{34}^{} T_{41}^j \,.
\end{equation}
In this case, there is an additional discrete $\bZ_{\mathrm{gcf}(8,N)}$ symmetry for even $N$.

\subsubsection*{Relationship between the different orientifolds}

One can show that the anomalies involving the continuous global symmetries match between all three theories considered above, where the nonvanishing anomalies (excluding discrete anomalies) are shown in table~\ref{tab:Y40anomalies}.
\begin{table}
\begin{equation*}
  \begin{array}{|c|c|} \hline
    \SU(2)^2 \, \U(1)_B & 2 N \\ \hline
    \U(1)_X^2 \, \U(1)_B & -4 N \\ \hline
    \SU(2)^2 \, \U(1)_R & -N^2-24 \\\hline
    \U(1)^2_B \, \U(1)_R & -4 \\\hline
    \U(1)^2_X \, \U(1)_R & -2(N^2+32) \\\hline
    \U(1)_R^3 & 3 N^2 - 164 \\\hline
    \U(1)_R & -20 \\\hline
  \end{array}
\end{equation*}
\caption[The anomalies of the three different $Y^{4,0}$ theories]{The nonvanishing anomalies of the three different $Y^{4,0}$ theories. We omit discrete anomalies for simplicity, as there are no discrete symmetries for odd $N$, whereas we argue that no dualities are possible for even $N$.\label{tab:Y40anomalies}}
\end{table}

For odd $N$ there are no discrete symmetries and therefore all three theories have matching global symmetry groups and anomalies. For even $N$ not divisible by eight, the global symmetry groups again match between all three theories. However, by removing $k$ D3 branes we can reduce $N \to N-2k$. Thus, consistency along the Coulomb branch rules out a duality between phase I and phase II for even $N$, since $\bZ_{\mathrm{gcf}(8,N)}\ne\bZ_{\mathrm{gcf}(4,N)}$ for $N=8p$.

This leaves open the possibility of a duality between the two orientifolds of phase I for even $N$. However, one can show that the operator spectra do not match in this case. Specifically, we can compare the baryons of minimum R-charge in both theories:
\begin{equation}
\mbox{
	\begin{tabular}{c|cc}
		 & I.a & I.b \\ \hline
		\rule{0pt}{3.5ex} Baryon & $A^{\frac{N+4}{2}}$ & $T_1^{\frac{N+2}{2}}$ or $T_3^{\frac{N+2}{2}}$ \\
		$Q_B$ & $1/2$ & $-1/2$ \\
		$Q_X$ & $0$ & $\frac{N+4}{2}$ or $-\frac{N+4}{2}$ \\
		$Q_R$ & $\frac{N}{4}-2$ & $\frac{N}{4}-3$
	\end{tabular}}
\end{equation}
Clearly the operators do not match each other, which is inconsistent with a duality between these two theories for even $N$. Moreover, in the phase II theory only integral $Q_B$ is possible, so these operators have no dual there either, consistent with the mismatch in discrete symmetries explained above. Thus, we conclude that there are no dualities between the different theories for even $N$.

For odd $N$, these issues do not arise, as the above operators are no longer well-defined, and only integral $Q_B$ is possible in all three theories. As an additional check that a duality can occur for this case, we again consider the baryons of minimal R-charge. For theory I.a, we find the baryons $P_{24}^{N-4} P_{12}^4 A^2$ and $P_{34}^{N-4} P_{13}^4 A^2$, which transform as $\left(\symm, -1, \pm(N-4), \frac{N-4}{2}\right)$ under $\SU(2)\times\U(1)_B\times\U(1)_X\times\U(1)_R$. Consistent with the proposed duality, we find that the theory I.b baryons $T_1^N (T_3 P_{34}^2 P_{41}^2)^2$ and $T_3^N (T_1 P_{12}^2 P_{23}^2)^2$ have the same charges under the global symmetries, as do the theory II baryons $P_{41}^N (P_{12}^2 P_{23} T_{23})^2$ and $P_{23}^N (P_{34}^2 P_{41} T_{41})^2$, where the F-term conditions play a nontrivial role in the latter two cases.

The duality between the two orientifolds of phase I is an intriguing new feature of this geometry which does not appear in the simpler examples we considered previously. We leave further discussion of it to a future work.

\section{Conclusions}
\label{sec4:conclusions}

In this paper, we showed that the $\mathcal{N}=1$ gauge theories
arising on D3 branes probing orientifolded Calabi-Yau singularities
exhibit a rich class of gauge theory dualities not previously explored
in the literature. We focused on a particular example of these
dualities, corresponding to the well-known $\bC_3/\bZ_3$ singularity,
providing extensive checks for the proposed duality, including anomaly
matching, matching of discrete symmetries, moduli space matching, and
matching of the superconformal indices. In some instances the matching
of various quantities between the two theories follows from, or would
imply, some remarkable mathematical identities, see for example
appendices~\ref{app:Specht} and \ref{sec:SCI-identity}. Together, the
success of these checks presents a compelling argument for the
existence of a duality.

In~\cite{transitions3}, we argue that this duality originates from the
$\SL(2,\bZ)$ self-duality of type IIB string theory, and is therefore
a close cousin of the more familiar Montonen-Olive duality of
$\mathcal{N}=4$ theories. In~\S\ref{subsec:limitations} we show that
$\SL(2,\bZ)$ then acts in the usual way on a particular combination of
holomorphic couplings which is constant along the RG flow and which
corresponds to the axio-dilaton of type IIB string theory. We conclude that the dual descriptions we find are different weakly
coupled limits of a single theory --- the theory of branes at the
orientifolded singularity --- valid for complementary ranges of axio-dilaton vevs. These features make it clear that this $\cN=1$
duality is of a different type than the more usually considered
Seiberg (infrared) duality. Rather, it is more closely analogous to
Montonen-Olive duality, differing only by the reduced supersymmetry
and consequently richer dynamics.

As the axio-dilaton corresponds to a holomorphic combination of
couplings which is RGE invariant, in general these theories will flow
to a complex fixed line parameterized by the axio-dilaton. (We have
demonstrated that this occurs in a specific example where part of the
fixed line is perturbatively accessible.) The $\SL(2,\bZ)$ duality
group therefore acts nontrivially on the fixed line, much as in the
$\mathcal{N}=1^*$ theories already understood in the
literature~\cite{Leigh:1995ep,Argyres:1999xu}, which are mass
deformations of $\mathcal{N}=2$ or $\mathcal{N}=4$ theories and
inherit their $\SL(2,\bZ)$ duality directly from that of the parent
theory. In certain special cases, however, the flows corresponding to
different values of the string coupling converge to a single fixed
point. In these cases, one of which we discuss in the text, the
$\SL(2,\bZ)$ duality gives rise to an infrared duality relating the
two dual theories, both taken at weak string coupling as in ordinary
Seiberg duality.

The orientifolded $\bC_3/\bZ_3$ singularity is but one example among many geometries that exhibit these dualities. We expect that D3 branes probing any orientifolded Calabi-Yau singularity will exhibit an $\SL(2,\bZ)$ duality so long as the O7 planes are compact, though in some cases it is only a self-duality. For example, the $dP_1$ singularity is a closely related geometry giving rise to a dual pair of gauge theories related to the $\bC^3/\bZ_3$ theories by Higgsing (corresponding to blowing down a two-cycle in the $dP_1$ base). These theories exhibit interesting dynamics, such as a quantum-deformed moduli space for the lowest rank example which we were able to completely match between the dual theories. It would be interesting to understand the dynamics of these theories for larger $N$. In~\cite{transitions2, transitions3} we provide infinite classes of geometries which generalize $\bC_3/\bZ_3$ and $dP_1$, all of which exhibit similar dualities.

In addition to $\SL(2,\bZ)$ dualities, more complicated geometries (such as the $\bF_0$ singularity)
also exhibit other interesting dualities. The simplest of these appear
to be closely related to Seiberg duality, at least from the
perspective of string theory, as we argue in~\cite{transitions3}. However,
from the field theory perspective, they are new (presumably infrared)
dualities not readily derivable from the Seiberg duals known in the
literature. We also find indications of further dualities whose string theoretic origin
is unclear, such as the duality relating the two orientifolds of phase
I of $Y^{4,0}$. It would be interesting to better understand the
nature and origin of these dualities.

We anticipate that further study of these dualities will lead to new
insights concerning both string theory and gauge theories. In
particular, on the gauge theory side, our work helps to substantially
expand the universe of known dualities to cases where product gauge
groups play a pivotal role, and radically broadens the contexts in
which $\SL(2,\bZ)$ dualities are seen to arise. The infinite variety
of Calabi-Yau singularities provides plenty of room for further study,
which could reveal further types of duality or further illuminate the
dualities we have considered here.

Finally, given a clear understanding of when dualities are expected to occur in string theory, it might be possible to construct examples of $\mathcal{N}=0$ dualities. Indeed, this has recently been attempted for the case where supersymmetry is broken by antibranes~\cite{Sugimoto:2012rt}.\footnote{Other interesting work
  somewhat related in spirit, although focusing on non-supersymmetric
  analogues of Seiberg duality, can be found in \cite{Armoni:2008gg},
  based on ideas reviewed in \cite{Armoni:2004uu}.} Our work suggests a related program of dualities from anti-branes at Calabi-Yau singularities, or from branes probing SUSY-breaking singularities, such as non-supersymmetric orbifolds. While string theory seems to suggest that both cases should lead to $\SL(2,\bZ)$ dualities, this seems extraordinary from the field theory perspective, making it a natural topic for further research.

\acknowledgments

We would like to thank P.~Aspinwall, M.~Berg, P.~Berglund, M.~Buican,
C.~Csaki, A.~Dabholkar, N.~Halmagyi, S.~Kachru, F.~Marchesano, L.~McAllister,
N.~Mekareeya, B.~Pioline, E.~Silverstein, G.~Torroba and A.~Uranga for
illuminating discussions and C.~Csaki, L.~McAllister, G.~Torroba and A.~Uranga for comments on the
manuscript. We particularly thank L.~McAllister and G.~Torroba for
initial collaboration and extensive discussions.  The work of B.H. and
T.W. was supported by the Alfred P. Sloan Foundation and by the NSF
under grant PHY-0757868. The work of B.H. was supported in part by a
John and David Boochever Prize Fellowship in Fundamental Theoretical
Physics. The work of T.W. was supported by a Research Fellowship
(Grant number WR 166/1-1) of the German Research Foundation
(DFG). B.H. gratefully acknowledges support for this work by the
Swedish Foundation for International Cooperation in Research and
Higher Education and by a Lucent Travel Award. We would like to thank
the organizers of String Phenomenology 2011, Strings 2011, and String
Phenomenology 2012 for providing stimulating environments in which
part of this work was carried out. B.H. would also like to thank the
organizers of the Nordita string phenomenology workshop, the 2012
Carg\`{e}se summer school on Gauge Theory and String Theory, and the
Simons Center Graduate Summer School on String Phenomenology for
likewise providing a stimulating and productive
environment. I.G.-E. thanks N.~Hasegawa for kind encouragement and
constant support.

\appendix

\addtocontents{toc}{\protect\setcounter{tocdepth}{1}}

\section{Quiverfolds}
\label{app:quiverfolds}

As the main focus of our paper is dualities relating gauge theories arising on the worldvolumes of D3 branes probing orientifolded Calabi-Yau singularities, it is useful to establish some general facts about these gauge theories.

While D-brane gauge theories are quiver gauge theories, the
introduction of O-planes leads to a slightly more general class of
theories which we refer to as ``quiverfold'' gauge
theories. Quiverfold gauge theories admit more general gauge groups
and matter content than quiver gauge theories. While quiver gauge
theories allow only $\SU$ gauge group factors as well as adjoint and
bifundamental $(\fund,\ov\fund)$ or $(\ov\fund,\fund)$ matter,
quiverfold gauge theories also allow $\SO$ and $\Sp$ groups, as well
as two-index tensor matter and bifundamental matter in the
$(\fund,\fund)$ or $(\ov\fund,\ov\fund)$ representations.

Such gauge theories cannot be described by standard quiver diagrams
(directed graphs), and we develop a more general diagrammatic notation
in~\S\ref{subsec:genquiverfolds}, which we call a ``quiverfold
diagram''. One can show that any connected quiverfold diagram which is
not a strict quiver can be thought of (in a precise way which we later
make clear) as the result of ``folding'' a quiver in half along a line
of $\bZ_2$ symmetry, which is the inspiration for the term
``quiverfold''.

Before discussing quiverfolds in~\S\ref{subsec:genquiverfolds}, we first motivate their introduction by ``deriving'' a set of rules for obtaining orientifold gauge theories from their parent (orientifold-free) quiver gauge theory in~\S\ref{subsec:orientifoldquiver} and applying these rules to a few simple examples in~\S\ref{subsec:quiverfoldexamples}. As shown in~\cite{transitions3}, these rules are equivalent to well-established results in the literature on orientifolding toric Calabi-Yau singularities using brane tilings~\cite{Franco:2007ii}. While the brane tiling method has some computational advantages relating to the superpotential, our approach (following~\cite{Wijnholt:2007vn}) is somewhat more intuitive, and we focus on it here for that reason, deferring further discussion of brane tiling methods to~\cite{transitions3}.

\subsection{Orientifolding a quiver gauge theory} \label{subsec:orientifoldquiver}

We consider a quiver gauge theory describing a collection of D-branes probing some background. Each node in the quiver corresponds to a stack of identical D-branes, with an associated $\U(N)$ gauge group. Arrows in the quiver, bifundamental matter in the quiver gauge theory, correspond to open strings stretched between the stacks of branes at their intersections.

To this picture, we now add orientifold planes (O-planes). The associated involution, $\sigma$, must map the background and the collection of branes onto itself (up to certain signs and orientations), and squares to the identity. Thus, the involution defines an order-two permutation on the nodes of the quiver. Moreover, the involution maps open strings to \emph{oppositely oriented} open strings. Thus, the involution also defines an order-two permutation on the arrows of the quiver, such that for any arrow $X: A\to B$ connecting node $A$ to node $B$, the arrow's orientifold image $X': B'\to A'$ connects $B'$ to $A'$, where $A'$ and $B'$ are the orientifold images of the nodes $A$ and $B$, respectively.

\begin{figure}
  \begin{center}
    \includegraphics[width=0.3\textwidth]{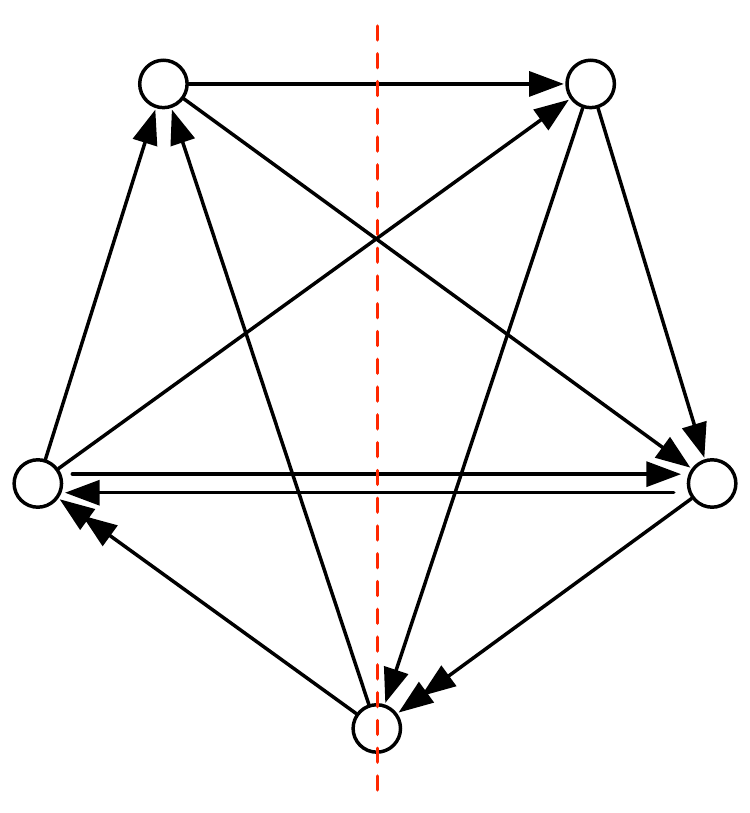}
  \end{center}
  \caption[An example of an involution of a quiver]{An example of an involution of a quiver. The quiver theory pictured here describes the toric $PdP_2$ singularity~\cite{Feng:2004uq}.}
  \label{fig:PdP2involution}
\end{figure}

The observations of the last paragraph may be summarized as follows:
\begin{quote}
{\bf Rule I:} The O-plane involution defines a $\mathbb{Z}_2$ automorphism of the quiver which reverses the directions of arrows.
\end{quote}
An example of the resulting involution is shown in figure~\ref{fig:PdP2involution}. Note that not every quiver has an involution, in the sense defined above. A necessary condition is that the quiver be isomorphic to its charge conjugate (the same quiver with the arrows reversed). This corresponds to the fact that not all brane configurations can be orientifolded, since the branes must then come in image pairs under the involution.

D-brane gauge theories generically come with a classical (tree-level) superpotential,\footnote{We restrict our attention to supersymmetric brane configurations and orientifolds.} which is determined by the geometry and brane configuration. Since these objects are appropriately covariant under the involution, we conclude that the superpotential must also be appropriately covariant: that is, $W\to W'$, where $W'$ is equivalent to $W$ up to some symmetry transformation. In the examples which follow, we shall see that the appropriate restriction is in fact:
\begin{quote}
{\bf Rule II:} The superpotential of the parent theory is invariant under the involution.
\end{quote}
Notice that if we impose the same requirement on the (generally unknown) K\"ahler potential, this implies that the corresponding gauge theory has a \emph{color-conjugation} symmetry.\footnote{Since in general there are multiple gauge groups, the theory can still be chiral (cf.~\cite{Csaki:1997aw}).} The orientifold theory results from identifying the chiral and vector superfields related by the involution. This can be restated as:
\begin{quote}
{\bf Rule III:} The orientifold gauge theory is derived from the parent theory by gauging the involution.
\end{quote}
Note that the above rules should only be interpreted at the classical level. For instance, the gauge group ranks compatible with anomaly cancellation are generally different in the parent and orientifold theories, corresponding to the tadpoles (RR charge) carried by the O-planes.

We have presented a heuristic argument (following~\cite{Wijnholt:2007vn}) for a set of rules relating the worldvolume theories of stacks of D-branes to the worldvolume gauge theories of their orientifolds. To the extent that these arguments hold, the above rules should be viewed as \emph{necessary} (but potentially insufficient) conditions which must be satisfied by consistent orientifold involutions. We now illustrate these arguments with a pair of examples.

\subsection{Examples} \label{subsec:quiverfoldexamples}

\subsubsection{\alt{$\cN=4$}{N=4} orientifolds}

We consider the worldvolume gauge theory of $N$ parallel D3 branes in flat space, which is $\cN = 4$ $\SU(N)$ super-Yang-Mills. This theory has an $\cN=1$ description with three adjoint chiral superfields $\Phi^i$, $i \in \{1, 2, 3\}$, and the superpotential:
\begin{align}
W = \frac{1}{3}\, \epsilon_{i j k} \Tr \Phi^i \Phi^j \Phi^k \,,
\end{align}
up to a superpotential coupling which can be removed by rescaling the fields. However, in this language only an $\SU(3) \times \U(1)_R$ subgroup of the $\SU(4)_R$ symmetry is manifest, where $\Phi^i$ transforms as $\fund_{+2/3}$.

We consider orientifolds of this theory, imposing the rules from the previous section. We first consider the action of the involution on the gauge bosons. It is well known that only (products of) $\U(N)$, $\SO(N)$ and $\Sp(N)$ gauge groups are possible in perturbative string theory. In particular, the involution must act on the gauge bosons as follows:
\begin{align}
A \to \pm M A^T M^{\dag}\,,
\end{align}
where $M$ must be unitary to leave the gauge kinetic term invariant. Since the involution squares to the identity, we find $M M^{*} = \pm1$ so that $M^T = \pm M$. In the case where $M$ is symmetric, it can be diagonalized by a gauge transformation, giving $M=\id$. The remaining unbroken gauge symmetry is $\SO(N)$, and we choose
\begin{align}
A \to - A^T
\end{align}
to ensure that the invariant gauge bosons correspond to the generators of $\SO(N)$. Conversely, if $M$ is antisymmetric, it can be put into the form
\begin{align}\label{eqn:symplectic}
\renewcommand{\arraystretch}{1}
M = \symp = \begin{pmatrix} 0 & 1 & 0 & 0 & \cdots \\ -1 & 0 & 0 & 0 & \\ 0 & 0 & 0 & 1 & \\ 0 & 0 & -1 & 0 & \\ \vdots & & & & \;\ddots \end{pmatrix}\,,
\end{align}
and the remaining unbroken gauge symmetry is $\Sp(N)$. We then choose
the involution
\begin{align}
A \to \symp A^T \symp
\end{align}
once again to ensure that the invariant gauge bosons correspond to the generators of $\Sp(N)$.

We now consider the action of the involution on the (adjoint) Weyl fermions $\psi^i$, $i \in 1\ldots 4$:
\begin{align}
\psi^i \to \Lambda^i_j\, \cI (\psi^j)^T \cI^{*}\,,
\end{align}
where $\cI$ acts on the gauge indices. Invariance under the remaining $\SO(N)$ or $\Sp(N)$ gauge symmetry requires that $\cI = \id$ or $\cI = \symp$, respectively, up to an overall factor which can be absorbed into $\Lambda^i_j$. Invariance of the kinetic term requires $\Lambda^i_j$ to be unitary, which can be diagonalized after an $\SU(4)_R$ transformation, taking the form $\Lambda^i_j = \diag(\pm_1 1, \pm_2 1, \pm_3 1, \pm_4 1)$. For each positive (negative) eigenvalue of $\Lambda^i_j$, the corresponding Weyl fermion projects down to its invariant symmetric (antisymmetric) component. To preserve at least $\cN = 1$ supersymmetry, at least one sign must be $-1$ ($+1$) to form a vector multiplet with the $\SO(N)$ ($\Sp(N)$) gauge bosons, which we take to be $(\pm_4)$ WLOG. In $\cN =1$ language, the remaining signs specify the action of the involution on the adjoint chiral superfields:
\begin{align}\label{eqn:C3PhiInvol}
\Phi^i \to \hat{\Lambda}^i_j\, \cI (\Phi^j)^T \cI^{*}\,,
\end{align}
where $\hat{\Lambda}^i_j = \diag(\pm_1,\pm_2,\pm_3)$. The superpotential transforms as:
\bea
W \to W' & = & \frac{1}{3} \epsilon_{i j k} \hat{\Lambda}^i_{i'} \hat{\Lambda}^j_{j'} \hat{\Lambda}^k_{k'} \Tr\left[ \cI (\Phi^{i'})^T \cI^{*} \cI (\Phi^{j'})^T \cI^{*} \cI (\Phi^{k'})^T \cI^{*}\right] \nonumber \\
& = & \frac{1}{3} \det(\hat{\Lambda}) (\pm_{\rm Sp})^3\, \epsilon_{i j k} \Tr \Phi^k \Phi^j \Phi^i \nonumber \\
& = & -(\pm_{\rm Sp}) \det(\hat{\Lambda})\, W \,,
\eea
where $(\pm_{\rm Sp})$ is $+1$ ($-1$) for an $\SO$ ($\Sp$) projection, so that $\cI \cI^{*} = \pm_{\rm Sp} \id$. Thus, invariance of the superpotential requires that
\begin{align}\label{eqn:N4signrule}
(\pm_{\rm Sp})(\pm_1)(\pm_2)(\pm_3) = -1\,.
\end{align}
This is our first example of a ``sign rule''~\cite{Franco:2007ii}: a restriction on the form of the involution, and thus the spectrum of the orientifold theory, due to the requirement that $W$ is invariant.

In $\cN = 4$ language, the above sign rule amounts to the requirement $\det \Lambda = 1$, since $\pm_4 = -(\pm_{\rm Sp})$. For an $\SO$ projection, the possibilities are $\Lambda = \diag(-,-,-,-)$ and $\Lambda = \diag(+,+,-,-)$, corresponding to the spectrum of an $\cN=4$ $\SO(N)$ gauge theory and an $\cN=2$ $\SO(N)$ gauge theory with a hypermultiplet in the symmetric representation, respectively. Similarly, for the $\Sp$ projection, the possibilities are $\Lambda = \diag(+,+,+,+)$ and $\Lambda = \diag(-,-,+,+)$, corresponding to the spectrum of an $\cN=4$ $\Sp(N)$ gauge theory and an $\cN=2$ $\Sp(N)$ gauge theory with a hypermultiplet in the antisymmetric representation, respectively.

By comparison, D3 branes are mutually supersymmetric with coincident O3 and O7 planes: $N$ D3 branes atop an O3$^-$ (O3$^+$) gives rise to an $\cN=4$ $\SO(N)$ ($\Sp(N)$) worldvolume gauge theory, whereas $N$ D3 branes atop an O7$^-$ (O7$^+$) gives rise to an $\cN=2$ $\Sp(N)$ ($\SO(N)$) worldvolume gauge theory, in agreement with the sign rule~(\ref{eqn:N4signrule}). This agreement relies on our choice of rule II as the correct restriction on the transformation of $W$ under the involution. Had we imposed $W \to -W'$ for instance, we would have obtained spectra with only $\cN = 1$ supersymmetry, which are not realized in string theory as the worldvolume gauge theory of a stack of D3 branes coincident with an O-plane in a flat background.

In fact, the geometric involution of the O$p$ brane can be computed directly from the action of the involution on the open string fields,~(\ref{eqn:C3PhiInvol}). We form gauge invariant single trace mesons:
\begin{align}
Z^{i_1 i_2 \ldots i_n} \equiv \Tr \Phi^{i_1} \Phi^{i_2} \ldots \Phi^{i_n} \,.
\end{align}
Upon imposing the F-term conditions, we obtain $[\Phi^i,\Phi^j]=0$, so that $Z^{i_1 i_2 \ldots i_n}$ is totally symmetric in its indices. Acting with the involution~(\ref{eqn:C3PhiInvol}), we obtain:
\begin{align}
Z^{i_1 i_2 \ldots i_n} \to \left[(\pm_{\rm Sp}) \hat{\Lambda}^{i_1}_{\, i_1'}\right] \left[(\pm_{\rm Sp}) \hat{\Lambda}^{i_2}_{\, i_2'}\right] \ldots \left[(\pm_{\rm Sp}) \hat{\Lambda}^{i_n}_{\, i_n'}\right] Z^{i_1' i_2' \ldots i_n'} \,,
\end{align}
modulo F-terms, where the extra signs $\pm_{\rm Sp}$ come from factors of $\symp^2=-1$ which appear in the trace for symplectic projections. Geometrically, $Z^i$ corresponds to the coordinates $z^i$ of the $\mathbb{C}^3$ in which the D3 branes are embedded. Thus, the geometric involution is simply:
\begin{align}
z^i \to (\pm_{\rm Sp}) \hat{\Lambda}^i_j z^j \,.
\end{align}
It is straightforward to check that this reproduces the O3 and O7 involutions for the $\cN=4$ and $\cN=2$ cases considered above. For example, choosing $\hat{\Lambda} = (-,+,+)$ with an $\SO$ projection, we obtain $z^1 \to - z^1$, $z^2 \to z^2$, $z^3 \to z^3$, corresponding to an O7 plane at $z^1=0$, whereas choosing $\hat{\Lambda} = (+,+,+)$ with an $\Sp$ projection, we obtain $z^i \to - z^i$, corresponding to an O3 plane at the origin.

To obtain the superpotential of the orientifold theory, we replace the fields with their projections:
\begin{align}
\Phi^i \to \phi^i \cI^{*}\,,
\end{align}
where invariance under the involution requires that
\begin{align}
\phi^i =\hat{\Lambda}^i_j\, (\phi^j)^T \,,
\end{align}
so that for $\hat{\Lambda} = \diag(\pm_1,\pm_2,\pm_3)$, $\phi^i$ is symmetric (antisymmetric) when $\pm_i$ is positive (negative), as previously noted. Applying this replacement to the superpotential, we obtain
\begin{align}
W = \frac{1}{6} \epsilon_{i j k} \Tr \phi^i \phi^j \phi^k\,,
\end{align}
where for $\Sp$ projections the trace implicitly includes factors of $\symp$ between each pair of fields, and we include an extra factor of $1/2$ by convention, the overall normalization being arbitrary up to field redefinitions. Written out explicitly, we obtain:
\begin{align}
W = \frac{1}{2} \Tr \phi^1 \phi^2 \phi^3-\frac{1}{2} \Tr \phi^3 \phi^2 \phi^1\,,
\end{align}
while $\Tr M = \Tr M^T$ implies that $\Tr \phi^3 \phi^2 \phi^1 = (\pm_{\rm Sp})(\pm_1)(\pm_2)(\pm_3) \Tr \phi^1 \phi^2 \phi^3$, where the first sign $(\pm_{\rm Sp})^3 =(\pm_{\rm Sp})$ comes from $\symp^T = -\symp$. Thus, imposing~(\ref{eqn:N4signrule}), the superpotential reduces to
\begin{align}\label{eqn:N4orientW}
W = \Tr \phi^1 \phi^2 \phi^3\,,
\end{align}
whereas imposing $W' = -W$ and following the same procedure, we would obtain a vanishing superpotential. Moreover, the superpotential~(\ref{eqn:N4orientW}) is exactly that required by the extended supersymmetry of the corresponding brane configurations.

\subsubsection{Orientifolds of \alt{$\bC^3/\bZ_3$}{C3/Z3}}
\label{subsec:C3Z3quiverfolds}

\begin{figure}
  \begin{center}
    \includegraphics[width=0.3\textwidth]{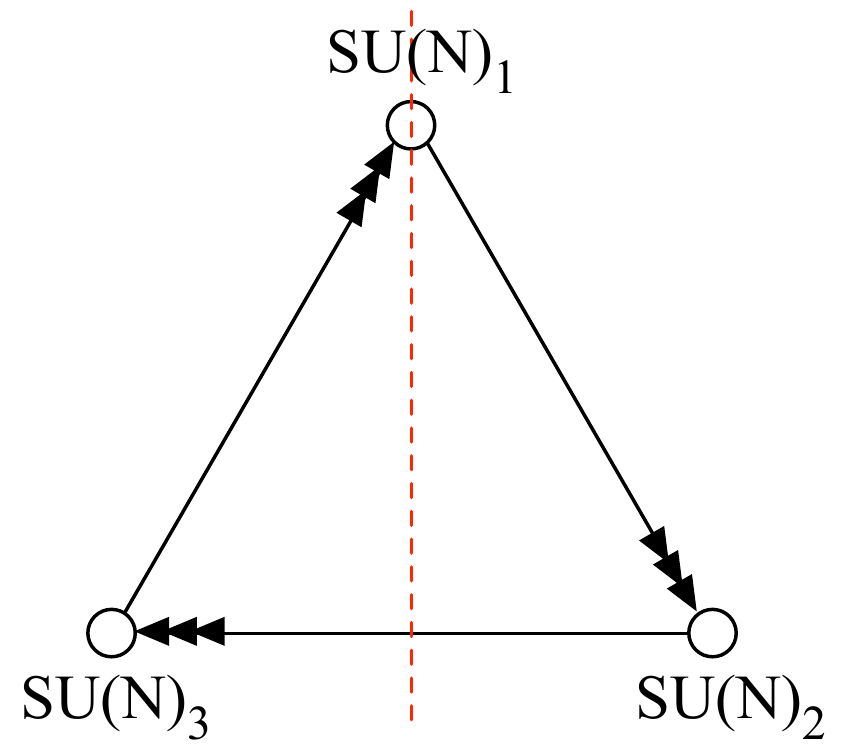}
  \end{center}
  \caption[An involution for the $\bC^3/\bZ_3$ quiver]{The quiver for $\bC^3/\bZ_3$, with the involution of interest indicated by the dashed line.}
  \label{fig:dP0quiver2}
\end{figure}

Next, we consider $N$ D3 branes probing the orbifold singularity
$\bC^3/\bZ_3$, with the orbifold action $z^i \to e^{2 \pi i/3}
z^i$. The resulting $\cN = 1$ quiver gauge theory, shown in
figure~\ref{fig:dP0quiver} (which we reproduce in
figure~\ref{fig:dP0quiver2} for convenience), is well known. The
corresponding superpotential is:
\begin{align}
W = \epsilon_{i j k} \Tr X_{12}^i X_{23}^j X_{31}^k\,,
\end{align}
up to a superpotential coupling which can be removed by rescaling the fields. An $\SU(3)\times \U(1)_R$ symmetry is manifest
under which the $X^i_{A B}$ transform as $\fund_{+2/3}$.

Applying the rules of~\S\ref{subsec:orientifoldquiver}, we search for orientifolds of this configuration. Inspecting the quiver, one can easily check that rule I implies that the involution must fix one node and exchange the other two. As the quiver has a $\bZ_3$ symmetry, we take the fixed node to be node~$1$ WLOG. The action of the involution on the chiral superfields is then:
\begin{align}\label{eqn:C3Z3invol}
X_{12}^i \to \Lambda^i_{\, j}\, \cI_1 (X_{31}^j)^T \delta_{32}^{*} \,,\quad
X_{23}^i \to \Sigma^i_{\, j}\, \delta_{23} (X_{23}^j)^T \delta_{23}^{*} \,, \quad
X_{31}^i \to (\Lambda^i_{\, j})^{\dag}\, \delta_{32} (X_{12}^j)^T \cI_1^{*} \,,
\end{align}
where $\Lambda$ and $\Sigma$ are unitary matrices, $\delta_{23} = \delta_{32}^T = \id$ breaks $\SU(N)_2 \times \SU(N)_3 \to \SU(N)$, and $\cI_1 = \id$ or $\symp$, depending on whether we choose an $\SO$ or $\Sp$ projection for the fixed node, respectively. Moreover, since the involution squares to the identity, $\Sigma^2 = \id$, so that $\Sigma$ is both unitary and Hermitian.

We compute the orientifold image of the superpotential:
\bea
W \to W' & = & \epsilon_{i j k}\, \Lambda^i_{\, i'} \Sigma^j_{\, j'} (\Lambda^k_{\, k'})^{\dag}\, \Tr \cI_1 (X_{31}^{i'})^T \delta_{32}^{*}\, \delta_{23} (X_{23}^{j'})^T \delta_{23}^{*} \, \delta_{32} (X_{12}^{k'})^T \cI_1^{*} \nonumber \\
& = & (\pm_{\rm Sp})\, \epsilon_{i j k}\, \Lambda^i_{\, i'} \Sigma^j_{\, j'} (\Lambda^k_{\, k'})^{\dag}\, \Tr X_{12}^{k'} X_{23}^{j'} X_{31}^{i'}\,.
\eea
Therefore, invariance of the superpotential requires:
\begin{align}
\epsilon_{i j k}\, \Lambda^i_{\, i'} \Sigma^j_{\, j'} (\Lambda^k_{\, k'})^{\dag} = -(\pm_{\rm Sp})\, \epsilon_{i' j' k'} \,.
\end{align}
In fact, this is only possible if $\Lambda = e^{i \theta} \Sigma$,\footnote{In general whenever $0 \ne \epsilon_{i j k} A^i_{i'} B^j_{j'} C^k_{k'} \propto \epsilon_{i' j' k'}$, then $A \propto B, C$.} where the phase factor can be removed by rotating $X_{12}^i \to e^{i \theta/2} X_{12}^i$ and $X_{31}^i \to e^{-i \theta/2} X_{31}^i$ (leaving the superpotential invariant). Thus, we take $\Lambda = \Sigma$, where the invariance of the superpotential requires
\begin{align}\label{eqn:C3Z3signrule}
\det \Sigma = -(\pm_{\rm Sp}) \,.
\end{align}
After an $\SU(3)$ transformation, we obtain $\Sigma = \diag(\pm_1,\pm_2,\pm_3)$, and the requirement that the superpotential be invariant takes the form of a sign rule:
\begin{align}\label{eqn:C3Z3signrule2}
(\pm_1) (\pm_2) (\pm_3) (\pm_{\rm Sp}) = -1 \,.
\end{align}
Thus, for an $\SO$ projection, there are two possible involutions $\Sigma = \diag(-,-,-)$ and $\Sigma = \diag(-,+,+)$ up to an $\SU(3)$ transformation. The spectrum of the latter theory turns out to be anomalous for any choice of gauge group ranks,\footnote{The anomaly can be cancelled by introducing noncompact ``flavor'' D7 branes into the geometry~\cite{Franco:2010jv}.} so we will focus on the first possibility. Similarly, for an $\Sp$ projection, $\Sigma = \diag(+,+,+)$ and $\Sigma = \diag(-,-,+)$ are possible, again up to an $\SU(3)$ transformation, with the latter being anomalous for any choice of ranks.

The anomalous orientifolds correspond to noncompact O7 planes, whereas the remaining possibilities correspond to compact O7 planes~\cite{Franco:2010jv}. We verify this by computing the geometric involution. Consider mesons of the form:
\begin{align}
Z^{i j k} \equiv \Tr X_{12}^i X_{23}^j X_{31}^k \,.
\end{align}
Upon imposing the F-term conditions, we find that $Z^{i j k}$ is totally symmetric in its indices. Applying the involution~(\ref{eqn:C3Z3invol}), we obtain:
\begin{align}
Z^{i j k} \to (\pm_{\rm Sp})\, \Sigma^i_{\, i'} \Sigma^j_{\, j'} \Sigma^j_{\, j'} Z^{i' j' k'} = [(\pm_{\rm Sp}) \Sigma^i_{\, i'}]\, [(\pm_{\rm Sp}) \Sigma^j_{\, j'}]\, [(\pm_{\rm Sp}) \Sigma^k_{\, k'}]\, Z^{i' j' k'} \,,
\end{align}
where the sign $\pm_{\rm Sp}$ comes from the $\symp^2 = -1$ which appears in the trace for $\Sp$ projections. The mesons $Z^{i j k}$ correspond to the coordinates $z^i z^j z^k$ of $\mathbb{C}^3/\mathbb{Z}_3$; thus, we read off the geometric involution
\begin{align}
z^i \to (\pm_{\rm Sp})\, \Sigma^i_j z^j \,.
\end{align}
From this, it is easy to check that the anomaly-free orientifolds, $(-,-,-)$ and $(+,+,+)$ for $\SO$ and $\Sp$ respectively, correspond to compact O7 planes, with the involution $z^i \to - z^i$, whereas the anomalous orientifolds, $(-,+,+)$ and $(+,-,-)$ for $\SO$ and $\Sp$ respectively, correspond to noncompact O7 planes, with the involution $z^1 \to -z^1$, $z^2 \to z^2$, $z^3 \to z^3$.

To derive the superpotential of the orientifold theory, we replace:
\bea
X_{12}^i & \to & \Sigma^i_j\, A^j \,, \nonumber \\
X_{23}^i & \to & \Sigma^i_k\, B^j\, \delta_{23}^{*} \,, \nonumber \\
X_{31}^i & \to & \delta_{32}\, (A^j)^T \cI_1^{*} \,,
\eea
where invariance under the involution requires that:
\begin{align}
B^i = \Sigma^i_{\, j}\, (B^j)^T \,,
\end{align}
so that for $\Sigma = \diag(\pm_1,\pm_2,\pm_2)$, $B^i$ is symmetric (antisymmetric) when $\pm_i$ is positive (negative). Applying these replacements to the superpotential, we obtain:
\begin{align}
W = \frac{1}{2} (\det \Sigma)\, \epsilon_{i j k} \Sigma^k_{\,l}\, \Tr A^i B^j (A^l)^T\, ,
\end{align}
where for $\Sp$ projections the use of $\symp$ in the trace is implicit.
Since $\Tr M = \Tr M^T$, this can also be written as:
\begin{align}
W = (\pm_{\rm Sp}) \frac{1}{2} (\det \Sigma)\, \epsilon_{i j k} \Sigma^k_{\,l}\,\Sigma^j_m\, \Tr A^l B^m (A^i)^T = -(\pm_{\rm Sp})(\det \Sigma)\, W\,.
\end{align}
Thus, as before, the sign rule~(\ref{eqn:C3Z3signrule}) is necessary to ensure that the superpotential of the orientifold theory does not vanish.

For the cases $\Sigma = \diag(-,-,-)$ and $\Sigma = \diag(+,+,+)$ for $\SO$ and $\Sp$ projections, respectively, the superpotential simplifies:
\begin{align}
W = \frac{1}{2}\, \epsilon_{i j k} A^i A^j B^k \, ,
\end{align}
where we leave the contractions of gauge indices implicit. The resulting theories have the same $\SU(3)\times\U(1)_R$ flavor symmetries as the parent quiver theory, and are discussed more thoroughly in~\S\ref{sec:C3Z3fieldtheory}.

\subsection{General Quiverfolds} \label{subsec:genquiverfolds}

In the simple examples discussed above, we applied the rules of~\S\ref{subsec:orientifoldquiver} in a straightforward (if tedious) fashion to rederive known results. We now discuss some general features of this program applied to arbitrary quiver gauge theories. Specifically, we show how to derive the gauge group and spectrum of the orientifold theory graphically using the quiver diagram, and define a suitable generalization of the quiver to represent these data.

For the purposes of this discussion, we mainly ignore the superpotential, though we emphasize that rule II is generally very restrictive, and not all involutions of the quiver will leave the superpotential invariant. An explicit computation to check that $W$ is invariant under the involution can be tedious, and for toric singularities the problem is well suited to brane tiling methods, as originally formulated in~\cite{Franco:2007ii} and reviewed in~\cite{transitions3}.

Rule I implies that the quiver of the parent theory possesses a $\bZ_2$ charge conjugation (arrow reversing) symmetry representing the involution in question. We embed the quiver in $\bR^2$ such that this symmetry is manifest as a reflection through a fixed line, as in figure~\ref{fig:PdP2involution}.\footnote{While this is always possible to do, in general there are many possible embeddings. For a fixed involution, all embeddings will give the same quiverfold, as discussed below.} In the resulting figure, fixed nodes must lie along the fixed line, and fixed edges will intersect it perpendicularly, whereas any unfixed edge which crosses the fixed line must intersect another edge (its image) at the point of crossing.

To obtain the gauge group and spectrum of the orientifold theory we cut the plane in two along the fixed line, discarding one half of it and labeling each node and perpendicular (fixed) edge along the boundary with a sign. The resulting diagram on the half-plane, which we call a ``quiverfold'', specifies the gauge group and spectrum of the orientifold theory as follows: each node away from the boundary (``whole'' node) corresponds to an $\SU$ gauge group, whereas each $+$ ($-$) node along the boundary (``half'' node) corresponds to an $\SO$ ($\Sp$) gauge group. Each arrow away from the boundary (``uncrossed'' (whole) edge) corresponds to bifundamental $(\fund,\bar{\fund})$ matter in the usual way, while each arrow intersecting the boundary obliquely is joined to its image arrow to form an edge (``crossed'' (whole) edge) with opposite orientations associated to each end, and corresponding to $(\fund,\fund)$ or $(\bar{\fund},\bar{\fund})$, depending on the orientation of the arrows. Finally, each $+$ ($-$) edge ending perpendicularly on the boundary (``half'' edge) corresponds to symmetric (antisymmetric) matter.

\begin{figure}
  \begin{center}
    \includegraphics[width=0.7\textwidth]{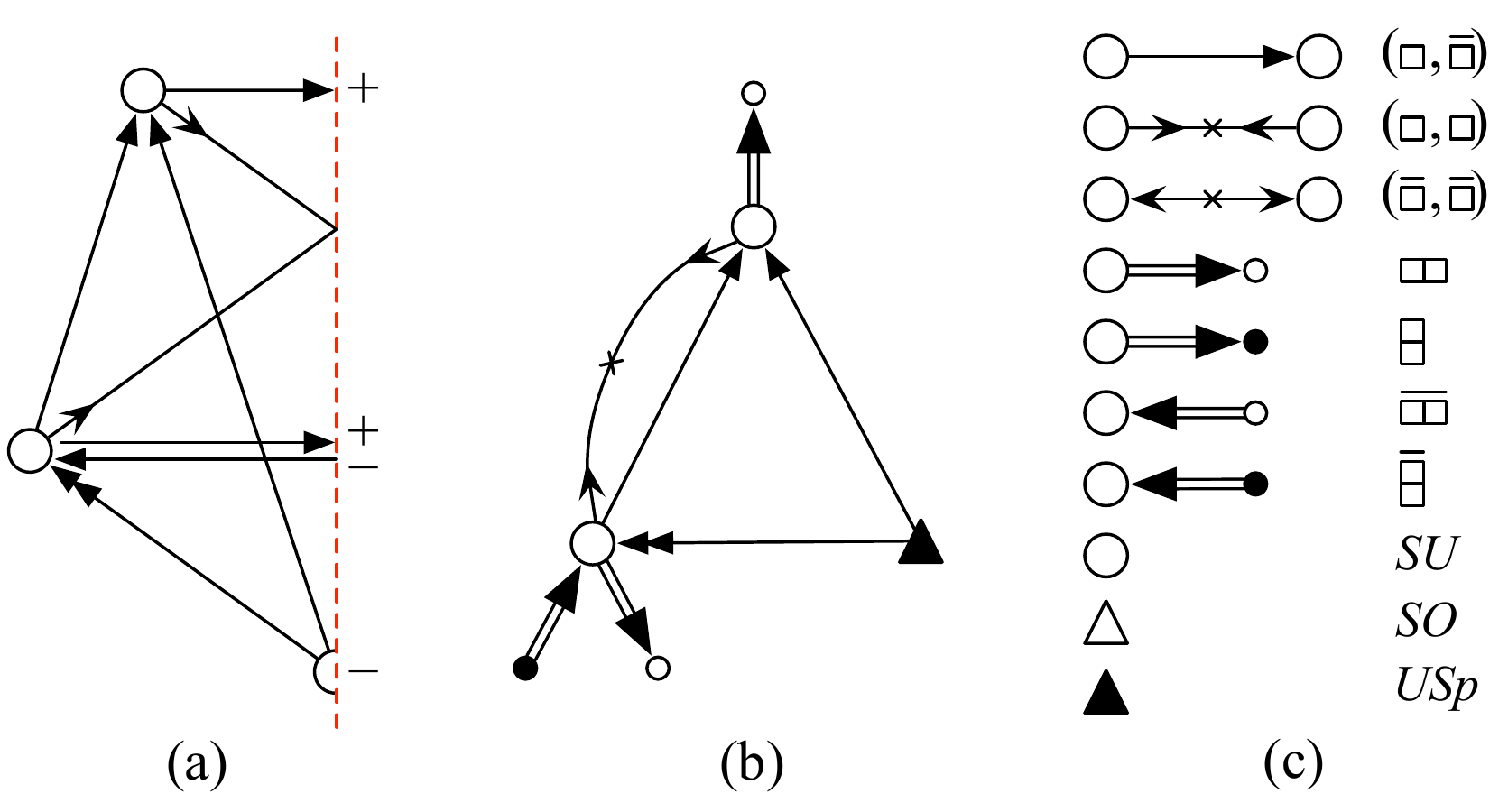}
  \end{center}
  \caption[An example of a quiverfold]{(a) An example of a quiverfold. The parent quiver is shown in figure~\ref{fig:PdP2involution}. (b) The quiverfold can be redrawn without the fixed line using appropriate symbols, defined in (c). }
  \label{fig:PdP2quiverfold}
\end{figure}

An example quiverfold is shown in figure~\ref{fig:PdP2quiverfold}(a). As shown in figure~\ref{fig:PdP2quiverfold}(b -- c), the quiverfold can be drawn without the boundary line by using appropriate symbols to denote the fixed elements and crossed edges. From this perspective, a quiverfold is just an ``enhanced'' quiver, with a few additional representations and gauge groups allowed. Just as the worldvolume gauge theory on intersecting D-branes can always be represented by a quiver gauge theory, orientifolds of these configurations can always be represented by a quiverfold (to the extent that rule I holds), which is then a very useful tool for concisely stating the gauge group and spectrum.

Note that some apparently different quiverfolds are isomorphic. In particular, any whole node of the quiverfold can be charge conjugated, yielding a new, equivalent quiverfold with different crossed and uncrossed edges; this corresponds to swapping the positions of a node and its image in the original $\bZ_2$ symmetric embedding of the quiver. In a strict quiver, there is no analogous operation: since crossed edges are not allowed, charge conjugation can only be applied to the quiver as a whole. Furthermore, not every edge of a quiverfold is directed at both ends, since arrows entering and exiting half nodes are equivalent. Thus, edges connecting a half node to a whole node have a single direction (they cannot be crossed), whereas edges connecting two half-nodes are undirected. Taking into account these isomorphisms,\footnote{There is moreover an isomorphism between a crossed edge connecting a whole node to itself (or a whole edge connecting a half-node to itself) and two half edges of opposite sign and like orientation connected to the node in question. While the involutions which give rise these configurations appear different, they are related by a nonabelian flavor symmetry of the parent theory, and the resulting spectra are the same.} it is straightforward to show that different $\bR^2$ embeddings of the same involution (with the same choice of fixed-element signs) lead to the same quiverfold. Moreover, given a quiverfold, it is possible to uniquely reconstruct the parent quiver and involution by embedding the quiverfold on the half-plane with fixed elements on the boundary, as above.

It should be emphasized that just as a quiver is a direct pictorial representation of a certain class of gauge theories (quiver gauge theories), a quiverfold is also a direct pictorial representation of a certain (somewhat broader) class of gauge theories, which we call quiverfold gauge theories. Just as gauge invariant (mesonic) operators are directed loops in the quiver diagram, gauge invariant (mesonic) operators are loops in the quiverfold,\footnote{If the loop includes a half-edge, it doubles back on itself at this point, reentering the same node it just exited.} subject to the requirement that the loop enter and exit each whole node on oppositely directed edges. However, in some cases the mesonic operator corresponding to such a loop vanishes due to symmetry (e.g.~it takes the form $\Tr M$ where $M$ is antisymmetric).

While quiverfolds are useful for computing and representing the gauge group and spectrum of a given orientifold, the set of involutions consistent with rule I is usually a superset of those involutions consistent with both rules I and II: as we saw in~\S\ref{subsec:quiverfoldexamples}, the invariance of the superpotential imposes important constraints, such as the sign rules~\eqref{eqn:N4signrule}, \eqref{eqn:C3Z3signrule2} and (in the latter case) the alignment of the flavor rotations $\Lambda^i_{\,j}$ and $\Sigma^i_{\,j}$.

It is possible to reformulate rule II graphically by describing the parent gauge theory and the involution in terms of a brane tiling, rather than a quiver diagram. We refer the interested reader to \cite{Franco:2007ii} for further details and references. As shown in~\cite{transitions3}, this approach is equivalent to the one outlined here. Regardless of the method used to apply these rules, quiverfold diagrams provide an intuitive and precise representation of the gauge group and spectrum of the orientifold gauge theory, much like quiver diagrams for D-brane gauge theories.

\section{Negative rank duality}
\label{app:negativerankdual}

In this appendix we review a fact about continuing $\SU(N)$,
$\SO(N)$ and $\Sp(N)$ groups to negative rank that turns out to be
very useful in the anomaly matching discussion in the main text. We
refer the reader to chapter~13 in \cite{Cvitanovic:2008zz} for more
details and further references. As we explain below, this continuation
relates for example an $\SU(-N)$ gauge theory to an
$\widetilde{\SU(N)}$ gauge theory and is often referred to as negative
rank duality although the two related theories are generically
\emph{not} dual in the physical sense. In particular the two related
gauge theories have generically different anomalies.

For an $\SU(N)$ gauge theory with matter in certain representation we can exchange symmetrization and antisymmetrization (i.e. reflect the Young tableau across the diagonal) and at the same time replace $N$ with $-N$. This leads to a new gauge theory we denote $\widetilde{\SU(N)}$. As was first noticed by \cite{King}, for $\SO(N)$ and $\Sp(N)$ theories we can likewise obtain a negative rank dual theory by exchanging symmetrization and antisymmetrization and replacing the $\SO(N)$ symmetric bilinear invariant $\delta_{ab}$ by the $\Sp(N)$ antisymmetric bilinear invariant $\symp_{ab}$ and replacing $N$ by $-N$: $\SO(-N) \cong \widetilde{\Sp(N)}$, $\Sp(-N) \cong \widetilde{\SO(N)}$.

In \cite{Cvitanovic:2008zz} it is proven that under these dualities any scalar quantity becomes the dual scalar quantity up to potentially an overall sign. In particular, if we have a matter field transforming in the representation $r$ which has a Young tableau with $p$ boxes and $\tilde{r}$ denotes the transposed tableau obtained by a flip across the diagonal, then the dimensions of the corresponding representations are related by\footnote{This statement only holds for representations with fixed, $N$ independent $p$. In particular we should think of the anti-fundamental representation of $\SU(N)$ as having $p=1$ and not $p=N-1$ and similarly for the adjoint representation we take $p=2$.}
\begin{align}\label{eq:dimensionnegrank}
d_N(r) = (-1)^p d_{-N}(\tilde{r})\,.
\end{align}
Thanks to the theorems of \cite{Cvitanovic:2008zz} that we mentioned above, the proof is simple since we only need to determine the overall sign $(-1)^p$: Any representation with $p$ boxes in the Young tableau has a leading $N$ scaling that is given by $N^p$ so that the overall sign under changing $N \rightarrow -N$ is $(-1)^p$, which gives the stated result.

Below we study the anomalies of negative rank dual theories of a
generic gauge theory (see \cite{Maru:1996fq} for related results). For
that we need the transformation properties of the Dynkin index $T(r)$
and anomaly coefficient $A(r)$ under the negative rank duality. These
are again determined by the leading $N$ scaling. Contrary to the
dimension the Dynkin index and anomaly coefficient of the fundamental
representation are independent of $N$. However, similarly to the
dimension any extra box in the Young tableau leads to an extra factor
of $N$ so that one finds
\begin{align}
T_N(r) = (-1)^{p-1} T_{-N}(\tilde{r})\,, \qquad A_N(r) = (-1)^{p-1} A_{-N}(\tilde{r})\,.
\end{align}
To prove this, we can again derive the leading $N$ scaling by calculating the Dynkin index and anomaly coefficient for the tensor product of $p$ fundamental representations. The Dynkin index $T(r)$ is defined by $(T^a_r)^m_n(T^b_r)^n_m = T(r) \delta^{ab}$, where the $T^a_r$ are the generators for the representation $r$. Taking the tensor product with another fundamental representation introduces a factor of $N$ in $T(r)$ and taking the tensor product of a fundamental with $(p-1)$ fundamental representation leads to the above result. Explicitly, for $\SU(N)$ we can choose one of the generators in the fundamental representation to be $T^1_{\fund} =\tfrac{1}{\sqrt{2((N-1)^2+N-1)}} \text{diag}(1,1,\ldots,1,-(N-1))$, which leads to $T(\fund)=\tfrac{1}{2}$. The leading $N$ scaling for any representation with $p$ boxes is the same as the leading $N$ scaling for the tensor product of $p$ fundamentals. Taking $p-1$-times the tensor product of the above generator with $\id_N$ we obtain a generator for the representation that is given by the tensor product of $p$ fundamentals and we find the leading $N$ term $T(r) \propto N^{p-1}$, which is true for all irreducible representation of $\SU(N)$ with $p$ boxes. Similarly one can explicitly work out the scaling for $\SO(N)$ and $\Sp(N)$. For the anomaly coefficient of $\SU(N)$, we us the following result from \cite{Banks:1976yg}: $A(r_1 \otimes r_2) = d(r_1) A(r_2) + d(r_2) A(r_1)$. Together with the fact that $A(\fund)$ is $N$ independent this leads to the leading $N$ scaling $A(r) \propto N^{p-1}$ which completes the proof.

We now show that for any gauge theory the negative rank dual is free of gauge anomalies if all chiral matter representations have dimensions that are even under the negative rank dual i.e. whenever for every chiral field the number of all boxes in the Young tableaux of all gauge theories we are dualizing is even. Furthermore, the global anomalies of the two theories are related by replacing the rank of each gauge group factor we are dualizing with its negative and adding an overall minus sign whenever the global anomaly involves a non-abelian gauge group that is also being dualized.\footnote{In the absence of an $\U(1)_R$ symmetry a negative rank dual theory is also anomaly free if all matter representation have dimensions that are odd under the negative rank dual. In that case all global symmetries pick up an extra overall minus sign.}

We take the combined gauge and global symmetry group to be $G=\U(1)_1 \times \ldots \times \U(1)_m \times G_1 \times \ldots \times G_n$, where $G_a$ are $\SU$, $\SO$ or $\Sp$ groups. We denote the chiral matters fields by $\chi$, the corresponding $\U(1)_i$ charges by $q_i^\chi$ \footnote{We assume for simplicity in the discussion below that the $q_i^\chi$ do not change sign under the negative rank duality. This condition can be relaxed so that for fixed $i$ the $q_i^\chi,\,\forall \chi$ change sign. This can lead to an extra overall minus sign in global anomalies involving $\U(1)_i$.} and the matter dimension by $d(\chi)=\prod_{a=1}^n d(r_\chi^{G_a})$ where $r_\chi^{G_a}$ denotes the representation of $\chi$ under the group $G_a$. The $\U(1)^3$ and $\U(1)$ anomalies are given by
\bea
\U(1)_i \,\U(1)_j \,\U(1)_k &=& \sum_\chi d(\chi) q_i^\chi q_j^\chi q_k^\chi\,,\\
\U(1)_i &=& \sum_\chi d(\chi) q_i^\chi\,,
\eea
where the sums are over all chiral superfields $\chi$. Whenever all chiral matter fields satisfy $d(\chi) = d(\tilde{\chi})$, then the above anomalies are unchanged after dualizing any of the $G_a$. The $G^2\, \U(1)$ and $G^3$ anomalies are
\bea
G_a^2 \,\U(1)_i &=& \sum_\chi \frac{d(\chi)}{d(r_\chi^{G_a})} T(r_\chi^{G_a})) q_i^\chi\,,\\
G_a^3 &=& \sum_\chi \frac{d(\chi)}{d(r_\chi^{G_a}) A(r_\chi(G_a))}\,.
\eea
If $G_a$ does not undergo a negative rank transition then the above anomalies are unchanged. In the case that $G_a$ undergoes a negative rank duality we use the fact that $T(r)/d(r) = -T(\tilde{r})/d(\tilde{r})$ and $A(r)/d(r) = -A(\tilde{r})/d(\tilde{r})$ to find that both of the anomalies above pick up an extra minus sign. In particular this means that all the gauge and mixed anomalies that do not involve the R-symmetry still vanish after the negative rank transition. In our examples the global non-abelian gauge groups will not undergo a negative rank transition so that none of the global anomalies pick up an extra minus sign. They are simply given by replacing the ranks of all the gauge groups that undergo the negative rank duality with their negative ranks.

Next we calculate the anomalies that involve the R-symmetry
\bea
\U(1)_R^3 &=& \sum_\chi d(\chi)(q_R^\chi-1)^3 + d(G_{gauge})\,,\\
\U(1)_i \, \U(1)_R^2 &=& \sum_\chi d(\chi) q_i (q_R^\chi-1)^2\,,\\
\U(1)_i\, \U(1)_j\, \U(1)_R &=& \sum_\chi d(\chi) q_i q_j (q_R^\chi-1)\,,\\
\U(1)_R &=& \sum_\chi d(\chi)(q_R^\chi-1) + d(G_{gauge})\,,\\
G_a^2\, \U(1)_R &=& \sum_\chi \frac{d(\chi)}{d(r_\chi^{G_a})} T(r_\chi^{G_a})(q_R^\chi-1) +T(\adj_{G_a})\,.
\eea
Above $d(G_{gauge})$ denotes the dimension of the entire gauge group (excluding the global symmetry group) and $T(\adj_{G_a})$ denotes the Dynkin index of the adjoint of $G_a$, if $G_a$ is part of the gauge group. If $G_a$ is part of the global symmetry group, then there are no gauginos that contribute and we have to set $T(\adj_{G_a})=0$. For the $\SU$, $\SO$ and $\Sp$ groups the group dimension has always even parity under the negative rank transition. Thus $d(G_{gauge})$ is even and as mentioned above $T(\adj_{G_a})$ is odd, if $G_a$ undergoes the negative rank transition since $p=2$. This means that only the last of the anomalies above picks up an overall minus sign if $G_a$ undergoes the negative rank transition. We thus conclude that all gauge and mixed anomalies vanish after the transition. In the case where none of the global non-abelian symmetry groups undergo a negative rank transition we can furthermore conclude that all anomalies of the negative rank dual theory are obtained by replacing the ranks of all gauge group factors that undergo the transition with their negative.

A simple example of two negative rank dual theories has already appeared above in~\S\ref{sec:C3Z3fieldtheory}. Both theories are related by taking the negative rank dual of both gauge group factors. The $\SO(N-4)\times \SU(N)$ extrapolated to negative $N$ is $\SO(-(N+4))\times \SU(-N)$ which dualizes to $\Sp(N+4)\times \SU(N)$. We also have to flip the Young tableaux so that the antisymmetric representation of $\SU(-N)$ becomes the symmetric representation of $\SU(N)$. The usefulness of this duality is that we did not have to calculate the anomalies in~\S\ref{subsec:classicchecks} for both theories, since they are related by changing the sign of $N$. Since the anomalies depend on $N(N-3)$ which becomes $N(N+3)$ we see that the two negative rank dual theories are \emph{not} dual in the physical sense since they have different anomalies. In this particular case the negative rank dual is however dual to the original theory after shifting the ranks of the gauge groups.

\section{Exactly dimensionless couplings} \label{app:exactlydimensionless}

Under certain assumptions, a sufficient condition for a
holomorphic coupling to be constant along the RG flow is that it be neutral
under all possible flavor symmetries, in particular those which are spurious and/or anomalous.

We focus first on the case where there are no nonabelian flavor symmetries.
Due to various nonrenormalization theorems (see e.g.~\cite{Terning:2006bq}),
holomorphic couplings are not perturbatively renormalized apart from
the one-loop running of holomorphic gauge couplings. Thus, in the
absence of nonperturbative renormalization of these couplings,
holomorphic couplings are independent of scale, provided we replace
the scale-dependent holomorphic gauge couplings $\tau(\mu)$ with the
holomorphic dynamical scale $\Lambda \equiv \mu e^{2 \pi i
  \tau(\mu)/b}$, where $b = 3 T(\adj) - T({\rm mat})$ is the one-loop
beta function coefficient (if $b=0$ then $\Lambda$ is ill-defined but
$\tau$ itself is independent of scale).

However, non-holomorphic couplings are not likewise protected against
renormalization, and in particular chiral superfields are subject to
wave-function renormalization through corrections to the K\"{a}hler
potential. Rescaling the chiral superfields to restore canonical
normalization leads to rescaling anomalies which alter the values of
the holomorphic couplings, leading to a nontrivial running for their
physical (canonically normalized) counterparts. In particular, the
rescaling may be realized as a complexification of a $\U(1)$ symmetry
under which the chiral superfield in question is charged, whereas the
corresponding $\U(1)$ may be spurious and/or anomalous, leading to a
rescaling of the corresponding spurions (superpotential couplings)
and/or the holomorphic dynamical scale(s) of the gauge
theory~\cite{Terning:2006bq}. However, if a certain holomorphic combination
of couplings is neutral under all of these $\U(1)$'s, then it is
unaffected by the rescaling, and therefore the corresponding physical
coupling is scale independent (has vanishing anomalous dimension).
Such a coupling is exactly dimensionless if and only if it is
classically dimensionless. This is readily shown to be equivalent to
the requirement that the coupling is neutral under the $\U(1)_R$ under
which all chiral superfields carry charge $+2/3$.

Thus, a holomorphic coupling corresponds to an exactly dimensionless physical (canonically normalized) coupling if it is neutral under all possible $\U(1)$ and $\U(1)_R$ symmetries\footnote{This is
  closely related to the criteria for an exactly marginal operator at
  the superconformal fixed point~\cite{Green:2010da}.} (since an arbitrary $\U(1)_R$ is a linear combination of an arbitrary $\U(1)$ with the ``canonical'' $\U(1)_R$ considered above), assuming that none of the constituent couplings are nonperturbatively renormalized. While the converse need not be true, the existence of an exactly marginal holomorphic coupling which violates these conditions imposes a nontrivial relation on the anomalous dimensions along the flow. Since anomalous dimensions typically cannot be computed exactly away from an infrared fixed point, computable examples without extended supersymmetry must satisfy these conditions.

If the gauge group is semi-simple,\footnote{If the gauge group contains a $\U(1)$ factor, then this argument still applies so long as we consider a global $\U(1)$ with a nonvanishing $\U(1)_{\rm gauge} \U(1)_{\rm global}^2$ anomaly to be a ``good'' $\U(1)$.} a straightforward counting argument gives the number $N_0$ of exactly dimensionless couplings of this type for a model with $N_G$ (simple) gauge groups, $N_W$ superpotential terms (each with a corresponding coupling), $N_{\chi}$ chiral superfields, and $N_{\U(1)}$ linearly independent ``good'' $\U(1)$ or $\U(1)_R$ symmetries (not broken by gauge anomalies or by the superpotential):
\begin{align}\label{eqn:numbermarginal}
N_0 = N_{\U(1)} + N_G + N_W - (N_{\chi}+1)\,.
\end{align}
The argument is as follows: there are $N_{\chi}+1$ linearly independent spurious and/or anomalous $\U(1)$ or $\U(1)_R$ symmetries in general, whereas the ``good'' $\U(1)$'s are those under which the $N_G + N_W$ holomorphic couplings are neutral, and can be represented by vectors of length $N_{\chi}+1$ which are annihilated by the $(N_G + N_W)\times (N_{\chi}+1)$ matrix of $\U(1)$ charges of the holomorphic couplings acting on the left. The rank of this matrix is therefore $N_{\chi}+1 - N_{\U(1)}$. By contrast, an exactly dimensionless coupling is a product of holomorphic couplings which is neutral under all the $\U(1)$'s, and can be represented by a vector of length $N_G + N_W$ which is annihilated by the same matrix acting on the right. Since row rank and column rank are equal, the number of linearly independent vectors of this type is $N_G + N_W - (N_{\chi}+1 - N_{\U(1)})$, reproducing the above formula.

These arguments must be modified to include any (potentially spurious) non-abelian flavor symmetries, since chiral multiplets with the same gauge quantum numbers are subject to kinetic mixing (unless forbidden by the global symmetries). In particular, the candidate combination of couplings must also be neutral under these non-abelian symmetries in addition to the $\U(1)$ and $\U(1)_R$ symmetries as a sufficient condition for exact marginality.

Let $G_F$ denote the semisimple component of the spurious flavor symmetries. Since only $G_F$-singlet combinations of couplings can appear in our candidate exactly dimensionless couplings and the holomorphic gauge couplings are all neutral under $G_F$, we need only consider $G_F$-invariant combinations of superpotential couplings. We can then treat $G_F$ as if it were gauged (without the corresponding gauge coupling). Thus, the above counting argument still holds, where now $N_{\chi}$ counts the number of irreducible $G_F$ multiplets, $N_W$ the number of independent $G_F$ invariant combinations of superpotential couplings, and $N_{\U(1)}$ counts the number of ``good'' $\U(1)$ or $\U(1)_R$ symmetries which commute with $G_F$.

\subsection{On nonperturbative effects}

So far we have ignored the possibility that the holomorphic couplings run due to nonperturbative effects. While it is not possible to exclude this in general, such effects are also constrained by nonrenormalization theorems, and are known to be absent in some simple cases, such as pure $\mathcal{N}=1$ super-Yang-Mills~\cite{ArkaniHamed:1997mj}.

In particular, for the gauge theories studied in this paper, we are interested in whether the string coupling $\axiodil$ (\ref{eqn:dP0axiodil}) --- which is not perturbatively renormalized by the above criteria --- can run nonperturbatively. A spurion analysis reveals that the exact (Wilsonian) beta function must take the form:
\begin{equation}
\mu \frac{d}{d \mu} \axiodil = f(\axiodil)
\end{equation}
where $f(\axiodil)$ is a holomorphic function satisfying $f(+i \infty) = 0$ due to the lack of perturbative running, and $f$ cannot depend on any other holomorphic couplings due to constraints imposed by the spurious and/or anomalous $\U(1)$ symmetries.

We first consider the $\SO$ theory for even $N$, where $\SL(2,\bZ)$ covariance requires that
\begin{equation}
f\left(\frac{a \tau + b}{c \tau + d}\right) = (c \tau + d)^{-2} f(\tau)
\end{equation}
Hence $f(\tau)$ is a modular form\footnote{In fact it is a cusp form, since $f(+i \infty) = 0$.} of weight $-2$. However, no such holomorphic modular form exists. Instead, such a modular form is necessarily \emph{meromorphic}, with poles in the upper half plane $\bH$ where the beta function blows up at finite coupling. Such poles signal a breakdown of the Wilsonian description, and are likely inconsistent. Analogous statements hold for odd $N$ (and for the $\Sp$ theory) where $\SL(2,\bZ)$ becomes $\Gamma_0(2)$ and $f$ is a level-two modular form. Hence, we conclude that $f(\tau) = 0$, and $\axiodil$ is not renormalized in either theory.

\section{Coulomb branch computation of the string coupling} \label{app:coulombbranch}

In this appendix, we provide a derivation of (\ref{eqn:dP0axiodil}) for completeness. A similar computation can be done for gauge theories arising from more complicated geometries.

To establish this result, we consider the $\SO(N-4+2 k) \times \SU(N+2 k)$ theory and switch on a mesonic vev, removing $k$ D3 branes from the orientifold plane and breaking the gauge group down to $\SO(N-4) \times \SU(N) \times \U(k)$, where the last factor corresponds to the $\mathcal{N}=4$ gauge theory on the $k$ D3 branes. The holomorphic gauge coupling of $ \U(k)$ is therefore equal to the ten-dimensional axio-dilaton, and by performing scale matching at each step of the computation, we can relate it to the couplings of the $\SO(N-4) \times \SU(N)$ theory, giving~(\ref{eqn:dP0axiodil}).

We now sketch the details of this argument. For simplicity, we routinely drop numerical factors throughout the computation, only keeping track of the dependence on the couplings. We aim to turn on a vev which breaks
\begin{equation}
  \SO(N - 4 + 2 k) \times \SU(N + 2 k) \longrightarrow
  \SO(N - 4) \times \SU(N) \times \U(k)\,,
\end{equation}
corresponding to removing D3 branes from the orientifold plane. In particular, a
suitable $B$ vev will break $\SU(N + 2 k) \rightarrow \SU(N)
\times \Sp (2 k)$, whereas an $A$ vev will then break $\SO(N -
4 + 2 k) \times \Sp (2 k) \rightarrow \SO (N - 4) \times \U(k)$, since Higgsing a bifundamental breaks $\SO(2 k) \times \Sp(2 k) \rightarrow \U ( k)$.
For a suitable normalization of the $\U(1)$ component, we have the decomposition
\begin{equation}
  \fund \rightarrow \fund_{\,+ 1} \oplus \ov{\fund}_{\,- 1}\,,
\end{equation}
for both $\SO ( 2 k) \rightarrow \U ( k)$ and $\Sp ( 2 k)
\rightarrow \U ( k)$. Thus, decomposing $A^i$ and $B^i$ into irreps of $\SO(N-4)\times\SU(N)\times\U(k)$, we find
\begin{eqnarray}
  A & \rightarrow& ( \fund, \ov{\fund}, 1) \oplus
  (\fund, 1, \fund_{\,+ 1} \oplus \ov{\fund}_{\,- 1}) \oplus
  (1, \ov{\fund}, \fund_{\,+ 1} \oplus \ov{\fund}_{\,- 1}) \nonumber \\
  &  &
  \oplus \left(1, 1, \symm_{\,+ 2} \oplus \asymm_{\,+2} \oplus \adj_0 \oplus \adj_0
  \oplus \overline{\symm}_{\,- 2} \oplus
  \ov{\asymm}_{\,- 2} \right)\,, \\
  B & \rightarrow & \left(1, \asymm, 1 \right) \oplus (1, \fund, \fund_{\,+ 1} \oplus \ov{\fund}_{\,- 1})
  \oplus \left(1, 1, \asymm_{\,+ 2} \oplus \ov{\asymm}_{\, - 2} \oplus \adj_0 \right)\,,
\end{eqnarray}
for each of the three $\SU ( 3)$ ``flavors'' of each field, where $\adj_0$ denotes the reducible $\U(k)$ adjoint representation of dimension $k^2$, containing both singlet and trace-free irreps.

We choose to turn on a vev for the singlet components of $A^3$ and $B^3$ only, which implies that only components
of these fields can be Higgsed.\footnote{One can show by explicit computation that a vev of this type satisfies the D-term conditions.}
We have broken $3 ( 2 N - 3) k + 5 k^2$ generators, therefore
\begin{equation}
  \frac{3 ( N + 2 k)  ( N + 2 k - 3)}{2} - 3 ( 2 N - 3) k - 5 k^2 =
  \frac{3 N ( N - 3)}{2} + k^2
\end{equation}
chiral superfields remain unHiggsed. The only way to get the correct scaling
with $N$ and $k$ is if the unHiggsed fields are
\begin{equation}
 ( \fund, \ov{\fund}, 1) \oplus \left(1, \asymm, 1 \right) \oplus (1, 1, \adj_0)\,,
\end{equation}
coming from $A^3$, $B^3$, and a linear combination of the two, respectively.
Thus, the matter content just below the Higgsing scale $v$ is precisely:
\begin{equation}
  \begin{array}{c|ccc|ccc}
    \mathrm{origin} & \SO(N - 4) & \SU(N) & \U(k) & \SU(2) & \U(1)_R' & \#\\
    \hline
    A & \fund & \ov\fund & 1 & \fund & 1 + \frac{2}{N}
    & 1\\
    A & \fund & \ov\fund & 1 & 1 & \frac{2}{N} & 1\\
    B & 1 & \asymm & 1 & \fund & 1 - \frac{4}{N} & 1\\
    B & 1 & \asymm & 1 & 1 & - \frac{4}{N} & 1\\
    \hline
    A & \fund & 1 & \fund_{\,+ 1} \oplus \ov{\fund}_{\,- 1} & \fund & 1 & 1\\
    A & 1 & \ov\fund & \fund_{\,+ 1} \oplus \ov{\fund}_{\,- 1} & \fund & 1 & 1\\
    B & 1 & \fund & \fund_{\,+ 1} \oplus \ov{\fund}_{\,- 1} & \fund & 1 & 1\\
    A & 1 & 1 & \symm_{\,+ 2} \oplus \ov{\symm}_{\,- 2} & \fund & 1 & 1\\
    A, B & 1 & 1 & \asymm_{\,+ 2} \oplus \ov{\asymm}_{\,- 2} & \fund & 1 & 2\\
    A^{\times 2}, B & 1 & 1 & \adj_0 & \fund & 1 & 3\\
    A / B & 1 & 1 & \adj_0 & 1 & 0 & 1
  \end{array}
\end{equation}
where the unbroken flavor symmetry is $\SU(2)\times\U(1)_R'$, with
\begin{equation}
  \U ( 1)_R' = \U ( 1)_R + \mathrm{diag}_{_{\SU ( 3)}} \left(
  \frac{1}{3}, \frac{1}{3}, - \frac{2}{3} \right) 
   + \mathrm{diag}_{_{\SU (
  N + 2 k)}} \left( \frac{2}{N + 2 k}, \ldots, - \frac{4 k}{N ( N + 2 k)},
  \ldots \right) \,.
\end{equation}
Note that, due to the unbroken global symmetries, the chiral superfields above and below the line cannot couple to each other at the renormalizable level.

One can check that the $\U(1) \subset \U (k)$ charged fields all receive masses at the
scale $\lambda v$ from the superpotential which descends from $\lambda A A B$, as do two of the three $\adj_0$ $\SU(2)$ doublets, leaving
\begin{equation}
  \begin{array}{c|ccc|cc}
    & \SO (N - 4) & \SU(N) & \U(k) & \SU(3) & \U(1)_R\\
    \hline
    A^i & \fund & \ov\fund & 1 & \fund & \frac{2}{3} + \frac{2}{N}\\
    B^i & 1 & \asymm & 1 & \fund & \frac{2}{3} - \frac{4}{N}\\
    \hline
    \Phi^i & 1 & 1 & \adj_0 & \fund & \frac{2}{3}
  \end{array}
\end{equation}
where we can now formally restore $\SU(3) \times \U(1)_R$ invariance,
and the superpotential now takes the form:
\begin{equation}
  W \sim \lambda\, \varepsilon_{i j k} \delta^{a b} A^i_{a ; m}
  A^j_{b ; n} B^{k ; m n} + \lambda\, \varepsilon_{i j k} \mathrm{Tr} [
  \Phi^i \Phi^j \Phi^k]\,,
\end{equation}
where the vev $\langle
\Phi^i \rangle = v^i \id$ breaks $\SU ( 3) \times \U ( 1)_R
\rightarrow \SU ( 2) \times \U ( 1)_R'$, but decouples from the other fields.
The $\U(k)$ gauge group factor decouples from the rest of the theory and flows to an $\mathcal{N}=4$ superconformal fixed point in the infrared. To make the enhanced supersymmetry manifest, we rescale $\Phi \to \lambda^{-1/3} \Phi$, setting the superpotential coupling to $1$ (up to a numerical factor) in the holomorphic basis.

To determine the gauge couplings of the low energy theory, we compute the beta function coefficients $b = 3
T_{\adj} - T_{\mathrm{mat}}$ above, between, and below the scales $v$ and
$\lambda v$ and apply the scale matching relations. Above or below both scales, we have:
\begin{equation}
b_{\SO} = - 18 \,, \quad b_{\SU} = 9\,,
\end{equation}
whereas between the two scales we find:
\begin{equation}
  b_{\SO}' = - 18 - 4 k \,, \quad  b_{\SU}' = 9 - 4 k\,.
\end{equation}
In either case, we have the scale matching relations
\begin{equation}
  \left( \frac{\Lambda}{v} \right)^b = \left( \frac{\Lambda'}{v}
  \right)^{b'} \,,\quad
  \left( \frac{\Lambda'}{\lambda v} \right)^{b'} = \left(
  \frac{\Lambda''}{\lambda v} \right)^b\,,
\end{equation}
so that
\begin{equation}
(\Lambda'')^b = \lambda^{b - b'} \Lambda^b\,.
\end{equation}
Thus, in net
\begin{equation} \label{eqn:SOSUmatching}
  \Lambda_{\SU(N)}^9 = \lambda^{4 k} \Lambda_{\SU(N + 2 k)}^9 \,, \quad \Lambda_{\SO(N-4)}^{- 18} = \lambda^{4 k} \Lambda_{\SO(N -4+ 2 k)}^{- 18} \,.
\end{equation}

Now consider the $\SU(k) \subset \U(k)$ factor.\footnote{We ignore the $\U(1) \subset \U(k)$ henceforward for
simplicity.} We have
\begin{equation}
  b_{\SU(k)} = - (6 ( N - 2) + 10 k)\,,
\end{equation}
between the scales $v$ and $\lambda v$, whereas scale matching at the scale
$v$ gives:
\begin{equation}
  \left( \frac{\Lambda_{\SU(k)}}{v} \right)^{- 6 ( N - 2) - 10 k} = \left( \frac{\Lambda_{\SO(N -4 + 2 k)}}{v} \right)^{- 18} \left( \frac{\Lambda_{\SU(N + 2 k)}}{v} \right)^{18}\,,
\end{equation}
since the index of embedding~\cite{Csaki:1998vv} for $\SU(k) \subset \SO(2 k)$ is $1$ whereas it is $2$ for $\SU(k) \subset \Sp (2 k) \subset \SU(2 k)$. Evaluating the holomorphic gauge coupling at the scale $\lambda v$, we obtain
\begin{align}
  \tau_{k, N} &= \frac{1}{2 \pi i} \ln \left( \frac{\Lambda_{\SU(k)}}{\lambda v}
  \right)^{- ( 6 ( N - 2) + 10 k)} \nn \\ &=  \frac{1}{2 \pi i} \ln \left[
  \lambda^{6 ( N - 2) + 10 k} \Lambda_{\SO(N -4 + 2 k)}^{- 18} \Lambda_{\SU(N + 2 k)}^{18}\right]\,,
\end{align}
which can be rewritten as
\begin{equation}
  \tau_{k , N} = \frac{1}{2 \pi i} \ln \left[ \lambda^{6 ( N - 2) - 2 k} \Lambda_{\SO(N-4)}^{- 18} \Lambda_{\SU(N)}^{18}\right]\,,
\end{equation}
using~(\ref{eqn:SOSUmatching}). Due to the vanishing of the beta function coefficient, the holomorphic gauge coupling does not run below the scale $\lambda v$. However, rescaling $\Phi^i$ to make $\mathcal{N}=4$ supersymmetry manifest alters $\tau$ due to a rescaling anomaly. We find:
\begin{equation} \label{eqn:tauresultSO}
  \hat{\tau} = \tau_{k, N} + \frac{1}{2 \pi i} \ln \lambda^{2 k} =
  \frac{1}{2 \pi i} \ln \left[ \lambda^{6 ( N - 2)} \Lambda_{\SO(N-4)}^{- 18} \Lambda_{\SU(N)}^{18}\right]\,.
\end{equation}
Note that the dependence on $k$ disappears. Moreover,~(\ref{eqn:tauresultSO}) is also independent of $N$,
which can be verified by applying~(\ref{eqn:SOSUmatching}).

Since the holomorphic gauge coupling on D3 branes probing a smooth background is just $\axiodil$ evaluated in that background, we interpret~(\ref{eqn:tauresultSO}) as the ten-dimensional axio-dilaton. Note that the result is independent of $v$, as expected from the constant axio-dilaton profile of the dual geometry at large $N$.

The computation for the $\Sp(\tN+4)\times\SU(\tN)$ theory is closely analogous, and we obtain the result
\begin{equation} \label{eqn:tauresultSp}
  \axiodil= \frac{1}{2 \pi i} \ln \left[\tilde{\lambda}^{6 ( \tN + 2)} \tilde{\Lambda}_{\Sp(\tN+4)}^{18} \tilde{\Lambda}_{\SU(\tN)}^{-18}\right]
\end{equation}
in place of~(\ref{eqn:tauresultSO}). However, at this point an important subtlety arises, since the factor inside the log is a perfect square. This can be rewritten as
\begin{equation} \label{eqn:tauresultSp2}
  \axiodil= \frac{1}{\pi i} \ln \left[\tilde{\lambda}^{3 ( \tN + 2)} \tilde{\Lambda}_{\Sp(\tN+4)}^{9} \tilde{\Lambda}_{\SU(\tN)}^{-9}\right]\,,
\end{equation}
but there is an ambiguity, since
\begin{equation} \label{eqn:tauresultSp3}
  \axiodil= \frac{1}{\pi i} \ln \left[\tilde{\lambda}^{3 ( \tN + 2)} \tilde{\Lambda}_{\Sp(\tN+4)}^{9} \tilde{\Lambda}_{\SU(\tN)}^{-9}\right] + 1
\end{equation}
is also consistent with~(\ref{eqn:tauresultSp}), depending on which sign we take for the square root. The resolution to this puzzle is that the two answers correspond to different types of O-planes, much like the distinction between O3$^+$ and $\widetilde{\rm O3}^+$ planes in the $\mathcal{N}=4$ examples discussed in~\S\ref{sec:MOduality}.

\section{Details of the superconformal index for \alt{$N=7$}{N=7}}
\label{app:SCIdetails}

In this appendix we discuss some technical details of the computation of the superconformal index for the $\SO(3)\times \SU(7) \leftrightarrow \Sp(8)\times \SU(4)$ dual pair. In~\S\ref{sec:B21} we present some technical details related to the calculation of the $\SU(3)$ representation of $\left(B_{(0)}^i\right)^{21}$ for the $\SO(3)\times \SU(7)$ theory whereas in~\S\ref{sec:SCIforSp} we present the rather lengthy results related to the calculation of the superconformal index for the $\Sp(8)\times \SU(4)$ theory (cf.~\S\ref{sec:SCI-t-expansion}).

\subsection{A note on computing \alt{$\left(B_{(0)}^i\right)^{21}$}{(B_0^i)^{21}} efficiently}\label{sec:B21}

In the simplest cases, the representation under the flavor group of the
gauge singlets contributing to the superconformal index can be
computed straightforwardly using a computer algebra program such as
\texttt{LiE} \cite{LiE}. However, the computation becomes more and more expensive as one
studies larger and larger baryons, and already for
$\left(B^i_{(0)}\right)^{21}$ direct computation becomes
intractable. One can then use a different and more efficient method,
which we now explain.

The first observation is that $B$ lives in a tensor product
representation $E\otimes F$ of $\SU(7)\times \SU(3)$. The $m$-th
symmetric tensor product (in our case $m=21$) representation of a
tensor product decomposes as \cite{Macdonald,Weyman}:
\begin{align}
  \Sym^m(E\otimes F) = \sum_{|\lambda|=m} L_\lambda E \otimes L_\lambda
  F \, .
\end{align}
Here we are summing over all partitions $\lambda$ of $m$ (i.e. all
standard Young tableaux with $m$ boxes), and $L_{\lambda}$ is the
Schur functor for $\lambda$. This expression already provides an
important simplification of the calculation, since $F$ is the
fundamental of $\SU(3)$, and thus $L_\lambda F$ is just the $\SU(3)$
representation described by the Young tableau $\lambda$. If $\lambda$
has more than 3 rows this terms vanishes, and we can ignore it in the
sum.

We are left with computing the number of singlets in $L_\lambda
E=L_\lambda (\bigwedge^2 f)$, with $f$ the fundamental representation
of $\SU(7)$. This can be done from general properties of plethysms. In
particular, denoting by $\mu$ the $1+1$ partition of 2 corresponding
to the antisymmetric $\bigwedge^2$, we can apply the formula
\cite{JaKe}:\footnote{One could alternatively use the formula in
  example I.8.9 of \cite{Macdonald}, in terms of generalized Kostka
  numbers. See also \cite{DoranI,DoranII} for similar formulas, and
  appendix~\ref{app:Specht} for a more analytic approach to the
  problem based on the discussion in \cite{DoranIII,DoranI,DoranII}.}
\begin{align}
  \label{eq:plethysm-formula}
  L_\lambda L_\mu = \frac{1}{m!}\sum_{|\kappa|=m} C({\kappa}) \,
  \chi^\lambda_\kappa \, \bigotimes_{i=1}^{\ell(\kappa)}
  \sA_{\kappa_i}(\mu)\, .
\end{align}
Here $C(\kappa)$ denotes the order of the elements of cycle class
$\kappa$ in the symmetric group $S_{|\lambda|}$, $\chi^\lambda_\kappa$
is the character $\chi^\lambda$ of elements of cycle type $\kappa$
evaluated in the representation of $S_{|\lambda|}$ associated to
$\lambda$, and $\ell(\kappa)$ is the number of parts (rows) of the
partition $\kappa$. This formula follows from well known facts, let us
give a quick proof. It is convenient to switch to the representation
in terms of symmetric polynomials \cite{Macdonald}, in which the left
hand side of~\eqref{eq:plethysm-formula} is given by $s_\lambda\circ
s_\mu$, with ``$\circ$'' is the plethysm operator, and $s_\lambda$ and
$s_\mu$ are the symmetric Schur functions indexed by the partitions
$\lambda$ and $\mu$ respectively. Decomposing $s_\lambda$ in terms of
power symmetric functions $p_\kappa$ indexed by the partition $\kappa$
we have \cite{Macdonald}:
\begin{align}
  s_\lambda = \frac{1}{m!} \sum_{|\kappa|=m} C(\kappa) \,
  \chi^\lambda_\kappa \, p_\kappa\, .
\end{align}
Formula \eqref{eq:plethysm-formula} now follows using $p_\kappa =
\prod_{i=1}^{\ell(\kappa)}p_{\kappa_i}$, the fact that $(ab)\circ c =
(a\circ c) (b\circ c)$, and the definition of plethysm with a
fundamental power symmetric polynomial: $p_n\circ \mu(x) = \mu(x^n)$.

The second simplification in the calculation now comes from observing
that the tensor product of Adams operators appearing in this formula
is actually independent of $\lambda$. It also happens to be the most
expensive part of the computation, so it just needs to be calculated
once. Making this manifest, the final formula we computed is
effectively:
\begin{align}
  \Sym^{m}(E\otimes F) = \frac{1}{m!}\sum_{|\kappa|=m} C({\kappa}) \,
  \left(\,\sum_{|\lambda|=m} \chi^\lambda_\kappa \, L_\lambda F\right)
  \left\langle \bigotimes_i^{\ell(\kappa)}
    \sA_{\kappa_i}(\mu)\right\rangle \, ,
\end{align}
where the brackets indicate taking the singlet part only.

\subsection{Check of the superconformal index calculation for \alt{$\Sp(8)\times \SU(4)$}{Sp(8)x SU(4)}}\label{sec:SCIforSp}
In this section we present some rather lengthy results related to the calculation of the superconformal index for the $\Sp(8)\times \SU(4)$ theory (cf.\~\S\ref{sec:SCI-t-expansion}).

The fields that contribute to the superconformal index for the $\Sp(\tilde{N}+4)\times \SU(\tilde{N})$ theory are shown in table~\ref{tab:SpNindexfields}.
\begin{table}
\begin{center}
\begin{tabular}{|c|c|c|c|c|}
  \hline
  Field & $\Sp(\tilde{N}+4)\times \SU(\tilde{N})$ & $\SU(3)$ & $t$ exponent & $\SU(2)_r$\\\hline
  $\tilde{A}^i_{(l)}$ & $(\fund,\ov{\fund})$ & $\fund$ & $\frac{2}{3}-\frac{2}{\tilde{N}}+l$ & $l+1$ \\\hline
  $\tilde{B}^i_{(l)}$ & $\lp1,\symm\rp$ & $\fund$ & $\frac{2}{3}+\frac{4}{\tilde{N}}+l$ & $l+1$ \\\hline
  $\bar{\psi}^{\tilde{A}}_{(l)}$ & $(\fund,\fund)$ & $\ov{\fund}$ & $\frac{4}{3}+\frac{2}{\tilde{N}}+l$ & $l+1$ \\\hline
  $\bar{\psi}^{\tilde{B}}_{(l)}$ &  $\lp1,\ov\symm\rp$ & $\ov{\fund}$ & $\frac{4}{3}-\frac{4}{\tilde{N}}+l$ & $l+1$ \\\hline
  $\lambda_{(l)}^{\Sp}$ & $(\symm,1)$ & 1 &$1+l$ & $l \oplus (l+2)$\\\hline
  $F_{(l)}^{\Sp}$ & $(\symm,1)$ & 1 &$2+ l$ & $(l+1) \oplus (l+1)$\\\hline
  $\lambda_{(l)}^{\SU}$ & $(1,\adj)$ & 1 &$1+l$ & $l \oplus (l+2)$\\\hline
  $F_{(l)}^{\SU}$ & $(1,\adj)$ & 1 &$2+ l$ & $(l+1) \oplus (l+1)$\\\hline
\end{tabular}
\end{center}
\caption[The fields which contribute to the SCI for $\Sp(\tilde{N}+4)\times \SU(\tilde{N})$]{The fields which contribute to the superconformal index for $\Sp(\tilde{N}+4)\times \SU(\tilde{N})$, where the $\SU(2)_r$ column denotes the representation under the $\SU(2)$ group generated by $\ov J_\pm, \ov J_3$.\label{tab:SpNindexfields}}
\end{table}
The gauge invariant contributions for the $\Sp(8)\times \SU(4)$ theory up to order $t^2$ are shown in tables~\ref{tab:Sp8indexinvariants1} and~\ref{tab:Sp8indexinvariants2}.
\begin{table}
\begin{center}
\begin{tabular}{|c|c|c|c|}
  \hline
  operator & \!$t$ ex.\! & $2 \bar{J}_3$ & $\SU(3)$ character\\\hline
  $(\tilde{A}_{(0)})^{4}$ & $\frac23$ & 0 & $\chi_{0,2}+\chi_{4,0}$ \\[1pt]\hline
  $(\tilde{A}_{(0)})^{8}$ & $\frac43$ & 0 & $5\chi_{0,4}+2\chi_{1,2}+5\chi_{2,0}+3\chi_{3,1}+3\chi_{4,2}+\chi_{8,0}$ \\[1pt]\hline
  $(\tilde{A}_{(0)})^{6} \psi_{(0)}^{\tilde{B}}$ & $\frac43$ & 0 & $\chi_{0,1} + 3\chi_{0,4} + 4\chi_{1,2} + 3\chi_{2,0}+ 3\chi_{3,1} + \chi_{4,2} + \chi_{5,0}$ \\[1pt]\hline
  $(\tilde{A}_{(0)})^{4} [\psi_{(0)}^{\tilde{B}}]^2$ & $\frac43$ & 0 & $2 \chi_{0,1} + 2\chi_{1,2} + \chi_{3,1} +\chi_{5,0}$ \\[1pt]\hline
  $(\tilde{A}_{(0)})^{2} [\psi_{(0)}^{\tilde{B}}]^3$ & $\frac43$ & 0 & $\chi_{0,1} $ \\[1pt]\hline
  $(\tilde{A}_{(0)})^{3} \tilde{A}_{(1)}$ & $\frac53$ & $\pm1$ & $\chi_{0,2} + \chi_{1,0} + 2 \chi_{2,1} + \chi_{4,0}$ \\[1pt]\hline
  $\lambda_{(0)}^{\Sp}(\tilde{A}_{(0)})^{4}$ & $\frac53$ & $\pm1$ & $\chi_{1,0} + \chi_{2,1}$ \\[1pt]\hline
  $\lambda_{(0)}^{\SU}(\tilde{A}_{(0)})^{4}$ & $\frac53$ & $\pm1$ & $\chi_{1,0} + 2\chi_{2,1}$ \\[1pt]\hline
  $\lambda_{(0)}^{\SU} \psi_{(0)}^{\tilde{B}} (\tilde{A}_{(0)})^{2}$ & $\frac53$ & $\pm1$ & $\chi_{1,0} + \chi_{2,1}$ \\[1pt]\hline
\end{tabular}
\end{center}
\caption[Gauge invariant contributions to the $\Sp(8)\times \SU(4)$ SCI, part I]{Gauge invariant contributions to the superconformal index for $\Sp(8)\times \SU(4)$ of order less than $t^2$, where $()^*$ denotes the symmetric tensor product and $[]^*$ the antisymmetric tensor product.\label{tab:Sp8indexinvariants1}}
\end{table}
Taking into account the factor $(-1)^F$ we find perfect agreement with \eqref{eq:SCIforN=7}.
\begin{table}
\begin{center}
\begin{tabular}{|c|c|c|c|}
\hline
  operator & \!$t$ ex.\! & $2 \bar{J}_3$ & $\SU(3)$ character\\\hline
  \multirow{3}{*}{$(\tilde{A}_{(0)})^{12}$} & \multirow{3}{*}{$2$} & \multirow{3}{*}{0} & $16+8 \chi_{0,3} + 8\chi_{0,6} + 22\chi_{1,1} +13\chi_{1,4}+42\chi_{2,2}$ \\*
  & & & $+\chi_{2,5}+12\chi_{3,0} +19 \chi_{3,3} +20 \chi_{4,1}+8\chi_{4,4}$\\*
  & & & $ + 8\chi_{5,2} + 15 \chi_{6,0}+ 2\chi_{6,3} + 4 \chi_{7,1}+3 \chi_{8,2}+\chi_{12,0}$\\\hline
  \multirow{3}{*}{$(\tilde{A}_{(0)})^{10} \psi_{(0)}^{\tilde{B}}$} & \multirow{3}{*}{$2$} & \multirow{3}{*}{0} & $16+28 \chi_{0,3} + 5\chi_{0,6} + 54\chi_{1,1} +27\chi_{1,4}+68\chi_{2,2}$ \\*
  & & & $+4\chi_{2,5}+37\chi_{3,0} +32 \chi_{3,3}+41 \chi_{4,1}+6\chi_{4,4}  $\\*
  & & & $+ 17\chi_{5,2} + 14 \chi_{6,0} + \chi_{6,3} + 6 \chi_{7,1}+ \chi_{8,2}+\chi_{9,0}$\\\hline
  \multirow{3}{*}{$(\tilde{A}_{(0)})^{8} [\psi_{(0)}^{\tilde{B}}]^2$} & \multirow{3}{*}{$2$} & \multirow{3}{*}{0} & $6+35 \chi_{0,3} + 52\chi_{1,1} +21\chi_{1,4}+46\chi_{2,2}+5\chi_{2,5}$ \\*
  & & & $+43\chi_{3,0} +22 \chi_{3,3}+35 \chi_{4,1}+\chi_{4,4} + 13\chi_{5,2} + 4 \chi_{6,0} $\\*
  & & & $+ \chi_{6,3} + 3 \chi_{7,1}+\chi_{9,0}$\\\hline
  \multirow{2}{*}{$(\tilde{A}_{(0)})^{6} [\psi_{(0)}^{\tilde{B}}]^3$} & \multirow{2}{*}{$2$} & \multirow{2}{*}{0} & $6+18 \chi_{0,3} + 26\chi_{1,1} +7\chi_{1,4}+22\chi_{2,2}+2\chi_{2,5}$ \\*
  & & & $+21\chi_{3,0} +7 \chi_{3,3}+14 \chi_{4,1}+3\chi_{5,2} + 2 \chi_{6,0} + \chi_{6,3}$\\\hline
  \multirow{2}{*}{$(\tilde{A}_{(0)})^{4} [\psi_{(0)}^{\tilde{B}}]^4$ } & \multirow{2}{*}{$2$} & \multirow{2}{*}{0} & $4+3 \chi_{0,3} + 9\chi_{1,1} +\chi_{1,4}+9\chi_{2,2}$ \\*
  & & & $+3\chi_{3,0} +\chi_{3,3} +2\chi_{4,1}+ \chi_{6,0} $\\\hline
  $(\tilde{A}_{(0)})^{2} [\psi_{(0)}^{\tilde{B}}]^5$ & $2$ & 0 & $1+2\chi_{1,1} +2\chi_{2,2}+\chi_{3,0}$ \\[1pt]\hline
  $[\psi_{(0)}^{\tilde{B}}]^6$ & $2$ & 0 & $\chi_{3,0}$ \\[1pt]\hline
  $[\tilde{\lambda}^{\Sp}_{(0)}]^2$ & $2$ & 0 & 1 \\[1pt]\hline
  $[\tilde{\lambda}^{\SU}_{(0)}]^2$ & $2$ & 0 & 1 \\[1pt]\hline
  $A_{(0)} \psi^A_{(0)}$ & $2$ & 0 & $1+\chi_{1,1}$ \\[1pt]\hline
  $B_{(0)} \psi^B_{(0)}$ & $2$ & 0 & $1+\chi_{1,1}$  \\[1pt]\hline
  $(A_{(0)})^2 B_{(0)}$ & $2$ & 0 & $1+\chi_{1,1}$  \\[1pt]\hline
\end{tabular}
\end{center}
\caption[Gauge invariant contributions to the $\Sp(8)\times \SU(4)$ SCI, part II]{Gauge invariant contributions to the superconformal index for $\Sp(8)\times \SU(4)$ of order $t^2$.\label{tab:Sp8indexinvariants2}}
\end{table}

\section{On the decomposition of certain generalized Specht modules}
\label{app:Specht}

One of the arguments presented in~\S\ref{sec:C3Z3fieldtheory}
for the agreement between the two dual theories relied on the matching
of the flavor representation of baryons with minimal $R$-charge
between the two descriptions of the theory. In particular, we could
argue that for all values of $N$ the baryon $\tilde A^{N-3}$ in the
$\Sp(N+1)\times\SU(N-3)$ theory transforms in the $\Sym^{(N-3)/2}(\symm)$
representation of the $SU(3)$ flavor group, where $\Sym^{k}(R)$ denotes
the $k$-th symmetric power of the representation $R$. We also argued,
and checked in a number of examples, that there is a corresponding
minimal $R$-charge baryon of the form $B^N$ on the
$\SO(N-4)\times\SU(N)$ side, transforming in the same representation
of the flavor group. The duality conjectured in this paper then
requires the group theoretical identity\footnote{In this appendix, as
in the rest of the paper, we will be assuming that $N$ is odd.}
\begin{align}
  \label{eq:A-B-group-identity}
  \left\langle \Sym^N\left(\asymm_{\SU(N)}\otimes
      \fund_{\SU(3)}\right)\right\rangle \cong \Sym^{(N-3)/2}(\symm)
\end{align}
to hold (as representations of the flavor $SU(3)$), where the angle
brackets denote taking the singlet part under $SU(N)$. In this
appendix we would like to demystify this expression somewhat by
reformulating it as an statement about representations of the
symmetric group $S_N$, and give some additional evidence for its
validity based on this new viewpoint. The interested reader can find
nice reviews of the required introductory material in
\cite{James,Macdonald,JaKe}. We will also make use of the generalized
Specht modules introduced by Doran in \cite{DoranIII} (see also
\cite{DoranI,DoranII}). We will show that in this
context~\eqref{eq:A-B-group-identity} follows from a conjectured
decomposition of certain generalized Specht module into ordinary
Specht modules.

We start by using the decomposition of the symmetric power of a tensor
product into a sum of ordinary tensor products
\cite{Macdonald,Weyman}:
\begin{align}
  \Sym^N(E\otimes F) = \sum_{|\lambda|=N}L_\lambda E \otimes L_\lambda F \, ,
\end{align}
where $L_\lambda$ is the Schur functor for the partition $\lambda$,
and the sum is over partitions of $N$. In our particular case we have
$E=\asymm_{\SU(N)}$ and $F=\fund_{\SU(3)}$. Taking the $SU(N)$ singlet
part:
\begin{align}
  \label{eq:sym-tensor-decomposition}
  \left\langle \Sym^N(E\otimes F) \right\rangle =
  \sum_{|\lambda|=N} \left\langle L_\lambda E\right\rangle \,
  L_\lambda F\, .
\end{align}
We thus see that we are left to enumerate the $\lambda$ for which
$L_\lambda\,\asymm$ contains $SU(N)$ singlets. As in
appendix~\ref{app:SCIdetails}, in order to do this it is convenient to
work with symmetric polynomials \cite{Macdonald} instead of directly
representations, so we rewrite~\eqref{eq:sym-tensor-decomposition} as:
\begin{align}
  \left\langle \Sym^N(E\otimes F) \right\rangle \cong \sum_{|\lambda|=N}
  \left\langle s_\lambda \circ e_2\right\rangle \, s_\lambda\, ,
\end{align}
where ``$\circ$'' denotes plethysm, $s_\lambda$ is the Schur symmetric
function indexed by the partition $\lambda$, and $e_2$ is the
elementary symmetric function or order 2 associated with the
antisymmetric. Expanding $s_\lambda$ into power symmetric polynomials
$p_\rho$:
\begin{align}
  \label{eq:sym-tensor-Schur}
  \left\langle \Sym^N(E\otimes F) \right\rangle \cong \frac{1}{N!}
  \sum_{|\lambda|=N}\sum_{|\rho|=N} C(\rho) \chi^\lambda_\rho
  \chi\left\langle p_\rho\circ e_2\right\rangle \, s_\lambda \, ,
\end{align}
where $\rho$ is a partition of $N$, and as
in~\eqref{eq:plethysm-formula} we have introduced the order $C(\rho)$
of the cycle class $\rho$ in $S_{|\rho|}$, and the character
$\chi^\lambda_\rho$ of elements of cycle type $\rho$ in the
representation indexed by $\lambda$. We can now use
\cite{GayCharacters,DoranI,DoranII}:
\begin{align}
  \label{eq:Doran-formula-e2}
  p_\rho\circ e_2 = \sum_{|\kappa|=2N} \chi^{\kappa',N}_\rho s_\kappa\,
  ,
\end{align}
where $\kappa$ runs over partitions of $2N$, and $\kappa'$ denotes the
transpose of $\kappa$. $\chi^{\kappa',N}_\rho$ is the character of
cycles of type $\rho$ in the generalized Specht module $S^{\kappa',N}$
\cite{DoranIII}, which we will describe further momentarily. (Notice
that the formula given in \cite{DoranI,DoranII} acts on $h_2$ rather
than $e_2$, but we can easily obtain \eqref{eq:Doran-formula-e2} by acting
with the involution $\omega$ exchanging $e_2$ and $h_2$
\cite{Macdonald}, which gives the transpose of $\kappa$.) There is a
single partition of $2N$ giving rise to a gauge singlet of $SU(N)$, it
is the partition $2N = 2 + 2 + \ldots \equiv 2^N$ (in standard
notation for partitions). Taking into account that the transpose
partition of $2^N$ is just $N^2$, we finally get:
\begin{align}
  \left\langle p_\rho\circ e_2\right\rangle =
   \chi^{N^2,N}_\rho\, .
\end{align}
Now, the generalized Specht module $S^{N^2,N}$ is a (in general
reducible) representation of the permutation group $S_N$, so let us
write $S^{N^2,N} \cong \bigoplus_\mu c_\mu S^\mu$ for its
decomposition into irreducible representations of $S_n$, the Specht
modules $S^\mu$, indexed by the partitions $\mu$ of $N$. Using
linearity of characters, we find that
\begin{align}
  \left\langle p_\rho\circ e_2\right\rangle = \sum_\mu c_\mu
  \chi^\mu_\rho \, .
\end{align}
Plugging this back in \eqref{eq:sym-tensor-Schur}, we obtain
\begin{align}
  \begin{split}
    \left\langle \Sym^N(E\otimes F) \right\rangle & \cong \sum_{|\mu|=N}
    c_\mu \sum_{|\lambda|=N} \left(\frac{1}{N!} \sum_{|\rho|=N} C(\rho)
      \chi^\lambda_\rho \chi^\mu_\rho\right) \, s_\lambda\\
    & = \sum_{|\mu|=N} c_\mu s_\mu\, ,
  \end{split}
\end{align}
where we have used orthogonality of characters to set the term in
parenthesis to $\delta_{\lambda\mu}$. We thus find the remarkably
simple result that the flavor representation of our baryon is just the
$\SU(3)$ representation associated to the generalized Specht module
$S^{N^2,N}$.\footnote{To each ordinary Specht module $S^\lambda$ we
  can associate in the usual way the $\SU(3)$ representation with
  Young tableau $\lambda$. Since $S^{N^2,N}$ is a sum of ordinary
  Specht modules we associate to it the corresponding sum of $\SU(3)$
  representations.} Furthermore we conjecture that the following
decomposition holds for all $N$:
\begin{align}
  \label{eq:conjectured-decomposition}
  S^{N^2,N} \cong \bigoplus_{k} S^\kappa\,,
\end{align}
where $\kappa$ runs over the three element partitions $\kappa_1 +
\kappa_2 + \kappa_3$ of $N$ such that all $\kappa_i$ are odd
numbers. Before presenting the evidence that we have found for this
conjecture, let us show that this
implies~\eqref{eq:A-B-group-identity}. The right hand side is given by
\cite{Macdonald,Weyman}:
\begin{align}
  \label{eq:sym-tensor-RHS-2}
  \Sym^{(N-3)/2}(\symm) \cong \bigoplus_{|\lambda|=N-3} S^{\lambda} \,,
\end{align}
where the parts of the partition $\lambda$ are all even numbers. Using
the fact that we are interested in representations of $SU(3)$, we can
restrict the sum to partitions with 3 parts at most (the rest vanish
as $SU(3)$ representations). We also notice that since $\kappa_i\in
2\bZ+1$ and $\kappa_i\geq 0$, we have $\kappa_i\geq 1$. Removing a
column of three boxes on the leftmost column of a Young tableau gives
rise to $SU(3)$ isomorphic representations, so we may just as well
send $\kappa_i\to \tilde\kappa_i = \kappa_i-1$, where now
$\tilde\kappa$ is a partition of $N-3$ with all parts even. We have
thus just obtained a natural isomorphism between
\eqref{eq:conjectured-decomposition} and \eqref{eq:sym-tensor-RHS-2}
as $\SU(3)$ representations, as we wanted.

Coming back to our conjecture~\eqref{eq:conjectured-decomposition}, we
have found various pieces of evidence for its validity. First of all,
direct computation (using \texttt{LiE} \cite{LiE}) shows that the identity
holds for all odd $N$ between 3 and 21. More conceptually, it is
possible to show (by using a straightforward modification of the
straightening procedure based on Garnir elements, for example) that
$S^{N^2,N}$ has a basis indexed by the semistandard tableaux of shape
$N^2$ and weight $2^N$. On the other hand it is well known that any
ordinary Specht module $S^\mu$ has a basis indexed by standard
tableaux of shape $\lambda$. So in order for the dimensions of the
corresponding modules to match it should hold that the number of
semistandard tableaux of shape $N^2$ and weight $2^N$ should be the
sum of the number of standard tableaux with shapes as in
formula~\eqref{eq:conjectured-decomposition}. This enumeration task is
well suited to a computer (we used SAGE \cite{sage}), and by direct
computation it is easy to see that the dimensions match up to $N=45$.

\section{A conjectured identity for elliptic hypergeometric integrals}
\label{sec:SCI-identity}

In this appendix we will reformulate the conjecture \eqref{eq:SCI-equality}, $\cI_{\Sp} = \cI_{\SO}$, in
terms of elliptic hypergeometric integrals,\footnote{We refer the reader to
  \cite{Spiridonov1,Spiridonov2} for the original works on
  hypergeometric integrals, and to \cite{Spiridonov3} for a nice
  review of the field.} giving rise to a
conjecture about elliptic hypergeometric functions that could perhaps
be proven along the lines of \cite{Rains} (we will not attempt to
prove it in this paper). One point of mathematical interest is that,
since the physical process behind our conjectured duality seems to be
qualitatively different from ordinary Seiberg duality (this is
particularly clear when formulated in string theory \cite{transitions2}),
one may expect that~\eqref{eq:SCI-equality} is a new fundamental
identity between elliptic hypergeometric functions, independent from
the one proven by Rains~\cite{Rains}.

It is by now an standard exercise to reformulate the superconformal
index in terms of elliptic hypergeometric functions (following
\cite{Dolan:2008qi}) so we will be somewhat brief. Let us start on the
$\Sp\times\SU$ side, which we will parametrize as
$\Sp(2M)\times\SU(L)$ (so one has $2M\equiv N+1$, $L\equiv N-3$,
assuming that the dual theory was $\SO(N-4)\times\SU(N)$). The
index~\eqref{eq:SCI-index} factorizes into:
\begin{align}
  \label{eq:SCI-USp-index}
  \cI_{\Sp}(t,x,f) = \int_{\Sp} \!\!\!\! [dz_1] \int_{\SU}
  \!\!  [dz_2]\,\,\, \cI_{\ov\sfund}(t,x,z_1,z_2,f)\,
  \cI_{\ssymm}(t,x,z_2,f)\, .
\end{align}
As in \cite{Sudano:2011aa}, we have absorbed the contribution to the
index coming from vector bosons into the integration measure. Explicit
expressions can be found in \cite{Sudano:2011aa}, and we reproduce
them here for the convenience of the reader, adapted to our notation:
\begin{align}
  \int_{\SU(N)}\!\!\!\!\!\!\!\!\! [dz] & \equiv\frac{1}{N!}\oint
  \left(\prod_{a=1}^{N-1}\frac{dz_a}{2\pi
      iz_a}(tx;tx)\left(\frac{t}{x};\frac{t}{x}\right)\right)\frac1{\prod\limits_{1\le
      b<c\le N}\Gamma(z_bz_c^{-1},z_b^{-1}z_c)}\bigg|_{\prod z_a=1}\,,\\
  \int_{\Sp(2N)}\!\!\!\!\!\!\!\!\!\!\!\![dz] & \equiv\frac{1}{N!}\oint
  \left(\prod_{a=1}^N\frac{dz_a}{4\pi
      iz_a}\frac{(tx;tx)\left(\frac{t}{x};\frac{t}{x}\right)}{\Gamma(z_a^2,z_a^{-2})}\right)\frac1{\prod\limits_{1\le
      b<c\le
      N}\Gamma(z_bz_c,z_bz_c^{-1},z_b^{-1}z_c,z_b^{-1}z_c^{-1})}\,,\\
  \int_{\SO(2N+1)}\!\!\!\!\!\!\!\!\!\!\!\!\!\![dz] & \equiv\oint
  \frac{1}{N!}\left(\prod_{a=1}^N\frac{dz_a}{4\pi
      iz_a}\frac{(tx;tx)\left(\frac{t}{x};\frac{t}{x}\right)}{\Gamma(z_a,z_a^{-1})}\right)\frac{1}{\prod\limits_{1\le
      b<c\le
      N}\Gamma(z_bz_c,z_bz_c^{-1},z_b^{-1}z_c,z_b^{-1}z_c^{-1})}\, .
\end{align}
where we have introduced the following standard special
functions:\footnote{It is common in the literature to introduce the
  new variables $p=tx$, $q=tx^{-1}$, and express the integrals in
  terms of these, but we will keep using the $t$ and $x$ variables we
  have been using so far.}
\begin{align}
  \label{eq:SCI-Gamma-series}
  \Gamma(u;t,x) & = \prod_{a,b\ge
    0}\frac{1-u^{-1}t^{a+b+2}x^{a-b}}{1-ut^{a+b}x^{a-b}}\,,\\
  \theta(u;y) & = \prod_{a\geq 0}(1-uy^a)(1-u^{-1}y^{a+1})\,,\\
  (u;y) & = \prod_{a\geq 0} (1-uy^a)\, .
\end{align}
Finally, we have also introduced the short-hand notation
\begin{align}
  \label{eq:SCI-Gamma}
  \Gamma(u) & \equiv \Gamma(u;t,x)\,,\\
  \Gamma(u_1,\ldots,u_k) & \equiv \prod_{i=1}^k \Gamma(u_i)\, .
\end{align}
The ordinary $\Gamma$ function (i.e. the generalization of the
factorial) will play no role in our discussion, so by $\Gamma(u)$ we
will always mean~\eqref{eq:SCI-Gamma}.

We are left to evaluate the contribution $\cI_{\ov\sfund}$ from the
bifundamental, and the contribution $\cI_{\ssymm}$ from the
symmetric. Let us start by $\cI_{\ov\sfund}$. This fields transforms
in the bifundamental of $\Sp(2M)\times \SU(L)$, and accordingly its
one-letter index is given by:
\begin{align}
  \label{eq:SCI-bifund-letter}
  i_{\ov\sfund}(t,x,f,z_1,z_2) = \frac{1}{(1-tx)(1-tx^{-1})}\left[t^r
    \chi_{\ov\sfund} - t^{(2-r)}\chi_{\sfund} \right]\, .
\end{align}
Here we have introduced the total character
$\chi_{\ov\sfund}=\chi_{\sfund}(z_1)\chi_{\ov\sfund}(z_2)\chi_{\sfund}(f)$,
and its conjugate $\chi_{\sfund}$. It is convenient to expand these
characters into elementary monomials:
\begin{align}
  \label{eq:SCI-eta}
  \begin{split}
    \chi_{\ov\sfund} & =
    \chi_{\sfund}(z_1)\chi_{\ov\sfund}(z_2)\chi_{\sfund}(f) \\
    & = \left(\sum_{a=1}^{M} (z_{1,a} +
      z_{1,a}^{-1})\right)\left(\sum_{b=1}^L
      \frac{1}{z_{2,b}}\right)\left(\sum_{c=1}^3 f_c\right)\\
    & = \sum_{a,b,c} \frac{z_{1,a}f_c}{z_{2,b}} +
    \sum_{a,b,c} \frac{z_{1,a}^{-1}f_c}{z_{2,b}}\\
    & \equiv \sum_q\eta_q\, .
  \end{split}
\end{align}
Where $\eta_q$ is a monomial in the expansion, and $q$ an unified
index. The subindices denote projection of the group elements into the
maximal torus, and for $\SU$ characters there are constraints of the
form $\prod z_{i,a}=1$, which we will not indicate explicitly in what
follows. Expanding the denominator in~\eqref{eq:SCI-bifund-letter}, we
have:
\begin{align}
  \begin{split}
    i_{\ov\sfund}(t,x,f,z_1,z_2) = \sum_{a,b\geq 0}\sum_q
    t^{a+b}x^{a-b} (t^r\eta_q - t^{(2-r)}\eta_q^{-1})\, .
  \end{split}
\end{align}
The plethystic exponent $\mathbb{E}_{\ov\sfund}$ in~\eqref{eq:SCI-index} then
becomes:
\begin{align}
  \begin{split}
    \mathbb{E}_{\ov\sfund} & =
    \sum_{k=1}^\infty\frac{1}{k}i_{\ov\sfund}(t^k, x^k,
    z_1^k, z_2^k, f^k) \\
    & = \sum_{k=1}^\infty\frac{1}{k}\sum_{a,b\geq 0}\sum_q
    t^{k(a+b)}x^{k(a-b)} (t^{kr}\eta^k_q - t^{k(2-r)}\eta_q^{-k})\\
    & = \sum_{a,b\geq 0}\sum_q
    \log\left(\frac{1-t^{2+a+b-r}x^{a+b}\eta_q^{-1}}{1-t^{a+b+r}x^{a-b}\eta_q}\right)\,,
  \end{split}
\end{align}
and the superconformal index~\eqref{eq:SCI-index}
\begin{align}
  \begin{split}
    \cI_{\ov\sfund}(t,x,z_1,z_2,f) & = \exp(\mathbb{E}_{\ov\sfund})\\
    & = \prod_q\prod_{a,b\geq
      0}\left(\frac{1-t^{2+a+b-r}x^{a+b}\eta_q^{-1}}{1-t^{a+b+r}x^{a-b}\eta_q}\right)
    \\
    & = \prod_q \Gamma(t^r \eta_q)\, ,
  \end{split}
\end{align}
where we have used the
definition~\eqref{eq:SCI-Gamma-series}. Expanding $\eta_q$ back from
its definition~\eqref{eq:SCI-eta}, we finally obtain:
\begin{align}
  \label{eq:SCI-USp-bifund}
  \cI_{\ov\sfund}(t,x,z_1,z_2,f) =
  \prod_{a=1}^M\prod_{b=1}^L\prod_{c=1}^3\Gamma\left(\frac{z_{1,a}f_c}{z_{2,b}},
    \frac{f_c}{z_{1,a}z_{2,b}}\right)\, .
\end{align}
The $\cI_{\ssymm}$ contribution can be computed similarly, one just
needs the group character for the symmetric of $\SU(N)$:
\begin{align}
  \chi_{\ssymm}(z_2) = \sum_{1\leq i < j \leq N} z_{2,i}z_{2,j} +
  \sum_{i=1}^N z_{z,i}^2\, .
\end{align}
Proceeding as above, we get:
\begin{align}
  \label{eq:SCI-USp-symm}
  \cI_{\ssymm}(t,x,z_2,f) =
  \prod_{c=1}^3\left(\prod_{1\leq i < j \leq L}\Gamma\bigl(t^r z_{2,i}
    z_{2,j}f_c\bigr)\right)
  \cdot \left(\prod_{i=1}^L\Gamma\bigl(t^r z_{2,i}^2 f_c\bigr) \right)\,.
\end{align}
Plugging~\eqref{eq:SCI-USp-bifund} and \eqref{eq:SCI-USp-symm}
into~\eqref{eq:SCI-USp-index} one gets an explicit integral expression
for the index in this phase.

Going to the dual theory, let us parametrize the gauge groups by
$\SO(2M+1)\times \SU(N)$ (we have $M=\frac{N-5}{2}$). The
superconformal index is now given by:
\begin{align}
  \label{eq:SCI-SO-index}
  \cI_{\SO}(t,x,f) = \int_{\SO} \!\!\!\! [dz_1] \int_{\SU}
  \!\!  [dz_2]\,\,\, \cI_{\ov\sfund}(t,x,z_1,z_2,f)\,
  \cI_{\sasymm}(t,x,z_1,z_2,f)\, .
\end{align}
In order to compute the contributions to the index from the
bifundamental and the antisymmetric we need the following group
characters:
\begin{align}
  \chi^{\SO(2M+1)}_{\sfund}(z_1) & = 1 + \sum_{a=1}^{M} (z_{1,a} +
  z_{1,a}^{-1})\,,\\
  \chi^{\SU(N)}_{\sasymm}(z_2) & = \sum_{1\leq i < j \leq N}
  z_{2,i}z_{2,j}\, .
\end{align}
Proceeding as above, one thus gets:
\begin{align}
  \cI_{\ov\sfund}(t,x,z_1,z_2,f) & = \prod_{b=1}^N\prod_{c=1}^3
  \Gamma\bigl(t^r z_{2,b}^{-1} f_c\bigr)
  \prod_{a=1}^M\Gamma\left(\frac{t^r z_{1,a}f_c}{z_{2,b}},\frac{t^r
      f_c}{z_{1,a}z_{2,b}} \right)\, ,\\
  \cI_{\sasymm}(t,x,z_2,f) & =
  \prod_{c=1}^3\left(\prod_{1\leq i < j \leq L}\Gamma\bigl(t^r z_{2,i}
    z_{2,j}f_c\bigr)\right)\, ,
\end{align}
and plugging these expressions into~\eqref{eq:SCI-SO-index} gives the
expression for the superconformal index in terms of elliptic
hypergeometric functions, as desired.

Now that we have the explicit expression in terms of elliptic
hypergeometric functions, one is left to prove the
identity~\eqref{eq:SCI-equality}. Given the physical interpretation of
our duality as a strong/weak duality, a preliminary first step would
be to prove the analogous index identity between $\SO(2N+1)$ and
$\Sp(2N)$ gauge groups in $\cN=4$. To our knowledge a complete proof
has not been found yet, although in \cite{Gadde:2009kb,Spiridonov:2010qv} it has
been shown that the relevant superconformal indices agree in a number
of simplifying limits.

\bibliographystyle{JHEP}
\bibliography{refsv2}

\end{document}